\documentstyle[epsf,psfig,twoside,makeidx]{pthesis}

\makeindex

\begin{document}
 \begin{frontmatter}
    \title{Non-perturbative States in Superstring Theories} 
    \author{Kwan-Leung Chan}
    \supervisor{Mirjam Cveti\v c}
    \graduategroup{Physics and Astronomy}
    \maketitle
    \legalname{Kwan-Leung Chan}
\thispagestyle{plain}
\begin{center}
        \textbf{ ACKNOWLEDGMENTS }
\end{center}

Before acknowledging the academic people, I would like to express my great appreciation and gratitude to my mother-in-law and my sister-in-law, for their continuous spiritual as well as financial support up to this very moment. Their support and encouragement is particularly important in my first year at Penn., when I had just left the wonderful city where I was born and raised, Hong Kong.

Within the academic circle, my dissertation advisor, Professor Mirjam Cveti\v c deserves my first and foremost acknowledgement. She guided me through the beautiful and abstract theoretical constructions of string theorists, helped me appreciate the works of the giants, initiated and inspired me to explore various challenging research directions. Without her talent and patience, this thesis would never have come to fruition.

Special thanks to Professor P. J. Steinhardt and Professor P. Nelson for their sincere and intelligent advices, both on physics and on the methodolgy of pursuing research. I would like to express my appreciation to Professor M. Cohen. His deep insight into seemingly obviouse physical problems and unlimited humor made my first year at Penn. both intellectually enlightening and enjoyable. I also need to say thanks to Professor R. W. Zurmuhle and Professor H. T. Fortune, for allowing me to work for the Experimental Nuclear Physics Group at Tandem at the Spring of 93.

I am obliged to mention the names of the people, who lighted up my desire for the search of the ultimate fundamental physical principles, and set up an intellectual role model when I was very young. They are Professor K. Young of the Chinese University of Hong Kong, Professor P. C. W. Fung of the University of Hong Kong, and Mr. K. C. Yeung of Po Leung Kuk C.F.A. No. 1 College.

My sincere thanks to my wife can hardly be expressed fully in words. She has given up so much - time, talent, a wonderful career at Hong Kong - just for taking care of someone who can never guarantee a bright future for her.

Nice words are never nice enough to express my appreciation to all the people who had been so important to me in the past half decade. 

May the Lord bless each one of them and lead them to a joyful and abundant life. To Him this dissertation is dedicated.
    \begin{abstract}

This thesis is devoted to the study of non-perturbative behavior of string theories. The study of non-perturbative states, that saturate the Bogomol'nyi-Prasad-Sommerfield (BPS) bound, constitutes the major part. The rest is on the phenomenology of the non-perturbative effects expected in string theories.

I first reviewed the features of BPS-saturated states. Then, a study of a class of four-dimensional BPS-saturated black-hole solutions of toroidally compactified heterotic string was presented. At the points of maximal symmetry of the two-torus moduli subspace, the dyonic black holes became massless. It led to the possibility of supersymmetry enhancement. 

I explicitly constructed a large class of BPS-saturated states in toroidally compactified type II string theory next. They corresponded to orthogonally intersecting BPS-saturated states in ten dimensions. With Kaluza-Klein monopole, they had four charges and preserved ${1 \over 8}$ of the $N=8$ supersymmetry. I found a simple map to associate each charge with the corresponding Killing spinor constraints. I also explicitly showed how the $N=4$ supersymmetries of toroidally compactified heterotic string were embedded into the $N=8$ supersymmetries of IIA superstring. 

A particular kind of static, non-orthogonally intersecting and non-threshold BPS-saturated states of type II string theories in ten dimensions was then studied. They were parametrised by four ${\it independent}$ charges with non-diagonal metrics. The metrics could be diagonalized when one charge was removed. However, the components of the configurations still intersected non-orthogonally. 

I started the phenomenological study on the non-perturbative behavior of compactified string theories by investigating the effect of constant threshold corrections on static, non-extreme, and electrically charged dilatonic black holes. Closed form solution for the perturbed black holes was obtained. Then, I studied the cosmological and phenomenological implications of a non-perturbatively induced superpotential in a $N=1$ supergravity theory. I used gaugino condensation to fix the dilaton dependence, and $T$-duality to fix the moduli dependence. Without any fine tuning of parameters, I obtained a supersymmetric vacuum with zero cosmological constant, which could be relevant to the cosmological moduli problem in string theories.

\end{abstract}
    \tableofcontents
  \end{frontmatter}

\pagenumbering{arabic}\pagenumbering{arabic}

\chapter{Introduction}
\label{chapter:Introduction}

The theme of this thesis is on the study of several aspects of non-perturbative states in string theories. The major part of this study focus on a special class of non-perturbative states, the Bogomol'nyi-Prasad-Sommerfield (BPS)-saturated states. I also discussed a few phenomenological aspects of the non-perturbative behavior of string theories.

The very basic assumption of string theory is to replace the notion of a `point' particle by a one-dimensional `string' \cite{WitPhysicsToday}. The diameter of a string is around $10^{-34}$m. This assumption leads to renormalizable theories that incorporate (and predict) gravity consistently. Geometry of string interactions also minimizes the arbitrariness in constructing consistent theories. However, there are five consistent ten dimensional string theories, instead of one unifying theory \cite{WitBook} \cite{LustLect}. 

The five string theories have clear distinction in their basic (perturbative) construction. The two type II theories consist only of closed oriented strings. Both have $N=2$ space-time supersymmetries. The two supersymmetries of the IIA string have different chiralities, whereas those of the IIB string have the same chirality. The type I theory consists of both open and closed unoriented strings, with $N=1$ supersymmetry. It contains the gauge group $SO(32)$. The corresponding charges attach to the ends of the open strings as Chan-Paton factors. The two heterotic strings also have $N=1$ space-time supersymmetry. The left (right) moving modes on both heterotic strings are just the same as that in type II strings. The right (left) moving parts consist of conformal field theories which make the whole string theories consistent quantum mechanically. The right (left) moving part of the $SO(32)$ heterotic string consists of world sheet fermions which provide the necessary operators to make the $SO(32)$ supermultiplets. Similarly, the right (left) moving part of the $E_8 \times E_8$ heterotic string allows the construction of the $E_8 \times E_8$ supermultiplets. 

Recent discoveries have shown that the distinction between the five string theories is only an artifact of our perturbative consideration of the theories. Conjectures of dualities \cite{WitVaDi}-\cite{HoraWit} relate the string theories non-perturbatively. We have come to the conclusion that there exists an eleven dimensional unifying M-theory, putting the five string theories on a common footing \cite{MT}. The BPS states play a crucial role in these discoveries. Their non-perturbative nature and resistance of quantum corrections make them a necessary component for studying the non-perturbative behavior of string theories.

In this chapter, I shall explain the meaning of Bogomol'nyi bounds and BPS-saturated states in subsection (1.1). In subsection (1.2), I shall briefly describe various types of BPS-saturated states in string theories. I shall demonstrate the importance of the BPS states more explicitly by describing several duality conjectures of string theories in subsection (1.3). The role of BPS states would be emphasized. Finally, I shall give a brief overview of the following chapters.

\section {Bogomol'nyi bound-Saturated States}

        I shall first study the definition of the Bogomol'nyl bound from the point of view of a non-supersymmetric field theory. In this context, some advantages of studying the states which saturate the bound shall be seen. These advantages remain in the supersymmetric generalization, which will be discussed later in this section. 

I shall consider the one dimensional field theory with the energy functional \cite{Bog},
\begin{equation}
E = \int^{\infty}_{- \infty} \left[ {1 \over 2} \left( {{d \phi} \over {dx}} \right)^2 + { \lambda \over 2} ( \phi^2 - F^2 )^2 \right] dx .
\label{BogE}
\end{equation}
The solution for $\phi$ can be obtained by solving the corresponding Euler-Lagrange equations. These are second order differential equations. With the following boundary conditions: $\phi (x) \rightarrow F$ as $x \rightarrow \infty$, and $\phi (x) \rightarrow -F$ as $x \rightarrow - \infty$, one discovers the domain wall solutions for the theory (\ref{BogE}). 

Another interesting way to solve the above problem is provided in \cite{Bog}. We can rewrite (\ref{BogE}) in the following suggestive form,
\begin{equation}
E = \int^{\infty}_{-\infty} \left[ {1 \over 2} \left( {{d \phi} \over {dx}} + \sqrt{\lambda} ( \phi^2 - F^2 ) \right)^2 - \sqrt{\lambda} { {d \phi} \over {dx}} ( \phi^2 - F^2 ) \right] dx.
\label{BogE2}
\end{equation}
Targeting at the domain wall solution, we make use of the boundary conditions right away and simplify (\ref{BogE2}) as follows, 
\begin{equation}
E = { {2 \sqrt{\lambda}} \over 3} F^2 | Q | + {1 \over 2} \int^{\infty}_{-\infty} \left[ {{d \phi} \over {dx}} + \sqrt{\lambda} (\phi^2 - F^2) \right]^2 dx,
\label{BogE3}
\end{equation}
where $Q=\phi(\infty) - \phi(-\infty)$ is the topological charge, and is equal to $2F$ in the above case.

From (\ref{BogE3}), we see that the minimum energy of the solution is,
\begin{equation}
E \ge { {2 \sqrt{\lambda}} \over 3} F^2 | Q | .
\label{Bogeq}
\end{equation}
The corresponding solution for $\phi$ satisfies a first order equation and makes the square bracket in the second term of (\ref{BogE3}) vanishes.

Therefore by rewriting the original energy functional (\ref{BogE}) into the form (\ref{BogE2}), we discover the following:(1) the energy of the state is proportional to the topological charge, (2) the minimum energy is determined even before any differential equations are solved, (3) only first order differential equations have to be solved, and (4) the stability of the solution is guaranteed as the energy functional is bounded from below. This lower bound of the energy is called the Bogomol'nyi bound. The advantages (1) to (4) remain in the supersymmetric generalization. 

Advantage (3) suggests that the differential equations determining the states which saturate the Bogomol'nyi bounds are integrable. Actually, a large class of supersymmetric black hole solutions for type II supergravity compactified on a six torus is explicitly found in Chapter 3. The set of coupled differential equations is indeed completely integrable and they are all first order.

In \cite{PS}, the exact classical solution for the theory with $SU(2)$ fields coupled to an $SU(2)$ Higgs field was found. The solution described dyons with both magnetic and electric charges. The corresponding mass saturated the Bogomol'nyl bound. States which saturate the Bogomol'nyl bound (often with supersymmetric theory) are called the Bogomol'nyi-Prasad-Sommerfield (BPS) states. 

I now consider the supersymmetric generalization of the Bogomol'nyi bound. Following \cite{WitOlive}, I study the four dimensional $N=2$ super Yang-Mills theory,
\begin{eqnarray}
L &=& \int d^4 x [ -{1 \over 4}F_{\mu \nu}^a F_{\mu \nu}^a + {1 \over 2} \bar{\Psi}^a_i i D {\Psi}^a_i + {1 \over 2} D_{\mu} A^a D_{\mu} A^a + {1 \over 2} D_{\mu} B^a D_{\mu} B^a   
\nonumber \\
&+& {1 \over 2} g^2 Tr[A, B][A, B] + {1 \over 2} i g \epsilon_{ij} Tr \left( [\bar{\Psi}^i, {\Psi}^j] A + [\bar{\Psi}^i, \gamma_5 {\Psi}^j] B \right) ] ,
\label{N2Wit}
\end{eqnarray}
where $\Psi_i, i=1,2$ are two Majorana fermions, $A$ and $B$ are scalar and pseudoscalar fields in the adjoint representation of the gauge group respectively. 

I choose $O(3)$ to be the gauge group. I assume a non-zero vacuum expectation value for $A$, but set the vev of $B$ to zero. The non-zero vev of $A$ spontaneously breaks $O(3)$ to $U(1)$. The electric and magnetic charges associated with the unbroken $U(1)$ shall determine the supersymmetric Bogomol'nyi bound.
                                                                
The supersymmetry charges $Q_{\alpha i}, i=1,2$ satisfy the algebra,
\begin{equation}
\{ Q_{\alpha i}, \bar{Q_{\beta j}} \} = \delta \gamma^{\mu}_{\alpha \beta} P_{\mu} + \epsilon_{ij} \left( \delta_{\alpha \beta} U + (\gamma_5)_{\alpha \beta} V \right) .  
\label{WitAl}  
\end{equation}
where
\begin{eqnarray}
U = \int d^3 x \partial_i ( A^a F_{0i}^a ),
\nonumber \\
V = \int d^3 x \partial_i ( A^a {1 \over 2} \epsilon_{ijk} F_{jk}^a ).
\label{UVWit}
\end{eqnarray}
If there are no solitons in the theory, $U$ and $V$ are identically zero and the algebra (\ref{WitAl}) reduces to the ordinary supersymmetry algebra without central charges. However, we can reasonably expect to have electric and magnetic charges associated with the unbroken $U(1)$ in the theory (\ref{N2Wit}). The electric charge $e$ and the magnetic charge $g$ are equal to ${U \over <A>}$ and ${V \over <A>}$ respectively. Therefore, from (\ref{WitAl}), the mass $M$ of a particle satisfies the equality
\begin{equation}
M \ge <A> \sqrt{e^2 + g^2} .
\label{SuBb}
\end{equation}
The lower bound of the mass is the supersymmetric version of the Bogomol'nyi bound. 

Some important properties of the states which saturate the supersymmetrized Bogomol'nyi bound can be seen from the algebra (\ref{WitAl}). The bound is saturated when the matrix $\{ Q_{\alpha i}, \bar{Q_{\beta j}} \}$ has zero eigenvalue(s). Therefore the states saturating the Bogomol'nyi bound is supersymmetric. As some linear combinations of $Q_{\alpha i}$ vanish (when sandwiched between the corresponding quantum states), the states that saturate the bound must form a representation of the supersymmetry algebra with smaller dimension. In fact, they form a four dimensional representation in the theory (\ref{N2Wit}), instead of the 16 dimensional representation for the states which do not saturate the bound \cite{Salam}. 

An important property of BPS states follows: the mass spectra of the BPS states are equal to their classical values and receive no quantum corrections irrespective of the size of the coupling constant in the theory. The reason is that the dimensionality of the representation of the BPS states cannot be changed by quantum effects.

In string theories, the compactified theories often have more than $N=1$ supersymmetry. Therefore they can have BPS states which break a fraction of the supersymmetry, and belong to representations of the supersymmetry algebra with various dimensions. For example, the toroidally compactified type II supergravities have $N=8$ supersymmetry in four dimensions. The BPS states which preserve $1 \over 2$ of the supersymmetry belong to the ultra-short ($16^2$ dimensional) multiplet. Those that preserve $1 \over 4$ supersymmetry belong to the short ($16^3$ dimensional) multiplet.

\section{BPS states in string theories and 11 dimensional supergravity}

        In this section, I shall describe a few important BPS states in string theories and the eleven dimensional supergravity. A brief review of some properties of these BPS states would help us appreciate more the work reported in the following chapters.

The BPS states that I am going to describe are the fundamental string, the solitonic five brane, the Kaluza-Klein monopole, and the D-branes in various string theories, and also the membrane and five-brane in eleven dimensional supergravity.
\footnote{A review of the status of solitons in superstring theory, with a comprehensive list of reference on the subject is given in \cite{Duffsolitons}.}
They all preserve $1 \over 2$ of the supersymmetry of the original theory. It is important to note that even though the underlying theory we are going to consider may only have $N=1$ supersymmetry, the Bogomol'nyi bound still exists because of the special geometry of the soliton solution in ten dimensions \cite{DiDi}.

\begin{center}
        {\bf The Fundamental String}
\end{center}

        The ten dimensional bosonic sector of the 3-form version of the $N=1$ supergravity theory is common to the effective low energy limit of all the five superstring theories,
\begin{equation}
I_{10}(string) = {1 \over 2} \int d^{10}x \sqrt{-g} \left( R - {1 \over 2} ( \partial \phi)^2 - {1 \over {2 \times 3!}} e^{- \phi} H^2 \right) ,
\label{Istring}
\end{equation}
where $\phi$ is the dilaton, and $H_{\mu \nu \rho}$ is the three-form field strength of the anti-symmetric tensor $B_{\mu \nu}$. It admits macroscopic string states as singular solutions \cite{Duffsolitons} \cite{SSS}
\footnote{In \cite{SSS}, the macroscopic string is considered as a solution of supergravity super-Yang-Mills theory, i.e., effective theory of heterotic strings. The same conclusion can be extended directly to the other string theories.}. The singular behavior means that the macroscopic string states should be considered as a source term in the theory,
\begin{equation}
S_2 = -T_2 \int d^2 \zeta \left( {1 \over 2} \sqrt{- \gamma} \gamma^{ij} \partial_i X^M \partial_j X^N g_{MN} e^{\phi \over 2} + {1 \over 2} \epsilon^{ij} \partial_i X^M \partial_j X^N B_{MN} \right) ,
\label{Isourcest}
\end{equation}
where $\gamma$ is the world sheet metric. It describes the coupling of a string to the metric, the antisymmetric tensor and the dilaton. The supergravity theory therefore consists of two pieces $S = I_{10} + S_2$
\footnote{The metric of the fundamental string can be obtained by double dimensional reduction of the supermembrane in eleven dimensional supergravity, to be discussed shortly.}.

This solution admits Killing spinors in the theory, and as such can be regarded as a supersymmetric bosonic background. The ADM mass per unit length $M$ of the macroscopic string saturates the corresponding Bogomol'nyi bound, as expected from supersymmetry. The Bogomol'nyi-type inequality is: $M \ge |Z|$, where $Z$ is given by the integral of the dual of the three form $H$ over a seven-sphere at infinity.

The fundamental string has been described only as a solution of string theories up to lowest order in $\alpha '$. It was argued in \cite{SSS} that the features of the solution should persist to all orders in $\alpha '$. In other words, there should exist superconformal field theories describing the fundamental string solution. 

The fundamental string solution is required by U-duality to be identified with the solitonic string \cite{HullU} that fills up the U-duality multiplet. This is consistent with the suggestion that the four dimensional heterotic string should be identified as an axion string \cite{Witcosmic}.

\begin{center}
        {\bf The Solitonic Five-brane}
\end{center}

        The fundamental string carries a Noether electric charge $Z$, and is singular right at the location of the string. It acts as a source in the original theory. 
There is also a solitonic five-brane solution \cite{Duffsolitons} to the theory without any source term, i.e., only $I_{10}$ from (\ref{Istring}) is involved. The five-brane is non-singular and carries a topological magnetic charge. It is closely associated with the fact that with the five-brane solution, the 3-form field strength $H$ is a harmonic form and cannot be written globally as the curl of the antisymmetric field $B_{\mu \nu}$, though it still satisfies the Bianchi identity.

The mass per unit 5-volume of the solitonic five-brane saturates a Bogomol'nyi-type bound, and is directly proportional to the magnetic charge it carries. It is supersymmetry, i.e., the solution admits Killing spinors. The electric charge of the fundamental string and the magnetic charge of the solitonic five-brane satisfies a generalization of the Dirac quantization rule \cite{DQ}
\footnote{That hints at a string/five-brane duality, in analogy with the electric/magnetic duality in $N=4$ supersymmetric guage theory. I shall not discuss this particular duality as it would involve the fundamental five-brane \cite{DuffE5b}.}
.

The metric of the five-brane, the three-form field strength, and the dilaton are, respectively,
\begin{equation}
ds^2 = ( 1 + {k_6 \over r^2} )^{-1/4} \eta_{\mu \nu} dx^{\mu} dx^{\nu} + ( 1 + {k_6 \over r^2} )^{3/4} \delta_{m n} dy^m dy^n  ,
\label{Sfbmetric}
\end{equation}
\begin{equation}
H = 2 k_6 \epsilon_3 ,
\label{SfbH}
\end{equation}
\begin{equation}
e^{2 \phi} = 1 + {k_6 \over r^2}  ,
\label{Sfbdilaton}
\end{equation}
where $k_6 = \kappa g_6 / \sqrt{2} \Omega_3$ with $\kappa, \Omega_3$ being the ten dimensional Planck mass and the volume of the unit 3-sphere. $r, g_6$, and $\epsilon_3$ are the transverse distance, magnetic charge, and the volume form of $S^3$, respectively.

A five-brane supersymmetric solution to the effective heterotic supergravity was found in \cite{Sthet}. It differs from the five-brane soliton discussed above by having a Yang-Mills instanton core. It is shown in \cite{Duffsolitons} that it is an exact solution in string theory, i.e., receives no $\alpha '$ corrections, by explicitly writting down the corresponding conformal field theory. 

\begin{center}
        {\bf The Kaluza-Klein Monopole}
\end{center}

        The Einstein-Hilbert action in five dimensions reduces to a theory of four dimensional gravity with a $U(1)$ gauge symmetry coupled to a Brans-Dicke type massless scalar field ($V$) when the five dimensional theory is compactified on a circle,
\begin{equation}
S = {-1 \over {16 \pi G}} \int d^4 x \sqrt{-g^4} V^{1 / 2} \left( R_4 + {1 \over 4} V F_{\mu \nu} F^{\mu \nu} \right) .
\label{Imonopole}
\end{equation}

Such a theory admits solitonic solutions \cite{GrossKK}. Among them is the magnetic monopole. It is regular, static and stable. Due to the repulsive interaction of the massless Brans-Dicke type scalar, these monopoles virtually exert no newtonian force on slowly moving test particles and seem as if they have no gravitational mass. 

Embedding the Kaluza-Klein monopole solutions to the ten dimensional supergravities which are the effective superstring theories \cite{HullU}, these solutions can be considered as consisting of the product of a self-dual Taub-NUT instanton with topology ${\mathcal{R}}^{4}$, a five-torus and a time-like $\mathcal{R}$. As this is the product of a five-metric and a five-torus, we can regard it as a five-brane solution of the ten dimensional theory wrapped on the five-torus. 

As for the exactness (with respect to $\alpha '$ correction) of the monopole solution, we have to notice that the Kaluza-Klein monopole solution is just a special case of the four parameter solution discussed in \cite{MCT4p}. That four parameter solution corresponds to a ten dimensional background which defines a conformal sigma model and is a particular case of a chiral null model with curved transverse part. The general theory about supersymmetric dyonic black holes in Kaluza-Klein theory compactified on a torus with various dimensions was studied in \cite{MCKK}. The magnetic Kaluza-Klein monopole is a special case in the class of solution discussed in that paper, and it must be supersymmetric. 

\begin{center}
        {\bf D-brane}
\end{center}

        The last non-perturbative BPS states in superstring theories we are interested in are the D-branes \cite{TasiD}. 

Consider adding open strings to type II closed superstring theories. The two dimensional superstring world sheet action allows us to impose an arbitrary number ($p+1$) of Neumann boundary conditions and a number ($9-p$) of Dirichlet boundary conditions on the open strings. Consequently, the end points of the open strings live on a ($p+1$) dimensional hyperplane, called the D-brane. Closed strings still propagate in the 10 dimensional space-time, and interact with the hyperplane through interaction with the open strings, thereby making the D-brane dynamical. Such a theory with open and closed strings and a D-brane is consistent as long as $p$ is even in IIA theory, and odd in IIB theory. 

The open string boundary conditions on the D-brane correlate the $N=2$ supersymmetries of the closed string theories, thereby reducing the supersymmetry to $N=1$. Thus the D-brane is a BPS state. The world-volume theory of a D-p-brane consists of the coupling between the brane and the corresponding Ramond-Ramond $p+1$-form potential, a $U(1)$ vector field, and $(9-p)$ scalar fields which describe the fluctuations of the brane. 

It was shown in \cite{POL} that D-branes are Ramond-Ramond charge carriers for type II superstring theories. The electric charge carried by a D-p-brane (a D-brane with $p$ spatial dimensions) and the magnetic charge carried by a D-p'-brane with $(p + p' = 6)$ satisfy the Dirac quantization condition with one unit of quanta. This strongly suggests that the D-branes are intrinsic to any non-perturbative formulation of the type II string theories, though I said I `add' open strings to the closed string theories at the beginning
\footnote{The type I string also contains a D-1-brane and a D-5-brane.}
. 

One can hardly underestimate the importance of the fact that D-branes are Ramond-Ramond charge carriers. There is no perturbative object in string theories that carries Ramond-Ramond charges. Perturbative objects only carries Neveu-Schwarz-Neveu-Schwarz charges. Only the field strengths, instead of the gauge potentials, of the Ramond-Ramond fields appear in the vertex operators which act on the string Hilbert space. On the other hand, string duality and the conjecture of the unifying eleven dimensional M-theory requires that no distinction should be made between Neveu-Schwarz-Neveu-Schwarz and Ramond-Ramond charges. The D-branes are also interesting objects to be studied, as they have a simple and exact description in conformal field theory. 

\begin{center}
        {\bf Membrane and five-brane in 11 dimensional supergravity}
\end{center}

        There is a lot of evidence to show that all the five superstring theories are just different perturbative expansions around different corners in the moduli space of the M-theory \cite{MT}. The short distance degree of freedom of M-theory is conjectured in \cite{Matrix} to be described by the large $N$ limit of the supersymmetric matrix quantum mechanical system which had been used to study the small distance behavior of D0 branes. The eleven dimensional supergravity \cite{Cremmers} is the low energy and large distance limit of the M-theory. As such, the soliton solutions of the eleven dimensional supergravity bear much implications to the non-perturbative behavior of string theories.

The bosonic field content of the $N=1$ eleven dimensional supergravity includes the metric and a three form field which is required by supersymmetry,
\begin{eqnarray}
S_G &=& \int d^{11} x [ {1 \over 2} \sqrt{-g} R - {1 \over 96} \sqrt{-g} F_{MNPQ} F^{MNPQ} 
\nonumber \\
&+& {1 \over {2 (12)^4}} \epsilon^{MNOPQRSTUVW} F_{MNOP} F_{QRST} A_{UVW} ],
\label{I11d}
\end{eqnarray}
where $F_{MNOP}$ is the field strength of the three-form field $A_{UVW}$. The supermembrane configuration (M-2-brane) solves exactly the field equations of the eleven dimensional supergravity. The corresponding metric and fields are,
\begin{equation}
ds^2 = \left( 1 + {K \over r^6} \right)^{-2/3} \eta_{\mu \nu} dx^{\mu} dx^{\nu}
+ \left( 1 + {K \over r^6} \right)^{1/3} \delta_{m n} dy^{m} dy^{n},
\label{MMemMetric}
\end{equation}
\begin{equation}
A_{\mu \nu \rho} = \pm {1 \over g_3} \epsilon_{\mu \nu \rho} \left( 1 + {K \over r^6} \right)^{-1}  ,
\label{MMemField}
\end{equation}
where $K ={ { {\kappa}^2 T} \over {3 {\Omega}_7}}$, and $\kappa, T, {\Omega}_7$ are the eleven dimensional Planck mass , tension of the membrane and the 7-volume respectively. The metric is singular on the worldvolume of the membrane, ${\it i.e.}$, at $r=0$ where $r$ is the transverse distance. That implies that the pure supergravity $S_G$ is coupled to a source term in which the membrane itself interacts with the bosonic fields of the supergravity \cite{Duff11mem},
\begin{eqnarray}
S_M = T \int d^3 \zeta [ &-&{1 \over 2} \sqrt{- \gamma} \gamma^{ij} \partial_i X^M \partial_j X^N g_{MN} + {1 \over 2} \sqrt{- \gamma} 
\nonumber \\
&+& {1 \over {3!}} \epsilon^{ijk} \partial_i X^M \partial_j X^N \partial_k X^P A_{MNP} ] .
\label{Isourceme}
\end{eqnarray}
Therefore the theory is $S = S_G + S_M$. The configuration is stabilized by Page charge \cite{Pagec} of the eleven dimensional supergravity. It is supersymmetric, i.e., there exist Killing spinors for the configuration. It breaks ${1 \over 2}$ of the supersymmetry, and the mass per unit area saturate the corresponding Bogomol'nyi bound.

There also exist the solitonic non-singular five-brane (M-5-brane) solution \cite{Guven} for the $N=1$ eleven dimensional supergravity without any source term, {\it i.e.}, only $S_G$ is involved. It preserves half of the supersymmetry, and the mass per unit 5-volume saturates the corresponding Bogomol'nyi bound, which is directly proportional to the magnetic charge of the five-brane. 

It should be noted that as there is no dual formulation of the $N=1$ eleven dimensional theory, the magnetic five-brane solutions cannot be obtained from the electric membrane by duality transformations (in contrast to the string/five-brane duality in type II string theories).

\section{BPS states and dualities}

        I shall list some of the major evidences in dualities in this section. The role of the BPS states will be emphasized. 

\begin{center}
        {\bf U-duality}
\end{center}

        The low energy effective theory of a type II string compactified on a six-torus is the four dimensional $N=8$ supergravity \cite{Cremmers}. The equations of motion are invariant under the group $E_{7}$, which has $SL(2,R) \times O(6,6)$ as the maximal subgroup. It was shown in \cite{HullU} that quantum effects break the $E_7$ group to a discrete subgroup $E_7(Z)$, and so is its maximal subgroup. Evidences showing that the $E_7(Z)$ symmetry, the U-duality group, is a symmetry of the full theory are investigated in \cite{HullU}. 

The consistency of U-duality requires that there must be soliton states carrying exactly the same quantum numbers as the fundamental Bogomol'nyi (supersymmetric) states. That is because the fundamental string excitations include additional Bogomol'nyi states which apparently cannot be belong to the U-duality multiplets with solitons, as the soliton multiplets are already complete (as shown by working out the mass formular). The BPS states we discussed in the previous section would be identified as the required states. 

We have to look for (28+28) soliton states, 28 electric and 28 magnetic, for consistency of U-duality. There are (12+12) solitonic states that carry Neveu-Schwarz-Neveu-Schwarz charges. There are 6 Kaluza-Klein (KK) monopoles accounting for 6 of the 12 magnetic solitons. Each of these KK monopoles associates with one different toroidal direction. Another 6 magnetic solitons come from wrapping 6 solitonic five-branes on the six-torus. There are 6 different ways of wrapping. The 12 electric solitons come from the fundamental strings. There are 6 winding modes of the fundamental string along the 6 different toroidal directions. The strings also have momentum in the 6 toroidal directions, giving rise to the source of the remaining 6 electric charges. 

The remaining (16+16) solitons have Ramond-Ramond charges, and must be carried by the D-branes
\footnote{In \cite{HullU}, the Ramond-Ramond charge carriers are considered to be black p-branes. The discovery of D-brane as Ramond-Ramond charge carriers \cite{POL} further strengthens the arguments made in \cite{HullU}.}
. I should only consider the IIA string for simplicity. The D-0-brane gives one electric soliton. The D-2-branes give 15 electric solitons as there are 15 ways for the membrane to wrap around the six-torus. Therefore the D-0-brane and the D-2-branes provides the 16 Ramond-Ramond electric solitons. The D-4-branes give 15 magnetic solitons as there are 15 ways to wrap the 4-branes on the six-torus. Finally, the D-6-brane wraps on the six-torus to give the final magnetic soltion. Therefore the D-4-branes and the D-6-brane provide the 16 Ramond-Ramond magnetic solitons. 

\begin{center}
        {\bf M/IIA Duality}
\end{center}

        We should see how the various BPS states in ten dimensional string theories help discover the hidden eleventh dimension of the underlying M-theory. 

The D-0-brane of type IIA string theory has mass proportional to $1 \over \lambda$ \cite{TasiD} where $\lambda$ is the coupling of the theory. The bound states \cite{WitBDb} of them provide infinitely many states with mass $M = c {n \over \lambda}$ with $n$ an integer. They carry the corresponding amount of Ramond-Ramond charges. As the D-branes are BPS states, we may go to the strong coupling limit where these states become light and contribute to the low energy behavior. A theory with these kinds of particles cannot be a local field theory in ten dimensions \cite{WitVaDi}. 

However, we can reproduce the spectrum by considering the strong coupling limit of IIA theory (with the D-0-branes) as an eleven dimensional theory on $R^{10} \times S^1$, and that the radius of the circle scales as $1 \over \lambda$. The Ramond-Ramond charge of the bound states of the D-0-branes would be identified as the electric charge associated with the Kaluza-Klein $U(1)$ gauge field of the eleventh dimension.

We may also use the BPS states to see the eleventh dimension in the weakly-coupled theory. The weakly-coupled IIA string can be obtained by wrapping the supermembrane around the circle $S^1$ (with radius $r$) in the eleven dimensional supergravity theory in the limit $r \rightarrow 0$ \cite{Duffsfm}. The Green-Schwarz action of the string follows from the Green-Schwarz action of the supermembrane. 

\begin{center}
        {\bf M/IIB Duality}
\end{center}

        The M-theory compactified on a two-torus ($T^2$) should be dual to the IIB string compactified on a circle ($S^1$), as required by the T-duality of IIA/IIB theories \cite{NewCon} \cite{TDuality} and the fact that the strong coupling limit of IIA string is M-theory `compactified' on a large circle. The matching of the corresponding $p$-branes in nine dimensions provide a consistency check \cite{SchPM}. In other words, it is precisely due to these BPS states that make the M/IIB duality possible.

The 0-branes in the nine dimensional theory come from wrapping the M-2-branes on $T^2$ from M-theory perspective, while from the perspective of the IIB theory, they come from wrapping the D-1-brane 
\footnote{The fundamental string and the D-1-brane in IIB theory are related by an $SL(2,Z)$ symmetry of the IIB theory \cite{Schwarz}.}
on $S^1$. The winding modes of IIB strings are identified with the Kaluza-Klein (KK) modes of $T^2$, while the KK modes of $S^1$ are identified with the winding modes of the M-2-brane on $T^2$. Such a matching implies the relation $L_B \sim A_M^{-3/4}$, where $L_B$ is the circumference of $S^1$, and $A_M$ is the area of $T^2$. Therefore if one lets $L_B \rightarrow 0$ while holding the shape of $T^2$ fixed (which is related to the IIB coupling), one ends up with the eleven dimensional M-theory. Clearly one can recover the ten dimensional IIB theory from $T^2$ compactified M-theory the other way around
\footnote{The M-theory relates IIA string by double dimensional reduction, while it relates IIB string through the two-torus and IIB is recovered only by shrinking the torus with specific shape. A more natural proposal to understand the IIB theory, in particular the $SL(2,Z)$ symmetry, is F-theory \cite{EvidF}.}. 

Similarly, wrapping M-2-brane on one toroidal direction gives nine dimensional 1-brane, which is the D-1-brane in IIB without wrapping. The M-2-brane gives the nine dimensional 2-brane when it does not wrap on the torus while D-3-brane in IIB can wrap on $S^1$ to give the same 2-brane. The M-5-brane can wrap on the torus, or just on one of the toroidal directions. That gives the nine dimensional 3-brane and 4-brane. The same nine dimensional branes can be obtained from the IIB theory by the D-3-branes and the D-5-brane wrapped on $S^1$ respectively. As for the nine dimensional 5-branes, they originate from the unwrapped M-5-brane. They correspond to the unwrapped KK 5-brane of the IIB theory (compare with the discussion on KK monopole in the previous section). The $SL(2,Z)$ family of 5-branes (mixing up solitonic five-brane and the D-5-brane) of the IIB theory corresponds to the KK 5-branes of the M-theory.

\begin{center}
        {\bf M/Heterotic Duality}
\end{center}

        It is conjectured that M-theory compactified on an interval $S^1 / Z_2$ is the strong coupling limit of the $E_8 \times E_8$ heterotic string \cite{WitHorhet}. Evidences include the vanishing of space-time gravitational anomalies, the strong coupling behavior of the string, and the vanishing of the world-volume gravitational anomalies. We can use the BPS states in the heterotic theory and that in the M-theory to describe the above conjecture as a consistency check.

The heterotic $E_8 \times E_8$ theory contains a fundamental string (the heterotic string \cite{Witcosmic}) and a solitonic five-brane. As the compactified eleven dimensional space on an interval is equivalent to two parallel ten dimensional boundaries seperated by the line interval, the closed heterotic string must be a cylindrical M-2-brane with one boundary attached to each boundary of the space-time \cite{SchMT} from the eleven dimensional perspective. There is one $E_8$ gauge group on each boundary. The weakly-coupling limit of the theory, i.e., the perturbative ten dimensional $E_8 \times E_8$ heterotic theory corresponds to a short cylinder with large radius. 

On the other hand, the solitonic 5-brane of the heterotic theory must exist as a closed surface in either boundary. From the M-theory perspective, it is just the M-5-brane. 

Some explicit relations between the two theories can be obtained by brane matching calculations like the M/IIB case. Let $L_1, L_2$ be the height and circumference of the cylinder used to compactified M-theory to nine dimensions. Let $L_o$ be the circumference of the circle on which the heterotic string is compactified. One finds that the heterotic coupling (${\lambda}_H$) is given by $L_1 / L_2$. Thus the strong coupling limit occurs when the cylinder is long. We also finds $L_o \sim (L_1 L_2)^{-3/4}$. Therefore the uncompactified heterotic theory is obtained by shrinking the cylinder to a point while keeping its shape fixed. We also see that diffeomorphism in eleven dimensions (i.e., $L_1 \leftrightarrow L_2$) implies strong/weak duality transformation of the $SO(32)$ (in nine dimensions) theory that relates the weakly-coupled heterotic limit to the strongly-coupled type I limit.
\footnote{Another very important reason for studying BPS states is to investigate quantum mechanics in Black Hole physics. A report on the investigation concerning the relation between the canonical entropy of black holes and string degrees of freedom, together with a list of important references can be found in \cite{Mald1}.}

\section {A brief overview}

        Having discussed the importance of BPS states in the previous sections, I shall begin with the BPS states for heterotic string \cite{CK} in Chapter 2. There we shall see that along the $T^2$ moduli space, some BPS states become massless at special point of the moduli space. These lead to symetry enhancement, and the speculated supersymmetry enhancement.

In Chapter 3, the study is extended to type IIA superstring theory \cite{KLC} . The BPS states are found by explicitly solving Killing spinor equations. 

The black hole solutions obtained in Chapter 3 have higher dimensional interpretation. They are 10 dimensional objects intersect orthogonally. In Chapter 4, a class of 10 dimensional non-perturbative states corresponding to non-orthogonal intersection of various string objects are studied. They are obtained by performing duality transformations on the supergravity solutions obtained from the chiral null model.

I studied the phenomenological implications of non-perturbative states in string theories in Chapter 5 and 6. They are two studies on threshold corrections. They describe the non-perturbative effects of the moduli field which specify the compactification scale of the superstring theories. In Chapter 5, I study the effect of a constant threshold correction to an electric dilatonic-Maxwell-Eintein black hole \cite{KLCthres}. In Chapter 6, I study the implications of an effective superpotential with non-trivial dependence on the dilaton and the moduli in a $N=1$ supergravity theory. 

\chapter{Massless BPS states of Heterotic String}
\label{chapter: Massless BPS states of Heterotic String on Six-torus}

\section{Introduction and Summary}

As emphasized in Chapter 1, BPS-saturated  states of string theories  provide a fruitful ground to  address non-perturbative aspects of string theory. In addition to their importance in establishing duality conjectures as described in Chapter 1, the BPS states of four-dimensional, toroidally compactified heterotic string theory include regular solutions, {\it i.e.}, those  with regular horizons, may shed light  on  quantum aspects of black hole physics, {\it e.g.}, on  statistical interpretation of  black hole entropy \cite{LW} \cite{CTII}. Those  that can become massless\cite{BEHR,CYII,KL} at certain points of moduli space may shed light on the nature of enhanced symmetries \cite{HTII,CYII} at special points of moduli space. Since the effective theory possessed $N=4$ supersymmetry, the  ADM mass  for these BPS-saturated states is protected from quantum corrections. In principle, one should be able to trust the BPS mass formula  even in the case where quantized charges   are of ${\cal O}(1)$.

In this Chapter, I report on a study of massless BPS-saturated states. In particular, I identify the  (quantized) charge vectors  and  the points (lines, hyper-surfaces) in the  moduli space for which the BPS-saturated states become massless. For the sake of simplicity I confine this study to the  two-torus ($T^2$) moduli sub-space. 
\footnote{The work generalizes  that of \cite{CYII}, where  the case of  the two-circle ($(S^1)^2$)-moduli sub-space was addressed.}

What I found is that within a four-dimensional, toroidally compactified heterotic string, I identify  (quantized)  charge vectors of electrically charged
BPS-saturated states (along with the  tower of $SL(2,Z)$ related dyonic  states), which preserve $1\over 2$ of  $N=4$ supersymmetry and become massless along  the hyper-surfaces of enhanced  gauge symmetry of the two-torus moduli sub-space. In addition, I identify   charge vectors  of  the  dyonic BPS-saturated states (along with the  tower of $SL(2,Z)$ related states), which  preserve ${1\over 4}$ of $N=4$ supersymmetry, and become massless at two points  with the  maximal gauge symmetry enhancement.

\section{BPS states in heterotic string on a six-torus}

The explicit form  of the generating solution for {\it all} the static, spherically symmetric BPS-saturated states in this class has been obtained in \cite{CTII,CYIII}.
\footnote{In Ref. \cite{CYIII} also all the non-extreme solutions were obtained. In Ref. \cite{CTII} it was shown that BPS-saturated  generating solution is an exact   target-space  background  solution of a conformal $\sigma$-model. } 
The generating solution is specified by five (electric and magnetic) charges  of the two  $U(1)_{a,b}$  Kaluza-Klein and two $U(1)_{c,d}$ two-form fields  associated with the two, say the first two, (toroidally) compactified  dimensions. The most general BPS-saturated state in this class is  parameterized by unconstrained 28 electric and 28 magnetic charges and is obtained  by  applying a subset of $T$-duality and $S$-duality transformations, which do not affect the 
four-dimensional space-time, on the generating solution. 

The ADM mass  for these states (BPS mass formula), which in general
preserve only  ${1\over 4}$ of supersymmetry, is  specified \cite{CYI,DLR},
in terms  of 28 electric and 28 magnetic charges.
For the purpose of studying the  moduli  (and
the dilaton-axion)  dependence of the BPS mass  
formula\cite{CYI,DLR} I rewrite  it  in  terms of {\it conserved}
 magnetic
($\vec\beta$)  and electric ($\vec\alpha$) 
charge vectors \cite{CTII}: \footnote{
I use the notation  and conventions,  as  specified  in  Refs. \cite{CYIII},
following {\it e.g.},  Ref. \cite{MS,SEN}.}
\begin{equation}
M_{BPS}^2 ={\textstyle{1\over 2} }e^{-2\phi_\infty} 
{\vec \beta}^T \mu_R{\vec \beta} +
{\textstyle{1\over 2}}e^{2\phi_\infty}{\vec {\tilde\alpha}^T}
 \mu_R{\vec {\tilde\alpha}}
+ \left [({\vec \beta}^T\mu_R{\vec \beta})
({\vec \alpha}^T\mu_R{\vec \alpha})-({\vec \beta}^T\mu_R
{\vec \alpha})^2\right]^{1\over 2},
\label{ME}
\end{equation}
where 
\begin{equation}
{\vec {\tilde \alpha}}\equiv{\vec \alpha}+\Psi_{\infty}{\vec\beta}, \ \ 
\mu_{R,L}\equiv M_{\infty}\pm L.
\end{equation} 
The charge vectors $\vec \alpha$ and $\vec \beta$   are related to the
physical magnetic  $\vec P$ and electric $\vec Q$ charges  in the
following way:
\begin{equation}
\sqrt{2}P_i=L_{ij}{\beta}_j\ , \ \ \ \ \sqrt{2}Q_i = e^{2\phi_{\infty}}
M_{ij\,\infty}(\alpha_j + \Psi_{\infty}\beta_j),\ \  (i=1,\cdots , 28)
\label{charges}
\end{equation}
where the subscript
 $\infty$ refers to the asymptotic ($r\to\infty$)
  value of the  corresponding fields.  Here, the  
 moduli matrix $M$ 
and the dilaton-axion field
$S\equiv\Psi+i{\rm e}^{-2\phi}$, transform covariantly
(along with the corresponding charge vectors) under the $T$- duality 
($O(6,22,Z)$) and
$S$-duality ($SL(2,Z)_S$), respectively,  while the BPS mass formula (\ref{ME})   remains
invariant under these transformations.

Note that when the magnetic  and electric  charges
are parallel in the $O(6,22)$ sense, {\it i.e.},  $\vec{\beta}\propto \vec{\alpha}$
(in the quantized theory the lattice charge  vectors should be relative
co-primes \cite{SEN}),
the BPS mass formula (\ref{ME})  is that of the
BPS-saturated states which  preserve ${1\over 2}$ of $N=4$ supersymmetry
(see, {\it e.g.}, \cite{SEN}).   In the case when the magnetic and
electric charges are not parallel, the mass is larger (the last term in
(\ref{ME}) is non-zero) and the
configurations preserve only $1\over 4$ of $N=4$ supersymmetry \cite{CYI}.
Note that states  preserving ${1\over 2}$ of $N=4$ supersymmetry belong to
the vector super-multiplets, while those  preserving ${1\over 4}$ of $N=4$
supersymmetry belong to the highest spin ${3\over
2}$-supermultiplets \cite{STRAD,KALL}. Thus, when the former [latter] states become
massless they may contribute to the enhancement of gauge 
symmetry\cite{HTII} [supersymmetry\cite{CYII}].

\section{Massless BPS states}

I shall discuss the massless BPS states in this section.

In the quantum theory  the charge vectors $\vec{\alpha}$, ${\vec\beta}$  are 
quantized. Following \cite{SEN},
one may attempt to constrain the allowed lattice charge vectors  by using
the constraints  for the elementary
 BPS-saturated string states of   toroidally compactified heterotic string,
 along with the  Dirac-Schwinger-Zwanziger-Witten (DSZW) \cite{WITTENIII}
 quantization condition.\footnote{In Ref. \cite{CTII} 
the charge quantization for the generating solution
is implied  by considering the conformal field theory  describing the throat region  of the
corresponding string solution.}
Purely electric BPS-saturated states (${\vec\beta}=0$)
 preserve $1\over 2$ of $N=4$ supersymmetry
and  have the same quantum numbers \cite{DR,CMP}
as the  elementary   BPS-saturated string states with
no excitations in the right-moving sector
 ($N_R={1\over 2}$). For the  electric   states  the 
quantized    charge vector $\vec\alpha$  is then  constrained to lie on an even
self-dual   lattice  $\Lambda_{6,22}$ 
with  the  following   norm (in the $O(6,22)$ sense)
\cite{SEN}:
\begin{equation}
{\vec \alpha}^TL{\vec\alpha}=2N_L-2=-2,0, 2, ... \ ,
\label{chl}
\end{equation}
where  the integer $N_L$  parameterizes the  level of the left-moving sector.
 
The  DSZW  charge quantization condition then  implies
an analogous  constraint for
${\vec \beta}^TL{\vec\beta}$; magnetic
 charge vectors $\vec \beta$ are  then  constrained  \cite{SEN} 
to lie on an even self-dual
   lattice  $\Lambda_{6,22}$  with the norm
\begin{equation}
{\vec \beta}^TL{\vec\beta}=2N_L-2=-2,0, 2, ... \ .
\label{chlII}
\end{equation}
Since I confine the analysis to the two-torus moduli  sub-space , the $T$-duality group reduces to $O(2,2)$. Then only  the $O(2,2)$  part of the   symmetric moduli metric $M$ is non-trivial and of the form:
\begin{equation}
M=\left ( \matrix{G^{-1} & -G^{-1}B \cr 
-B^T G^{-1} & G + B^T G^{-1}B} 
\right ), \ \ \  L =\left ( \matrix{0 & I_2\cr
I_2 & 0} \right )
\label{modulthree}
\end{equation} 
where $G \equiv [{G}_{mn}]$ ($(m,n)=1,2$), $B\equiv B_{12}$ 
 are the four moduli of
the two-torus and $L$ is an $O(2,2)$ invariant matrix.

The four moduli fields can be expressed in terms
of two complex fields $T$ and $U$, (see, {\it e.g.}, 
 \cite{TDuality} and references therein):
\begin{equation}
{T \equiv \sqrt{{\cal G}} + i B}, \ \ \ 
U \equiv {{\sqrt{\cal G} - i G_{12} }\over G_{11}}
\end{equation}
where ${\cal G} \equiv {\rm det}(G_{mn})$.  The $T$ and $U$ fields transform 
 covariantly under  $PSL(2,Z)_T $ and ${ PSL(2,Z)}_U$,
respectively, {\it i.e.}, the subgroups of the
duality  group  $O(2,2,Z)={ PSL(2,Z)}_T \times { PSL(2,Z)}_U
\times Z_{2\, T\leftrightarrow U} $ \cite{TDuality}.

In order to address massless BPS states which preserve $1 \over 2$ of
supersymmetry, I first concentrate on purely electrically charged configurations
($\vec\beta=0$) with  the  electric lattice  charge vector 
${\vec \alpha}\equiv (\alpha_a,\alpha_b;\alpha_c,\alpha_d)$ 
 whose norm  is:  
\begin{equation}
{\vec\alpha}^TL{\vec\alpha}=-2.
\label{elc}
\end{equation}
Namely, only the states  with  the  electric charge 
norm  (\ref{elc})  can become massless\cite{HTII} along the lines
(hyper-surfaces) of moduli space
for which:
\begin{equation}
{\vec\alpha}^T\mu_R{\vec\alpha}=0.
\label{zmc1}\end{equation}
It turns out that  (\ref{zmc1}) is satisfied along the following hyper-surfaces,
along with the following 
 accompanying electric charge vectors ${\vec \alpha}$:\footnote{In the following we
 suppress the subscript $\infty$  for the asymptotic values of the  moduli
 fields.}  
\begin{equation}
{{\cal L}_1: U = T\Leftrightarrow(G_{11},G_{22},G_{12},B)=(1,G_{22},-B,B);
\ \ {\vec\alpha}= {\vec {\lambda}}_{1\,\pm}\equiv\pm (1,0,-1,0),}
\label{line1}
\end{equation}

\begin{equation}
{{\cal L}_2: U = {1 \over T} \Leftrightarrow 
(G_{11},G_{22},G_{12},B)=(G_{11},1,B,B)};\ \ 
{\vec\alpha}={\vec {\lambda}}_{2\,\pm}\equiv\pm (0,1,0,-1),
\label{line2}
\end{equation}

\begin{equation}
{{\cal L}_3: U = T - i \Leftrightarrow
(G_{11},G_{22},G_{12},B)=(1,G_{22},1-B,B)}; \ \ 
{\vec {\alpha}}={\vec{\lambda}}_{3\,\pm}\equiv\pm (1,1,-1,0),
\label{line3}
\end{equation}

$$
{{\cal L}_4: U = {T \over {i T + 1} } \Leftrightarrow
(G_{11},G_{22},G_{12},B)=(G_{11},-1+2B+G_{11},-1+B+G_{11},B);}$$
\begin{equation}
{\vec {\alpha}}={\vec{\lambda}}_{4\,\pm}\equiv\pm (1,0,-1,1),
\label{line4}
\end{equation}
Those  are the {\it same}  four hype-surfaces of the two-torus moduli sub-space  
(in the fundamental domain),  for which
the gauge symmetry
of toroidally compactified  heterotic string is 
enhanced due to the Halpern-Frenkel-Ka\v c mechanism, {\it i.e.},  those are
the hyper-surfaces where the perturbative string states (with
$N_R={1\over 2}$), which have the same quantum numbers as electrically charged
BPS-saturated states, become massless.
Thus, on the heterotic side these electrically charged 
BPS-states  are identified with the
elementary string excitations. 

Along  each of the  hyper-surfaces ${\cal L}_{1,2,3,4}$ these  electrically
charged massless BPS-saturated states,  which are
 scalar components of the  vector super-multiplets,
contribute to the enhancement of the  gauge symmetry from 
$[U(1)_a\times U(1)_b\times U(1)_c\times U(1)_d]$  (at generic points of moduli
space)  to 
 $[U(1)_b\times U(1)_d \times U(1)_{a+c}\times
SU(2)_{a-c}]$, $[U(1)_a\times U(1)_c \times U(1)_{b+d}\times SU(2)_{b-d}]$,
$[U(1)_d\times U(1)_{a+c} \times U(1)_{a-2b-c}\times SU(2)_{a+b-c}]$ and 
$[U(1)_b\times U(1)_{a+c}\times U(1)_{a-c-2d}\times SU(2)_{a-c+d}]$, respectively.

 At  the point $U=T=1$, {\it i.e.}, $(G_{11},G_{22},G_{12},B)=(1,1,0,0)$ (the self-dual point of
the two-circle),   ${\cal L}_1$  and  ${\cal L}_2$  meet and 
the enhanced  gauge symmetry is 
$[U(1)_{a+c}\times U(1)_{b+d}\times SU(2)_{a-c}\times
SU(2)_{b-d}]$. At  the point $T=U^*=e^{i {\pi \over 6}}$,
{\it i.e.}, $(G_{11},G_{22},G_{12},B)=(1,1,{1 \over 2},{1 \over 2})$, 
 ${\cal L}_2$, ${\cal  L}_3$  and ${\cal L}_4$ 
 meet and the enhanced  gauge symmetry  is $[U(1)_{a+c}\times
U(1)_{a-2b-c-2d}\times SU(3)_{b-d,2a+b-2c+d}]$.  Here
the subscript(s) for the
non-Abelian gauge factors  ($SU(2)$, $SU(3)$) denote the linear 
combinations  of the  Abelian  generators  that determine
the diagonal  generator(s) of the non-Abelian factors.

Due to the $SL(2,Z)_S$ symmetry, along with
each of the charge vectors  ${\vec \alpha}$ (as specified in
(\ref{line1})-(\ref{line4})), there is a  tower of dyonic configurations  
(including the $Z_2$  related  purely magnetic states)
with $p{\vec \beta}=q{\vec\alpha}$, where  $p$ and $q$ are the relative
co-primes \cite{SEN}. 
 These dyonic configurations 
 become massless at the same points of moduli space as
purely electric configurations. 

I now address massless dyonic states  
whose electric and magnetic charge vectors {\it are not parallel}. 
These states only preserve $1\over 4$ of $N=4$ supersymmetry.  The necessary
condition for them to become massless is that
 both the  electric and the  magnetic  charge vector
 norms satisfy:
 \begin{equation}
{\vec\alpha}^TL{\vec\alpha}=-2, \ \  
 {\vec\beta}^TL{\vec\beta}=-2.
\label{lcv}\end{equation}
These BPS-saturated states become massless
at  the points of moduli space  for which now  the following {\it three} constraints are satisfied: 
\begin{equation}  
{\vec\alpha}^T\mu_R{\vec\alpha}=0,\ \ {\vec\beta}^T\mu_R{\vec\beta}=0,\ 
\ {\vec\beta}^T\mu_R{\vec\alpha}=0, 
\label{const}
\end{equation} 
By explicit calculation I found that  the three constraints (\ref{const})
 are satisfied only at  the  following  two points:
\begin{equation}
T=U=1,  \ \ \ \ \ \ \ \ \ \ \ \ ({\vec \alpha},{\vec \beta})=({\vec\lambda}_{1\,\pm}, 
{\vec\lambda}_{2\,\pm}), \ \ 
\label{dy1}\end{equation}
\begin{equation}
T=U^*=e^{i {\pi \over 6}}, \ \ \ \ \ \ \ \ 
({\vec \alpha},{\vec \beta})=({\vec\lambda}_{i\,\pm},{\vec\lambda}_{j\,\pm}), 
\ \  [(i,j)=2,3,4,\ \ i< j].
\label{dy2}
\end{equation}
The charge assignments  for the four  massless dyonic  BPS-saturated  states
 (\ref{dy1}) at the self-dual point of the  two-circle were found in Ref. \cite{CYII}. At the point $T=U^*=e^{i {\pi \over 6}}$ there are
twelve massless dyonic BPS-saturated  states (\ref{dy2}).  In addition, there
is an  infinite $SL(2,Z)_S$ related tower of  massless states (including the
$Z_2$ related states with  electric and magnetic charge vectors  in
(\ref{dy1}) and (\ref{dy2}) interchanged).  Since these states
 belong to the highest
spin $3\over 2$-supermultiplet, they may contribute to the enhancement of
supersymmetry there
\footnote{The possibility of getting supersymmetry enhancement is discussed in \cite{MCHullse}.}
.
Note that   dyonic states  (\ref{dy1}) and (\ref{dy2}) {\it are not} in the perturbative spectrum of
toroidally compactified heterotic  string.

\section{Comments}

A few comments are in order. The  discussed BPS-saturated states become massless at special points  and hyper-surfaces of moduli space, regardless of the strength of the dilaton-axion coupling.\footnote{Note, that the BPS mass formula (\ref{ME}) is semi-positive definite for any asymptotic value of the axion field $\Psi_\infty$. This result is due to the fact that the  lattice vectors satisfy 
${\vec\alpha}^T\mu_R{\vec\alpha}\ge 0,\ \ {\vec\beta}^T\mu_R{\vec\beta}\ge0$,\ 
and $({\vec\alpha}^T\mu_R{\vec\alpha})( {\vec\beta}^T\mu_R{\vec\beta}) -
({\vec\beta}^T\mu_R{\vec\alpha})^2\ge 0$ everywhere in the moduli space.} 
Note also that all the discussed  states are singular four-dimensional
 solutions. Namely, for the solutions to be regular, {\it i.e.},  with the (Einstein frame) event  horizon, the norms of the lattice charge vectors  have to  satisfy the following constraints \cite{CTII}:
\begin{equation}
{\vec \alpha }^TL{\vec\alpha }> 0  \ ,\  \ \ \
{\vec \beta}^TL{\vec\beta}> 0\ , \  \ \  \
({\vec \beta}^T L {\vec \beta})
({\vec {\alpha}}^T L{\vec {\alpha}})-({\vec \beta}^T L  {\vec
{\alpha}})^2 > 0.
\label{norm}
\end{equation}
Since the norms  (\ref{elc}), (\ref{lcv}) of the massless BPS states 
 are negative, all the above
solutions  are singular  from the four-dimensional point of view.

The above solutions were obtained as semi-classical solutions of the toroidally compactified heterotic string; they are parameterized in terms of  classical bosonic fields  of heterotic string and (quantized) lattice  charge vectors, consistent with the  heterotic string constraints and  the DSZW quantization condition.  It is important to address the stability of these configurations, as well as to identify these semi-classical  solutions in terms of the $D-$brane\cite{POL} solutions of Type IIA  string.

\chapter{BPS states in type II superstrings}
\label{chapter:BPS states in type II superstrings}

\section{Introduction}

In the previous chapter, I studied a special class of BPS states of the heterotic string. In this chapter, I shall systematically construct a large class of four dimensional supersymmetric black hole solutions of toroidally compactified type IIA superstring theory by explicitly solving Killing Spinor equations. They correspond to orthogonally intersecting configurations in ten dimensions. With the Kaluza-Klein monopole, they are parametrized by four charges and preserve ${1 \over 8}$ of the $N=8$ supersymmetry. I shall find a simple map to associate each charge with the corresponding Killing spinor constraints. The embedding of the $N=4$ supersymmetry of toroidally compactified heterotic string into the $N=8$ supersymmetry of IIA superstring will be explicitly shown. I shall also find explicitly the configurations with only Ramond-Ramond charges, and those with both Neveu-Schwarz Neveu-Schwarz charges and Ramond-Ramond charges, including the dilaton and the internal metrics. The T-dual of these configurations shall be shown to satisfied the Killing spinor equations as well.

As explained in the Introduction (Chapter 1), supersymmetric black hole solutions with masses saturating the corresponding Bogomol'nyi bounds are essential in understanding the non-perturbative aspects of string theories. They are necessary to establish the proposed non-perturbative duality conjectures \cite{WitVaDi} - \cite{HoraWit} \cite{Font} \cite{Hull} of the five superstring theories, and also the unifying nature of the underlying M-theory \cite{MT}. These BPS-saturated states can also affect the low energy behaviour of the theories by becoming massless in special region of moduli space which parametrizes the underlying string vacua \cite{CYII} \cite{HTII} \cite{Strominger} - \cite{Beh}. That leads to gauge symmetry enhancement as well as supersymmetry enhancement. Furthermore, they provide a convenient background to study quantum gravity from the string perspective. In particular, the Bekenstein-Hawking entropies of certain four and five dimensional black holes of string theories were shown to match the corresponding degeneracy provided by microscopic string degree of freedom \cite{Mald1}. 

Four dimensional black holes are of particular interest. Not only because our observable world (with present technology) is four dimensional, there is the phenomena of having massless four dimensional black holes at maximally symmetrical point in the moduli space of the toroidally compactified heterotic string \cite{CK}\cite{Gibbons1}-\cite{Gibbons2}. Furthermore, it remains a challenging problem to work out the string degree of freedom in four dimensional black holes with Ramond-Ramond charges only and show explicitly that it matches the expected thermodynamical entropies.

Though the most general static spherically symmetric four dimensional black hole solutions of the effective supergravity of toroidally compactified heterotic string have been found in \cite{MCT4p} \cite{CTII} \cite{CYIII} \cite{CYI}, there is no systematic study on the four dimensional black hole solutions of toroidally compactified type II superstring theories. In addition to the black holes with four independent Neveu-Schwarz Neveu-Schwarz (NS-NS) charges like those of the heterotic string, there are type II black holes with Ramond-Ramond (RR) charges, and black holes with both NS-NS and RR charges. In \cite{Gaunt}, eleven dimensional orthogonally intersecting configurations without Kaluza-Klein monopole were studied. Upon compactification they can lead to four dimensional black hole solutions with three charges only. The spinor constraints were shown for cases with only two charges. In \cite{KT}, two black holes with four charges were obtained, but the spinor constrains were not shown
\footnote{There has been much progress in understanding non-orthogonally intersecting ten dimensional configurations \cite{Dymem}-\cite{CB}. They should lead to more general four dimensional configurations. My work on orthogonally intersecting configurations is only a step forward towards understanding four dimensional type II black holes.}.

This Chapter is an attempt to fill in this gap. I shall find static spherically symmetric four dimensional supersymmetric black hole solutions of the toroidally compactified type IIA superstring theory by explicitly solving the Killing spinor equations (KSEs). I will turn off all the scalar fields except the dilaton and the diagonal internal metric elements. That amounts to restricting my study only to orthogonally intersecting configurations in ten dimensions. 
\footnote{In \cite{CY4}, same kind of black hole solutions were studied. However, only the special cases when either the Kaluza-Klein fields or the three-form fields of the underlying $N=1$ 11 dimensional supergravity were turned on. The resulting black holes could only have two charges.} 

I start with the $N=1$ 11 dimensional supergravity (SG) theory in Section (3.2). I compactify this supergravity theory on the seven-torus $T^7$ \cite{Cremmers} \cite{Huq} and obtain the effective four dimensional action. Then I express the same four dimensional action in terms of fields of the compactified type IIA superstring on the six-torus $T^6$. The ten dimensional IIA supergravity is obtained by compactifying the $N=1$ 11 dimensional theory on a circle. The field redefinition rules that relate the four dimensional fields from SG on $T^7$ and IIA on $T^6$ are used in Section (3.3) to get the KSEs of IIA superstring from the KSEs of SG, which are obtained directly from the 11 dimensional gravitino transformation under supersymmetry. I shall put down a very simple set of rules for assigning a spinor constraint to each non-zero charge in Section (3.3.3). The spinor constraints determine the pattern of supersymmetry breaking. This set of supersymmetry breaking rules have been verified in all cases considered in this paper.

In Section (3.4), I solved the KSEs obtained in Section (3.3). Only the NS-NS charges are turned on in Section (3.4.1). The spinor degree of freedom for the $N=4$ supersymmetry are carried by spinors which from 10 dimensional perspective involve ${\bf both}$ the left Killing spinor and the right Killing spinor of the $N=2$ type IIA superstring theory. It is very different from the case of heterotic string \cite{CYI}. There all the Killing spinors originate from the same 10 dimensional spinor, which is the Killing spinor of the $N=1$ heterotic string.

In Section (3.4.2), I solve the KSEs with RR charges only. A solution with charges U-dual to the NS-NS charges of the configuration found in Section (3.4.1) is explicitly obtained. It corresponds to two D-2-brane orthogoanlly intersecting two D-4-branes in ten dimensions. I show that the configurations T-dual to this solution are also solutions of the KSEs. One of these configurations corresponds to a D-0-brane coupled to the intersection of three intersecting D-4-branes \cite{Larsen}. The other configuration corresponds to a D-6-brane containing three intersecting D-2-branes. These classical configurations are composed of a large number of D-branes and provide a consistency check of the D-brane intersection rules \cite{Dnotes}, which are defined microscopically.

In Section (3.4.3), I find solutions with both NS-NS charges and RR charges. The first solution that I explicitly obtain corresponds to a bound state of a D-2-brane, a D-4-brane, a gravitational wave running along the intersection of the D-branes, and a Kaluza-Klein monopole. I will also show that the configurations T-dual to the above configuration are also solutions of the KSEs. One of them corresponds to a bound state of a D-0-brane, a D-4-brane, a (winding) fundamental string lying orthogonally to the D-4-brane, and a Kaluza-Klein monopole. The monopole is supported on a direction orthogonal to both the D-4-brane and the fundamental string \cite{Johnson}. The other configuration corresponds to a bound state of a D-6-brane containing a D-2-brane which intersects a solitonic 5-brane, and a gravitational wave travelling on the intersection \cite{Tseytlin} \cite{Mald2}. The patterns of supersymmetry breaking are especially interesting in this case and will be studied in detail. 

In all cases, the BPS-saturated states with three to four charges preserve $N=1$ supersymmetry, those with two charges preserve $N=2$ supersymmetry, and those with only one charge preserve $N=4$ supersymmetry. Spinor constraints allow no more than four non-zero charges for the BPS-saturated states.

\section{Effective action from 11-d supergravity on $T^7$ in IIA language}

In this section, I derive the field redefinition rules between the 4d actions obtained from compactifying the N=1, d=11 SG on $T^7$ and that from N=2A, d=10 on $T^6$. That can simplify the way to obtain KSEs of the compactified IIA superstring in Section III. Most material in this section has been described in \cite{CY4}. This section is included here for the sake of completeness and for establishing notations.

The bosonic Lagrangian density of the $N=1$ 11 dimensional supergravity (SG) is \cite{Cremmers}:
\begin{equation}
{\cal L} = -{1\over 4}E^{(11)}[{\cal R}^{(11)} + {1\over {12}}
F^{(11)}_{MNPQ}F^{(11)\ MNPQ} - {8\over {12^4}}
\varepsilon^{M_1 \cdots M_{11}}F_{M_1 \cdots M_4}F_{M_5 \cdots M_8}
A_{M_9 M_{10} M_{11}}] ,
\label{L11d}
\end{equation}
where $E^{(11)} \equiv {\rm det}\, E^{(11)\ A}_M$ is the determinant of the Elfbein, ${\cal R}^{(11)}$ is the 11d Ricci scalar, and $F^{(11)}_{MNPQ} (\equiv 4\partial_{[M}A^{(11)}_{NPQ]})$ is the field strength associated with the three-form field $A^{(11)}_{MNP}$. The metric signature is ($+--\cdots -$), and ($A,B,...$), ($M,N,...$) denote flat and curved 11d indices respectively.

Dimensional reduction of the 11d SG to 4d on $T^7$ is achieved by  
the standard Kaluza-Klein (KK) Ansatz for the Elfbein and a consistent truncation of the other 11d fields. The field content of the resulting 4d theory includes the dilaton $\varphi \equiv {\rm ln}\,{\rm det}\,e^a_i$ with $e^a_i$ being the 4d vierbein, 7 KK Abelian gauge fields $B_{\mu}^i$, 35 pseudo-scalars $A_{ijk}$, 21 pseudo-vectors $A_{\mu\,ij}$ and 7 two-forms $A_{\mu\nu\,i}$. The two-forms $A_{\mu\nu\,i}$ are equivalent to (axionic) scalar fields $\varphi^i$ after making duality transformation in four dimensions. In order to ensure that the fields $A_{\mu\,ij}$ and $A_{\mu\nu\,i}$ are scalars under the internal coordinate transformation $x^i \to x^{\prime\ i} = x^i + \xi^i$, and transform as $U(1)$ gauge fields under the gauge transformation: $\delta A^{(11)}_{MNP} = \partial_M \zeta_{NP} + \partial_N \zeta_{PM} + \partial_P \zeta_{MN}$, one have to define new canonical 4-d fields:
\begin{equation}
A^{\prime}_{\mu\,ij} \equiv A_{\mu\,ij} - B^k_{\mu} A_{kij},\ \ \ \
A^{\prime}_{\mu\nu\,i} \equiv A_{\mu\nu\,i} - B^j_{\mu}A_{j\nu\,i} 
-B^j_{\nu}A_{\mu\,ji} + B^j_{\mu} B^k_{\nu} A_{jki} .
\label{cantensor}
\end{equation}
The bosonic action (\ref{L11d}) is reduced to the following effective 4-d action:
\begin{equation}
{\cal L} = -{1\over 4}e[{\cal R} - {1\over 2}\partial_{\mu} \varphi 
\partial^{\mu} \varphi +{1\over 4}\partial_{\mu} g_{ij} \partial^{\mu}
g^{ij} - {1\over 4} e^{\varphi}g_{ij}G^i_{\mu\nu}G^{j\,\mu\nu} 
+{1\over 2}e^{\varphi}g^{ik}g^{jl}F^4_{\mu\nu\,ij}
F^{4\,\mu\nu}_{\ \ \ \ kl} + \cdots ] ,
\label{L4df11d}
\end{equation}
where ${\cal R}$ is the 4-d Ricci scalar. The Einstein-frame 4d metric is $g_{\mu\nu} = \eta_{\alpha\beta}e^{\alpha}_{\mu}e^{\beta}_{\nu}$, and $G^i_{\mu\nu} \equiv \partial_{\mu} B^i_{\nu} - \partial_{\nu}B^i_{\mu}$, $F^4_{\mu\nu\,ij} \equiv F^{\prime}_{\mu\nu\,ij} + G^k_{\mu\nu}A_{ijk}$. The dots ($\cdots$) denote the terms involving the pseudo-scalars $A_{ijk}$ and the two-form fields $A_{\mu\nu\,i}$.

I am going to express the same Lagrangian (\ref{L4df11d}) in terms of fields from toroidally compactified IIA supergravity. The zero slope limit of IIA 10d superstring theory can be obtained by dimensional reduction of the 11d SG on a circle $S^1$ \cite{Huq}. The field content of the resulting 10d theory include the 10d dilaton $\Phi (\equiv {3 \over 2} \ln$ (radius of the circle)) which is expected from NS-NS sector of the IIA superstring theory, the Zehnbein $e^{(10)\, \breve{\alpha}}_{\breve{\mu}}$, and a KK Abelian guage field $B_{\breve{\mu}}$ corresponding to a one-form in RR sector
\footnote{I do not assume a diagonal 11d metric of the SG theory. I only assume a diagonal internal metric for the 10d superstring theory.}.  
Here, the breve denotes the 10d space-time vector index. And the 3-form $A^{(11)}_{MNP}$ is decomposed into $A_{\breve{\mu}\breve{\nu}\breve{\rho}}$ and $A_{\breve{\mu}\breve{\nu} 11} (\equiv A_{\breve{\mu}\breve{\nu}})$, with $A_{\breve{\mu}\breve{\nu} \breve{\rho}}$ being identified as a Ramond-Ramond (RR) 3-form and $A_{\breve{\mu}\breve{\nu}}$ the NS-NS 2-form. The 11d bosonic action (\ref{L11d}) becomes the following 10d, $N=2$ supergravity:
\begin{equation}
{\cal L} = {\cal L}_{NS} + {\cal L}_R , 
\label{N210d}
\end{equation}
with 
\begin{eqnarray}
{\cal L}_{NS} &=& -{1\over 4}e^{(10)}e^{-2\Phi}[{\cal R} + 
4\partial_{\breve{\mu}}\Phi \partial^{\breve{\mu}}\Phi - 
{1\over 3}F_{\breve{\mu}\breve{\nu}\breve{\rho}}
F^{\breve{\mu}\breve{\nu}\breve{\rho}}], 
\nonumber \\
{\cal L}_R &=& -{1 \over 4}e^{(10)}[{1\over 4}G_{\breve{\mu}\breve{\nu}}
G^{\breve{\mu}\breve{\nu}} +{1\over {12}}F^{\prime}_{\breve{\mu}\breve{\nu}
\breve{\rho}\breve{\sigma}} F^{\prime\, \breve{\mu}\breve{\nu}\breve{\rho}
\breve{\sigma}} - {6 \over {(12)^3}}\varepsilon^{\breve{\mu}_1 \cdots 
\breve{\mu}_{10}} F_{\breve{\mu}_1 \cdots \breve{\mu}_4}
F_{\breve{\mu}_5 \cdots \breve{\mu}_8} A_{\breve{\mu}_9 \breve{\mu}_{10}}],
\label{LNSR}
\end{eqnarray}
where $\cal R$ is the 10d string frame Ricci scalar, $F_{\breve{\mu}\breve{\nu}\breve{\rho}} \equiv 3\partial_{[\breve{\mu}}
A_{\breve{\nu}\breve{\rho}]}$, $G_{\breve{\mu}\breve{\nu}} \equiv 
2\partial_{[\breve{\mu}}B_{\breve{\nu}]}$, $F^{\prime}_{\breve{\mu}\breve{\nu}
\breve{\rho}\breve{\sigma}} \equiv 4\partial_{[\breve{\mu}}A_{\breve{\nu}
\breve{\rho}\breve{\sigma}]} - 4F_{[\breve{\mu}\breve{\nu}\breve{\rho}}
B_{\breve{\sigma}]}$, and $\varepsilon^{\breve{\mu}_1 \cdots \breve{\mu}_{10}} 
\equiv \varepsilon^{\breve{\mu}_1 \cdots \breve{\mu}_{10} 11}$.  
The ferminoic sector includes the Majorana gravitino $\psi_{\breve{\mu}}$ and the dilatino $\psi_{11}$. They originate from the 11d gravitino $\psi^{(11)}_M$ of SG, {\it i.e.}, $\psi^{(11)}_M = (\psi_{\breve{\mu}}, \psi_{11})$. Each of these spinors can be split into two Majorana-Weyl spinors with different chiralities.

I dimensionally reduced the 10d IIA supergravity by using the well-known KK ansatz for the Zehnbein and truncate the theory consistently. As assumptions, I set all scalar fields except the 10d dilaton and the ``internal'' metric elements corresponding to the Sechsbein to zero. The field content of the resulting 4d theory includes the string frame 4d Vierbein $e^{\alpha}_{\mu}$, the 6d Sechsbein $\bar{e}^a_m$, the 4d NS-NS dilaton $\phi \equiv \Phi - {1 \over 2} {\rm ln}\, {\rm det}\, \bar{e}^a_m$ (parameterizing the {\it string coupling}), 6 NS-NS KK Abelian gauge fields $\bar{B}^m_{\mu}$ ($m=1,...,6$), 6 NS-NS gauge fields ${\bar A}_{\mu\, n}$ originating from the 10d two-form NS-NS fields, one RR gauge field ${\bar B}_{\mu}$ originating from the 10d RR one-form, and 15 RR gauge field ${\bar A}_{\mu\, mn}$ originating from the 10d RR three-form fields. It should be noted that as I have turned off the scalar fields associated with the 10d $U(1)$ gauge field $B_{\breve{\mu}}$, ${\it i.e.}$, the internal metric
coefficients $g_{m7}$ of the 11d SG, the $SO(7)$ symmetry among the 7 KK 
gauge fields and among the 21 three-form  gauge fields of SG breaks down separately to the $SO(6)$ symmetry, which {\it do not mix} the gauge fields of RR 
and NS-NS sectors. For the RR sector, the RR vector transforms as a singlet of $SO(6)$, and the 15 gauge fields $\bar{A}_{\mu\, mn}$ transform as 
{\bf 15} antisymmetric representation of $SO(6)$. For the NS-NS sector, the 6 KK gauge fields, as well as the 6 gauge fields $\bar{A}_{\mu\, n}$, transform as a ${\bf 6}$ vector representation of $SO(6)$. 

The string-frame 4d bosonic action for IIA superstring is:
\begin{eqnarray}
{\cal L}_{II} = &-&{1\over 4}e[e^{-2\phi}({\cal R} + 4\partial_{\mu}
\phi \partial^{\mu}\phi + {1\over 4}\partial_{\mu}\bar{g}_{mn} 
\partial^{\mu}\bar{g}^{mn} - {1\over 4}\bar{g}_{mn}\bar{G}^m_{\mu\nu}
\bar{G}^{n\, \mu\nu} -\bar{g}^{mn}{\bar F}_{\mu\nu\, m}
{\bar F}^{\mu\nu}_{\ \ n}) 
\nonumber \\ 
&+&{1\over 4}e^{{\sigma}}\bar{G}_{\mu\nu}\bar{G}^{\mu\nu} + 
{1\over 2}e^{{\sigma}} \bar{g}^{mn}\bar{g}^{pq}
\bar{F}_{\mu\nu\, mp}\bar{F}^{\mu\nu}_{\ \ nq}] , 
\label{LIIAS}
\end{eqnarray}
where ${\sigma} \equiv {\rm ln}\, {\rm det}\, \bar{e}^a_m$ 
(parameterizing the volume of 6-torus), $\bar{g}_{mn} \equiv \eta_{ab} 
\bar{e}^a_m \bar{e}^b_n = -\bar{e}^a_m \bar{e}^a_n$, and $\bar{G}^m_{\mu\nu} 
\equiv \partial_{\mu} \bar{B}^m_{\nu} - \partial_{\nu} \bar{B}^m_{\mu}$.
Here, the field strengths $\bar{F}_{\mu\nu\, m}$, $\bar{G}_{\mu\nu}$ and 
$\bar{F}_{\mu\nu\, mn}$ are for the 6 NS-NS gauge fields, the RR gauge field originates from the 10d RR one-form, and the 15 RR gauge field originating from the 10d RR three-form fields, respectively. With the Weyl rescaling $g_{\mu\nu} \to g^E_{\mu\nu} = e^{-2\phi}g_{\mu\nu}$, I obtain the Einstein-frame action:
\begin{eqnarray}
{\cal L}_{II} = &-&{1\over 4}e^E[{\cal R}^E - 2\partial_{\mu}\phi 
\partial^{\mu} \phi + {1\over 4}\partial_{\mu}\bar{g}_{mn} 
\partial^{\mu}\bar{g}^{mn} - {1\over 4}e^{-2\phi}\bar{g}_{mn}
\bar{G}^m_{\mu\nu}\bar{G}^{n\,\mu\nu}-e^{-2\phi}\bar{g}^{mn}
{\bar F}_{\mu\nu\, m}{\bar F}^{\mu\nu}_{\ \ n} 
\nonumber \\ 
&+& {1\over 4}e^{\sigma}\bar{G}_{\mu\nu}\bar{G}^{\mu\nu} 
+ {1\over 2}e^{\sigma}\bar{g}^{mn}\bar{g}^{pq}
\bar{F}_{\mu\nu\, mp}\bar{F}^{\mu\nu}_{\ \ nq}],
\label{LIIAE}
\end{eqnarray}
where $e^E \equiv \sqrt{-{\rm det}\,g^E_{\mu\nu}}$ and ${\cal R}^E$ 
is the 4d Einstein-frame Ricci scalar defined in terms of the metric $g^E_{\mu\nu}$.

By keeping track of the field decomposition and redefinitions and comparing the two different descriptions of the same 4d Lagrangian, {\it i.e.}, one corresponding to 11d $\to$ 10d $\to$ 4d (\ref{LIIAE}) and the other corresponding to 11d $\to$ 4d (\ref{L4df11d}), one can express the fields in the 4d action of 11d SG in terms of those of IIA superstring,
\begin{eqnarray}
\varphi &=& -{4 \over 3} \phi + {1 \over 3} \sigma,\ \ \   
{\rm ln} \, e_7^{\hat 7} = {2\over 3}\phi + {1\over 3} \sigma,\ \ \  
e_m^{\hat m} = e^{ -{1 \over 3} \phi - {1 \over 6} \sigma } {\bar{e}}_m^{\hat m}, 
\nonumber \\
B^m_{\mu} &=& {\bar{B}}^m_{\mu}, \ \ \ \  B_{\mu}^7 = \bar{B}_{\mu} , \ \ \ 
A_{\mu m n} = \bar{A}_{\mu m n}, \ \ \ \  A_{\mu\,m7} = \bar{A}_{\mu\,m}, 
\label{Rel}
\end{eqnarray}

where $m,n = 1,...,6$. Flat indices are hatted, and the bar on Sechsbein has been dropped.

\section{Killing Spinor Equations}

The supersymmetry transformation of the gravitino field $\psi^{(11)}_M$ 
of the $N=1$ 11d SG (before compactification) in a bosonic background is \cite{Cremmers}: 
\begin{equation}
\delta \psi^{(11)}_M = D_M\, \varepsilon +{i\over 144} (\Gamma^{NPQR}_
{\ \ \ \ \ M} - 8\Gamma^{PQR}\delta^N_M)F_{NPQR}\,\varepsilon ,
\label{STG11d}
\end{equation}
where $D_M\,\varepsilon = (\partial_M + {1\over 4}\Omega_{MAB}
\Gamma^{AB})\,\varepsilon$ is the gravitational covariant derivative 
on the spinor $\varepsilon$, and $\Omega_{ABC} \equiv -\tilde{\Omega}_
{AB,C} + \tilde{\Omega}_{BC,A} - \tilde{\Omega}_{CA,B}$ 
($\tilde{\Omega}_{AB,C} \equiv E^{(11)\,M}_{[A}E^{(11)\,N}_{B]}
\partial_N E^{(11)}_{MC}$) is the spin connection defined in terms of 
the Elfbein. The gravitino transformation (\ref{STG11d}) expressed in terms of 4d canonical fields obtained by compactifying SG directly on $T^7$ are
\begin{eqnarray}
\delta \hat {\psi}_{\mu} &=& \partial_{\mu} \varepsilon + {1\over 4} 
\omega_{\mu\beta\gamma} \gamma^{\beta\gamma} \varepsilon - 
{1 \over 4}e^{\alpha}_{\mu}\eta_{\alpha[\beta}e^{\nu}_{\gamma ]}
\partial_{\nu} \varphi \gamma^{\beta\gamma} \varepsilon + 
{1 \over 8}(e^l_b \partial_{\mu}e_{lc} - e^l_c 
\partial_\mu e_{lb})\gamma^{bc} \varepsilon 
\nonumber \\
& &+ {i \over {24}}e^{\varphi \over 2} 
F_{\nu\rho\,ij}\gamma^{\nu\rho}_{\ \ \mu}\gamma^{ij} \varepsilon -
{i \over 6}e^{\varphi \over 2}F_{\mu\nu\,ij}
\gamma^{\nu}\gamma^{ij}\varepsilon + {1 \over 4} e^{\varphi \over 2} e_{ib} G^i_{\mu \alpha} \gamma^{\alpha 5} \gamma^b \varepsilon, 
\nonumber \\
\delta \psi_k &=& -{1\over 4}e^{\varphi \over 2}(\partial_{\rho} e_{kb} + 
e^c_k e^l_b \partial_{\rho} e_{lc})\gamma^{\rho 5} 
\gamma^b \varepsilon +{i \over {24}}e^{\varphi}F_{\mu\nu\,ij}
\gamma^{\mu\nu 5}\gamma^{ij}_{\ \  k}\varepsilon - {i\over 6}
e^{\varphi}F_{\mu\nu\,kl}\gamma^{\mu\nu 5}\gamma^l\varepsilon 
\nonumber \\
& &+ {1 \over 8} e^{\varepsilon} g_{kn} G^n_{\beta \alpha} \gamma^{\alpha \beta} \varepsilon
\label{STG411}
\end{eqnarray}
where $\delta \hat {\psi}_{\mu} \equiv \delta \psi_{\mu} - B^m_{\mu} \delta \psi_m$,  $\omega_{\mu\beta\gamma}$ is the spin-connection defined in terms 
of the Vierbein $e^{\alpha}_{\mu}$ and $[a\,\cdots\,b]$ denotes antisymmetrization of the corresponding indices.  For the 11d gamma matrices, which satisfy the $SO(1,10)$ Clifford algebra $\{ \Gamma^A, \Gamma^B \} = 2\eta^{AB}$, one have used the following representation: 
\begin{equation}
\Gamma^{\alpha} = \gamma^{\alpha} \otimes I , \ \ \ \ \ \ 
\Gamma^a = \gamma^5 \otimes \gamma^a , 
\label{DFGAMMA}
\end{equation}
where $\{ \gamma^{\alpha}, \gamma^{\beta} \} = 2\eta^{\alpha\beta}$, 
$\{ \gamma^a , \gamma^b \} = -2\delta^{ab}$, $I$ is the $8 \times 8$ 
identity matrix and $\gamma^5 \equiv i\gamma^0 \gamma^1 \gamma^2 
\gamma^3$. Multiple indices of the gamma matrices are antisymmetrized, ({\it e.g.}, $\gamma^{\alpha\beta} \equiv \gamma^{[\alpha}\gamma^{\beta]}$) and the gamma matrices with curved indices are defined by multiplying with the Vierbein, {\it e.g.}, $\gamma^{\mu} \equiv e^{\mu}_{\alpha} \gamma^{\alpha}$. With the representation of the gamma matrices (\ref{DFGAMMA}), the spinor index $A$ of an 11d spinor, $\varepsilon^A$, can be decomposed into $A = ({\bf a}, {\bf m})$, {\it i.e.}, $\varepsilon^A = \varepsilon^{({\bf a}, {\bf m})}$, where ${\bf a}=1,...,4$ is the spinor index for a four component 4d spinor and ${\bf m}=1,...,8$ is the index for the spinor representation of the group $SO(7)$.

The supersymmetry transformations of the gravitinos and modulinos given in (\ref{STG411}) is a simple sum of the corresponding transformations from the Kaluza-Klein sector and the 3-form fields sector \cite{CY4}. That is not surprising because the effective Lagrangian (\ref{L4df11d}) (after setting $A_{ijk}$ and $A_{\mu \nu\, i}$ to zero) is just a simple sum of the Ricci scalar, the kinetic energy of the scalar fields, and the kinetic energies of the gauge fields. The effective Lagrangian contains no terms that describe any mixing between the two types of gauge fields. However, careful examination of the definitions of field strengths and tedious manipulation are required to verify explicitly the expected supersymmetry transformation rules.

I shall evaluate the KSEs in the following subsections under the assumptions of time-independence and spherical symmetry. I will turn off all the scalar fields except the dilaton and the diagonal internal metric elements.

\subsection{From SG perspective}

In this subsection, I express the KSEs in terms of 4d fields from the compactified $N=1$ 11d SG. With spherical symmetry, the 4d space-time metric can be taken as 
\begin{equation}
g_{\mu\nu}{\rm d}x^{\mu}{\rm d}x^{\nu} = \lambda (r){\rm d}t^2
- \lambda^{-1} (r) {\rm d}r^2 - R(r)({\rm d}\theta^2 + {\rm sin}^2 \theta
{\rm d}\phi^2) .
\label{MET4d}
\end{equation}
Field strengths $G_{\mu \nu}^i$ and $F_{\mu \nu ij}$ for KK $U(1)$ gauge field and that from three-form fields, respectively, have the following non-zero components:
\begin{eqnarray}
G^i_{tr} = {{g^{ij} Q_j} \over {R e^{\varphi}}}, \ \ \ \ \ 
G^i_{\theta \phi} = P^i {\rm sin} \theta ,
\nonumber \\
F_{trij} = {{g_{ik} g_{jl} Q^{kl}} \over {R e^{\varphi}}}, \ \ \ \ \ 
F_{\theta \phi ij} = P_{ij} {\rm sin} \theta ,
\label{PQ}
\end{eqnarray}
where $Q_i$ ($P^i$) and $Q^{ij}$ ($P_{ij}$) are the physical electric (magnetic) charges
\footnote{To simplify the formulae, I have assumed that the internal metric and the dilaton approach unity and zero respectively as r $\rightarrow \infty$. This can always be done through manifest $SL(7,R)$ symmetry and S-duality.}.
The internal metric, $g_{ij}$, is proportional to $\delta_{ij}$ by assumption. The supersymmetry transformations of the gravitinos and modulinos, (\ref{STG411}), can thus be simplified:
\begin{eqnarray}
{i \over 4} \left( - {\bf P}^m \mp i {\bf Q}_m \right) \varepsilon_{ul} - {1 \over 2} R \sqrt{\lambda} ( \ln e_m^{\hat m} )' \gamma^m \varepsilon_{lu}+ {1 \over 12} ( \pm {\bf P}_{ij} - i {\bf Q}^{ij} ) \gamma^{ij}_{\ \ m} \varepsilon_{ul}  
\nonumber \\
 + {1 \over 3} ( \mp {\bf P}_{mi} + i {\bf Q}^{mi} ) \gamma^i \varepsilon_{ul} = 0 ,
\label{DSM11}
\end{eqnarray}
\begin{equation}
\mp {R \over \sqrt{\lambda}} \left( \lambda ' - \lambda \varphi ' \right) \varepsilon_{ul} - {\bf Q}_i \gamma^i \varepsilon_{lu} + {1 \over 3}( {\bf P}_{ij} \mp 2 i {\bf Q}^{ij} ) \gamma^{ij} \varepsilon_{lu} = 0 ,
\label{DST11}
\end{equation}
\begin{equation}
{i \over 2} \sqrt{R} \varepsilon_{ul} - {i \over 4} \sqrt{\lambda} ( R' - R \varphi ' ) \varepsilon_{ul} + {1 \over 4} {\bf P}^i \gamma^i \varepsilon_{lu} + {1 \over 12} ( - {\bf Q}^{ij} \mp 2 i {\bf P}_{ij} ) \gamma^{ij} \varepsilon_{lu} = 0 ,
\label{DSTH11}
\end{equation}
\begin{equation}
R \sqrt{\lambda} \partial_r \varepsilon_{ul} \pm {1 \over 4} {\bf Q}_i \gamma^i \varepsilon_{lu} + {1 \over 12}( \mp {\bf P}_{ij} + 2 i {\bf Q}^{ij} ) \gamma^{ij} \varepsilon_{lu} = 0 ,
\label{DSR11}
\end{equation}
\begin{equation}
(\varepsilon^{1, {\bf m}}_{u, \ell}, \varepsilon^{2, {\bf m}}_{u,\ell}) = 
e^{i\sigma^2 \theta /2} e^{i\sigma^3 \phi /2}(a^{1, {\bf m}}_{u,\ell}(r), 
a^{2, {\bf m}}_{u,\ell}(r)) , 
\label{DSTHS11}
\end{equation}
where
\begin{eqnarray}
{\bf{Q}}_m \equiv e^{- {\varphi \over 2}} e^m_{\hat m} Q_m,\ \ \ \ \ {\bf{P}}^m \equiv e^{\varphi \over 2} e_m^{\hat m} P_m ,
\nonumber \\
{\bf{Q}}^{mi} \equiv e^{- {\varphi \over 2}} e_m^{\hat m} e_i^{\hat i} Q^{mi}, \ \ \ \ \ {\bf{P}}_{mi} \equiv e^{\varphi \over 2} e^m_{\hat m} e^i_{\hat i} P_{mi} ,
\label{DFBPQ11}
\end{eqnarray}
$\varepsilon^{\bf m}_{u, \ell}$ are the upper (or lower) two components of the 4d four-component spinor $\varepsilon^{\bf m}$ {\it i.e.}, $(\varepsilon^{\bf m})^T = (\varepsilon^{\bf m}_u, \varepsilon^{\bf m}_\ell )$, and $a^{\bf m}_{u,\ell}(r)$ are the corresponding two-component spinors
\footnote{I also call the quantities $\varepsilon^{\bf m}_u$ and $\varepsilon^{\bf m}_l$ two-component ${\it spinors}$. That only means that they are the upper and lower two components of the 4d spinors, $\varepsilon^{\bf m}$, respectively. The upper (lower) two components of the 4d Majorana spinors do not have definite transformation properties under the 4d Lorentz group.}
that depend on the radial coordinate $r$ only
\footnote{Note that I suppress the index ${\bf m}$ of the spinors in all of the equations in this paper.}.
The KSEs (\ref{DSR11}) and (\ref{DSTHS11}) determine the radial and angular dependence of the spinors. As all information of the fields and the constraints on the spinors are contained in (\ref{DSM11}) to (\ref{DSTH11}), I shall not elaborate equations (\ref{DSR11}) and (\ref{DSTHS11}) any further. It should be noted that $\gamma^i$ act only on the index $\bf m$. I have used an explicit representation of the 4d gamma matrices to write the KSEs in terms of relations between upper and lower two components of the 4d space-time spinors. 

\subsection{From IIA perspective}

In order to rewrite the KSEs (\ref{DSM11})-(\ref{DSTH11}) in IIA language, I define the projection operators
\begin{equation}
P_+ \equiv {1 \over 2} \left( 1 - i \gamma^7 \right), \ \ \ \ \
P_- \equiv {1 \over 2} \left( 1 + i \gamma^7 \right),
\label{PP}
\end{equation}
and project the two components of $\varepsilon$,
\begin{equation}
\varepsilon = \varepsilon^+ + \varepsilon^- ,
\label{ESPM}
\end{equation}
where
\begin{equation}
\varepsilon^+ \equiv P_+ \varepsilon,\ \ \ \ \ \varepsilon^- \equiv P_- \varepsilon .
\label{DFESPM}
\end{equation}
The two types of spinors, $\varepsilon^{{\bf m}+}$ and $\varepsilon^{{\bf m}-}$, originate from the two spinors associated with the $N=2$ supersymmetry of the IIA superstring in 10 dimensions
\footnote{It can be shown that the chirality of a 10d Majorana-Weyl spinor is labelled by the corresponding eigenvalue of $\gamma^7$ which only acts on the index ${\bf m}$, when the 10d Lorentz group is reduced to a direct product of the 4d Lorentz group and the 6d rotation group.}.

With (\ref{Rel}), I rewrite the SG charges defined in (\ref{DFBPQ11}) in terms of the charges of IIA superstring as:
\begin{eqnarray}
{\bf{Q}}_i = ( 1 - \delta^7_i ) {\bf{Q}}^{NK}_i - \delta^7_i {\bf{Q}}^{RK},\ \ \ \ \ {\bf{P}}^i = ( 1 - \delta^7_i ) {\bf{P}}^i_{NK} + \delta^7_i {\bf{P}}_{RK}
\nonumber \\
{\bf{Q}}^{ij} = ( 1 - \delta^7_i ) ( 1 - \delta^7_j ) {\bf{Q}}^{ij}_{RF} - ( 1 - \delta^7_i ) \delta^7_j {\bf{Q}}^i_{NF} + ( 1 - \delta^7_j ) \delta^7_i {\bf{Q}}^j_{NF}
\nonumber \\
{\bf{P}}_{ij} = ( 1 - \delta^7_i ) ( 1 - \delta^7_j ) {\bf{P}}_{ij}^{RF} + ( 1 - \delta^7_i ) \delta^7_j {\bf{P}}_i^{NF} - ( 1 - \delta^7_j ) \delta^7_i {\bf{P}}_j^{NF}
\label{PQ11NSR}
\end{eqnarray}
where
\begin{eqnarray}
{\bf{Q}}^{NK}_i \equiv e^{\phi} \bar{e}^i_{\hat i} Q^{NK}_i, \ \ \ \ \ {\bf{Q}}^{RK} \equiv e^{- {1 \over 2} \sigma} Q^{RK}
\nonumber \\
{\bf{P}}_{NK}^i \equiv e^{- \phi} \bar{e}_i^{\hat i} P_{NK}^i, \ \ \ \ \ {\bf{P}}_{RK} \equiv e^{{1 \over 2} \sigma} P_{RK}
\nonumber \\
{\bf{Q}}^{i}_{NF} \equiv e^{\phi} \bar{e}_i^{\hat i} Q_{NF}^i, \ \ \ \ \ {\bf{Q}}^{ij}_{RF} \equiv e^{- {1 \over 2} \sigma} \bar{e}^{\hat i}_i \bar{e}^{\hat j}_j Q^{ij}_{RF}
\nonumber \\
{\bf{P}}_{i}^{NF} \equiv e^{- \phi} \bar{e}^i_{\hat i} P^{NF}_i, \ \ \ \ \ {\bf{P}}_{ij}^{RF} \equiv e^{ {1 \over 2} \sigma} \bar{e}_{\hat i}^i \bar{e}_{\hat j}^j P^{RF}_{ij}
\label{PQNSR}
\end{eqnarray}
$N, R, K, F$ indicate that the charge is from NS-NS sector, RR sector, with KK origin, and with 1-form, 3-form (in RR sector) or 2-form (in NS-NS sector) origin respectively. The index, $i$, indicates that the charge is associated with the $i$ th compactified dimension. From (\ref{PQ}), $G^7_{tr} \rightarrow - {{Q_7} \over r^2}$ as $g_{77} \rightarrow -1$ asymtotically. As ${\bar B}_{\mu} \equiv B^7_{\mu}$ from (\ref{Rel}) and $Q^{RK}$ is defined to be the charge of the gauge field ${\bar B}_{\mu}$, ${\it i.e.}$ ${\bar G}_{tr} \rightarrow {Q^{RK} \over r^2}$, $Q^{RK} = -Q_7$. Similarly, $Q^{NF}_i = - Q^{i7}$ as $F_{tri7}$ contains the extra factor $g_{77}$, compared with the definition of $Q^i_{NF}$.

With equations (\ref{Rel}) and (\ref{PQ11NSR}), I act $P_+, P_-$ on both sides of (\ref{DSM11}) and get
\footnote{I shall normalize the charges such that $(Q_{NF}, P^{NF}, Q_{RF}, P^{RF}) \rightarrow {1 \over 2} (Q_{NF}, P^{NF}, Q_{RF}, P^{RF})$, to make the equations more symmetrical.}:
\begin{eqnarray}
- R \sqrt{\lambda} ( \ln \bar{e}^{\hat{m}}_m )' \varepsilon^+_{ul} =
{1 \over 2} \left[ \left( \mp {\bf{Q}}^{NK}_m - {\bf{Q}}^m_{NF} \right) + i \left( - {\bf{P}}^m_{NK} \mp {\bf{P}}^{NF}_m \right) \right] \gamma^m \varepsilon^-_{lu} 
\nonumber \\
+ {1 \over 4} \left[ {1 \over 2} \left( \mp {\bf{P}}^{RF}_{ij} - i {\bf{Q}}^{ij}_{RF} \right) \gamma^{ij} + 2 \left( \pm {\bf{P}}^{RF}_{mi} + i {\bf{Q}}^{mi}_{RF} \right) \gamma^{mi} + {\bf{P}}_{RK} \pm i {\bf{Q}}^{RK} \right] \varepsilon^+_{lu}
\nonumber 
\end{eqnarray}
\begin{eqnarray}
- R \sqrt{\lambda} ( \ln \bar{e}^{\hat{m}}_m )' \varepsilon^-_{ul} =
{1 \over 2} \left[ \left( \mp {\bf{Q}}^{NK}_m + {\bf{Q}}^m_{NF} \right) + i \left( - {\bf{P}}^m_{NK} \pm {\bf{P}}^{NF}_m \right) \right] \gamma^m \varepsilon^+_{lu} 
\nonumber \\
+ {1 \over 4} \left[ {1 \over 2} \left( \mp {\bf{P}}^{RF}_{ij} - i {\bf{Q}}^{ij}_{RF} \right) \gamma^{ij} + 2 \left( \pm {\bf{P}}^{RF}_{mi} + i {\bf{Q}}^{mi}_{RF} \right) \gamma^{mi} - {\bf{P}}_{RK} \mp i {\bf{Q}}^{RK} \right] \varepsilon^-_{lu} ,
\label{NSRM}
\end{eqnarray}
where $m =$ 1,...6. From $\delta \hat {\psi}_7 = 0$, 
\begin{eqnarray}
R \sqrt{\lambda} ( \phi + {1 \over 2} \sigma )' \varepsilon^+_{ul} =
{1 \over 4} \left[ 3 \left( - {\bf{P}}_{RK} \mp i {\bf{Q}}^{RK} \right) + {1 \over 2} \left( \pm {\bf{P}}^{RF}_{ij} + i {\bf{Q}}^{ij}_{RF} \right) \gamma^{ij} \right] \varepsilon^+_{lu} 
\nonumber \\
+ {1 \over 2} \left( \pm i {\bf{P}}^{NF}_i + {\bf{Q}}^{i}_{NF} \right) \gamma^{i} \varepsilon^-_{lu}
\nonumber 
\end{eqnarray}
\begin{eqnarray}
R \sqrt{\lambda} ( \phi + {1 \over 2} \sigma )' \varepsilon^-_{ul} =
{1 \over 4} \left[ -3 \left( - {\bf{P}}_{RK} \mp i {\bf{Q}}^{RK} \right) + {1 \over 2} \left( \pm {\bf{P}}^{RF}_{ij} + i {\bf{Q}}^{ij}_{RF} \right) \gamma^{ij} \right] \varepsilon^-_{lu}
\nonumber \\
 - {1 \over 2} \left( \pm i {\bf{P}}^{NF}_i + {\bf{Q}}^{i}_{NF} \right) \gamma^{i} \varepsilon^+_{lu} ,
\label{NSR7}
\end{eqnarray}
With (\ref{NSR7}), I can write (\ref{DST11}) and (\ref{DSTH11}) respectively as:
\begin{eqnarray}
R \sqrt{\lambda} \left( \ln {\lambda} + 2 {\phi} \right)' \varepsilon^+_{ul} =  {1 \over 4} \left[ \left( \pm {\bf P}^{RF}_{ij} - i {\bf Q}^{ij}_{RF} \right) \gamma^{ij} + 2 \left( - {\bf P}_{RK} \pm i {\bf Q}^{RK} \right) \right] \varepsilon^+_{lu}
\nonumber \\
+ \left( \mp {\bf Q}^{NK}_i + {\bf Q}^i_{NF} \right) \gamma^i \varepsilon^-_{lu} ,
\nonumber \\
R \sqrt{\lambda} \left( \ln {\lambda} + 2 {\phi} \right)' \varepsilon^-_{ul} = {1 \over 4} \left[ \left( \pm {\bf P}^{RF}_{ij} - i {\bf Q}^{ij}_{RF} \right) \gamma^{ij} - 2 \left( - {\bf P}_{RK} \pm i {\bf Q}^{RK} \right) \right] \varepsilon^-_{lu} 
\nonumber \\
+ \left( \mp {\bf Q}^{NK}_i - {\bf Q}^i_{NF} \right) \gamma^i \varepsilon^+_{lu} ,
\label{NSRT}
\end{eqnarray}
\begin{eqnarray}
R \sqrt{\lambda} \left( \ln {\lambda} - 2 {\phi} \right)' \varepsilon^+_{ul} = {1 \over 4} \left[ \left( \pm {\bf P}^{RF}_{ij} - i {\bf Q}^{ij}_{RF} \right) \gamma^{ij} + 2 \left( - {\bf P}_{RK} \pm i {\bf Q}^{RK} \right) \right] \varepsilon^+_{lu}
\nonumber \\
+i \left( {\bf P}_{NK}^i \mp {\bf P}_i^{NF} \right) \gamma^i \varepsilon^-_{lu},
\nonumber \\
R \sqrt{\lambda} \left( \ln {\lambda} - 2 {\phi} \right)' \varepsilon^-_{ul} = {1 \over 4} \left[ \left( \pm {\bf P}^{RF}_{ij} - i {\bf Q}^{ij}_{RF} \right) \gamma^{ij} - 2 \left( - {\bf P}_{RK} \pm i {\bf Q}^{RK} \right) \right] \varepsilon^-_{lu} 
\nonumber \\
+i \left( {\bf P}_{NK}^i \pm {\bf P}_i^{NF} \right) \gamma^i \varepsilon^+_{lu},
\label{NSRTH}
\end{eqnarray}
Therefore NS-NS charges relate spinors with different chiralities (from the 10d view point), while RR charges relate spinors with the same chirality.

\subsection{Spinor constraints}

In this subsection, I shall put down a map to associate a spinor constraint with each charge in the KSEs. The patterns of supersymmetry breaking are governed by the spinor constraints. Each spinor constraint reduces half of the spinor degree of freedom, leading to the breaking of half of the supersymmetry. From (\ref{NSRM}) to (\ref{NSRTH}), one sees that each charge is associated with a particular spinor constraint when only that particular charge is non-zero. For example, if only $Q^{NK}_1$ is non-zero there will be two sets of consistent KSEs. One set relates $\varepsilon^+_u$ and $\varepsilon^-_l$, with the spinor constraint: $\eta_* \varepsilon^+_u = \gamma^1 \varepsilon^-_l$, where $\eta_*$ = $\pm 1$. The other set relates $\varepsilon^-_u$ and $\varepsilon^+_l$, with the spinor constraint: $\eta_* \varepsilon^-_u = \gamma^1 \varepsilon^+_l$. 

Here I collect all the eight spinor constraints associated with the eight different types of charges: 
\begin{eqnarray}
Q^{NK}_m: \ \ \ \ \ \eta_1 \varepsilon^+_u = \gamma^m \varepsilon^-_l, \ \ \ \ \ \eta_1 \varepsilon^-_u = \gamma^m \varepsilon^+_l,
\nonumber
\end{eqnarray}
\begin{eqnarray}
Q^{NF}_m: \ \ \ \ \ \eta_2 \varepsilon^+_u = \gamma^m \varepsilon^-_l, \ \ \ \ \ \eta_2 \varepsilon^-_u = - \gamma^m \varepsilon^+_l,
\nonumber
\end{eqnarray}
\begin{eqnarray}
P^{NK}_m: \ \ \ \ \ \eta_3 \varepsilon^+_u = i \gamma^m \varepsilon^-_l, \ \ \ \ \ \eta_3 \varepsilon^-_u = i \gamma^m \varepsilon^+_l,
\nonumber
\end{eqnarray}
\begin{eqnarray}
P^{NF}_m: \ \ \ \ \ \eta_4 \varepsilon^+_u = i \gamma^m \varepsilon^-_l, \ \ \ \ \ \eta_4 \varepsilon^-_u = -i \gamma^m \varepsilon^+_l,
\nonumber
\end{eqnarray}
\begin{eqnarray}
Q^{RF}_{ij}: \ \ \ \ \ \eta_5 \varepsilon^+_u = i \gamma^{ij} \varepsilon^+_l, \ \ \ \ \ \eta_5 \varepsilon^-_u = i \gamma^{ij} \varepsilon^-_l,
\nonumber
\end{eqnarray}
\begin{eqnarray}
P^{RF}_{ij}: \ \ \ \ \ \eta_6 \varepsilon^+_u = \gamma^{ij} \varepsilon^+_l, \ \ \ \ \ \eta_6 \varepsilon^-_u = \gamma^{ij} \varepsilon^-_l,
\nonumber
\end{eqnarray}
\begin{eqnarray}
Q^{RK}: \ \ \ \ \ \eta_7 \varepsilon^+_u = - i \varepsilon^+_l, \ \ \ \ \ \eta_7 \varepsilon^-_u = i \varepsilon^-_l,
\nonumber
\end{eqnarray}
\begin{equation}
P^{RK}: \ \ \ \ \ \eta_8 \varepsilon^+_u = - \varepsilon^+_l, \ \ \ \ \ \eta_8 \varepsilon^-_u = \varepsilon^-_l .
\label{GSC}
\end{equation}
where $\eta_i = \pm 1$. The bosonic configurations in general depend on the $\eta$s. 

After explicitly solving the KSEs, I find that the spinor constraints associated with each multi-charged configuration considered in this paper are precisely the same as those expected from (\ref{GSC}). Each non-zero charge brings a spinor constraint with the same form as that from (\ref{GSC}) independently
\footnote{One implication of these spinor constraints is that there is no supersymmetric configuration that contains both $P^{RK}$ and $P^{NK}$ or both $P^{RK}$ and $Q^{NF}$, under my working assumptions.}. 
Therefore, I have found a very simple set of supersymmetry breaking rules (\ref{GSC}) for the 4d black hole solutions of the toroidally compactified IIA supergravity.

\section{BPS configurations}

I shall find explicit solutions to the KSEs in this Section. In Section (3.4.1), only NS-NS charges are turned on, the RR charges are turned off. Then in Section (3.4.2), only RR charges are turned on. In Section (3.4.3), configurations with charges from both NS-NS and RR sectors shall be discussed.

\subsection{Neveu-Schwarz-Neveu-Schwarz sector}

With the RR charges set to zero, the KSEs (\ref{NSRM}) - (\ref{NSRTH}) can be rewritten as two sets of consistent KSEs. One set relates the spinors: ${\varepsilon}^+_l$ to ${\varepsilon}^-_u$, and the other set relates ${\varepsilon}^+_u$ to ${\varepsilon}^-_l$. These two sets of KSEs can be written together as follows:
\begin{equation}
-R \sqrt{\lambda} ( \ln \bar{e}^{\hat{m}}_m )' \varepsilon_{ul} =
{1 \over 2} \left[ \pm \left( \eta_* {\bf{Q}}^{NK}_m - {\bf{Q}}_m^{NF} \right) + i \left( -{\bf{P}}_m^{NK} + \eta_* {\bf{P}}^{NF}_m \right) \right] \gamma^m \varepsilon_{lu}
\label{HETME}
\end{equation}
\begin{equation}
R \sqrt{\lambda} \left( {\phi} + {1 \over 2} \sigma \right)' \varepsilon_{ul} =
{1 \over 2} \left( - i \eta_* {\bf{P}}^{NF}_i \pm {\bf{Q}}_{i}^{NF} \right) \gamma^{i} \varepsilon_{lu}
\label{HETD}
\end{equation}
\begin{equation}
R \sqrt{\lambda} \left( \ln {\lambda} + 2 {\phi} \right)' \varepsilon_{ul} = \pm \left( \eta_* {\bf Q}^{NK}_i + {\bf Q}_i^{NF} \right) \gamma^i \varepsilon_{lu}
\label{HETP}
\end{equation}
\begin{equation}
R \sqrt{\lambda} \left( \ln {\lambda} - 2 {\phi} \right)' \varepsilon_{ul} =i  \left( {\bf P}^{NK}_i + \eta_* {\bf P}_i^{NF} \right) \gamma^i \varepsilon_{lu}, \label{HETM}
\end{equation}
where $\eta_*$ = 1 when $({\varepsilon}_u, {\varepsilon}_l) \equiv ({\varepsilon}^+_l, {\varepsilon}^-_u)$, and $\eta_*$ = -1 when $({\varepsilon}_u, {\varepsilon}_l) \equiv ({\varepsilon}^+_u, {\varepsilon}^-_l)$. Note that the two components of $\delta \hat {\psi}_{7}$ of SG with opposite chiralities from the 10d view point has been identified with the two dilatinos of the IIA superstring in 10 dimensions. 

The structure of this set of KSEs is identical to that of the heterotic string \cite{CYI}. Actually, one can reproduce the KSEs of the toroidally compactified heterotic string with the above KSEs of the toroidally compactified IIA superstring at each $\eta_*$ with the following maps:
\begin{equation}
\varepsilon^+_l \rightarrow \varepsilon_u, \ \ \ \
\varepsilon^-_u \rightarrow \varepsilon_l, \ \ \ \, \eta_* = 1,
\label{MAP1}
\end{equation}
\begin{equation}
\varepsilon^+_u \rightarrow \varepsilon_u, \ \ \ \
\varepsilon^-_l \rightarrow \varepsilon_l, \ \ \ \, \eta_* = -1,
\label{MAP2}
\end{equation}
\begin{equation}
\eta_* Q^{NK}_m \rightarrow Q^{(1)}_m, \ \ \ \ 
\eta_* P^{NF}_m \rightarrow P^{(2)}_m, \ \ \ \, \eta_* = \pm 1,
\label{MAP12}
\end{equation}
where the quantities on the left of $\rightarrow$ belong to the compactified IIA superstring, and the quantities on the right belong to the compactified heterotic string. The superscript 1 and 2 indicates the origin of the charges, ${\it i.e.}$, KK gauge fields or two-form fields, respectively.

In the case of heterotic string, which has $N=1$ in 10d and $N=4$ in 4d, the KSEs relate the upper and lower components of the 4d spinors which originate from the same 10d Majorana-Weyl spinor with a definite chirality. While in the case of IIA superstring, which has $N=2$ in 10d and $N=8$ in 4d, the KSEs relate the upper(lower) components of the 4d spinors which originate from a 10d Majorana-Wely spinor of certain chirality to the lower(upper) components of spinors which originate from another 10d Majorana-Wely spinor of opposite chirality. 

From (\ref{HETP}), non-zero $( \eta_* {\bf Q}_i^{NK} + {\bf Q}_i^{NF})$ gives a spinor constraint of the form : $\varepsilon_l = - \eta_* \eta_+ \gamma^i \varepsilon_u$. From (\ref{HETM}), non-zero $( {\bf P}_j^{NK} + \eta_* {\bf P}_j^{NF})$ gives a spinor constraint of the form : $\varepsilon_l = -i \eta_- \gamma^j \varepsilon_u$, where $\eta_{\pm}$ can be equal to $+1$ or $-1$. Therefore the maximum number of charges allowed by constraints on spinors is four, with one electric NK and one electric NF charge from the same compactified dimension, and one magnetic NK and one magnetic NF charge from another compactified dimension
\footnote{I follow the same line of argument used in finding the generating solution for the supersymmetric, spherically symmetric solutions in Abelian Kaluza-Klein theory \cite{MCKK}.}.
Without loss of generality, I choose the non-zero charges to be: $P^{NK}_1, P^{NF}_1, Q^{NK}_2, Q^{NF}_2$. Solving (\ref{HETME}) to (\ref{HETM}), I get the fields:
\begin{eqnarray}
{ \lambda = { r^2 \over { \left[ ( r + \eta_p P^{NK}_1 ) ( r + \eta_* \eta_p P^{NF}_{1} ) ( r + \eta_q Q^{NK}_{2} ) ( r + \eta_* \eta_q Q^{NF}_{2} ) \right]^{1 \over 2} }}}
\nonumber 
\end{eqnarray}
\begin{eqnarray}
{ R = \left[ ( r + \eta_p P^{NK}_1 ) ( r + \eta_* \eta_p P^{NF}_{1} ) ( r + \eta_q Q^{NK}_{2} ) ( r + \eta_* \eta_q Q^{NF}_{2} ) \right]^{1 \over 2} }
\nonumber 
\end{eqnarray}
\begin{eqnarray}
e^{2 \phi} = \left[ { { (r+\eta_p P^{NK}_1) (r+ \eta_* \eta_p P^{NF}_1) } \over { (r+\eta_q Q^{NK}_2) (r+ \eta_* \eta_q Q^{NF}_2) } } \right]^{1 \over 2}
\nonumber 
\end{eqnarray}
\begin{eqnarray}
e^{\sigma} = \left[ { { (r+\eta_q Q^{NK}_2) (r+ \eta_* \eta_p P^{NF}_1) } \over { (r+\eta_* \eta_q Q^{NF}_2) (r+\eta_p P^{NK}_1) } } \right]^{1 \over 2}
\nonumber 
\end{eqnarray}
\begin{eqnarray}
e^{\hat 1}_1 = \left( { {r+\eta_* \eta_p P^{NF}_1} \over {r+ \eta_p P^{NK}_1} } \right)^{1 \over 2}
\nonumber 
\end{eqnarray}
\begin{eqnarray}
e^{\hat 2}_2 = \left( { {r+ \eta_q Q^{NK}_2} \over {r+ \eta_* \eta_q Q^{NF}_2} } \right)^{1 \over 2}
\nonumber 
\end{eqnarray}
\begin{equation}
e^{\hat m}_m = 1, \ \ \ \ \ m = 3,...,6 .
\label{HETSOL}
\end{equation}
I have dropped the bars on the internal metric components. Note that I always define the radial distance in such a way that the horizons of all the black holes considered in this paper are at $r=0$.

This configuration has a 10d interpretation as a gravitational wave travelling on a (winding) fundamental string which lies on a solitonic 5-brane \cite{MCT4p}, and bounded with a magnetic monople. 

Each non-zero magnetic charge ($P^{NK}_1$, $P^{NF}_1$, or both) and each non-zero electric charge ($Q^{NK}_2$, $Q^{NF}_2$, or both) creates the following spinor constraint respectively:
\begin{equation}
\varepsilon_u = i \eta_- \gamma^1 \varepsilon_l, \ \ \ \ \ \
\varepsilon_u = \eta_* \eta_+ \gamma^2 \varepsilon_l .
\label{NSSC}
\end{equation}
These constraints are the same as expected from (\ref{GSC}) with suitable definitions of the $\eta$s. 

The mass of the black hole from (\ref{HETSOL}) is
\begin{equation}
M = {1 \over 4} \left[ \eta_- \left( P^{NK}_1 + \eta_* P^{NF}_1 \right) + \eta_+ \left( Q^{NK}_2 + \eta_* Q^{NF}_2 \right) \right]
\label{NSMASS}
\end{equation}
The eight different combinations of $\eta$s correspond to the positive and negative values of the four central charges \cite{Kall} \cite{CH}: $|Q_R + P_R|, |Q_R - P_R|, |Q_L + P_L|, |Q_L - P_L|$, where $P_{R,L} \equiv P^{NK}_1 \pm P^{NF}_1$ and $Q_{R,L} \equiv Q^{NK}_2 \pm Q^{NF}_2$. The mass of a BPS state is equal to the maximum of the central charges, thus $\eta$s have to be chosen to maximize the mass given by (\ref{NSMASS}). Consequently there is no massless solution unless all charges vanish, i.e., no gauge or supersymmetry enhancement. This is in contrast with the case of $N=4$ heterotic string \cite{CK}. The difference is made by the extra freedom of maximizing the mass with $\eta_*$, which is a result of the $N=8$ supersymmetry.

With three $\eta$s and four non-zero charges, one always get a solution with a naked singularity when an odd number of charges are(is) negative. For example, a configuration with the charges that satisfy the inequalities: $P^{NK}_1, P^{NF}_1, Q^{NK}_2 \gg -Q^{NF}_2 > 0$, have all the $\eta$s equal to one in order to maximize the mass. Therefore, there is a naked singularity at $r = - Q^{NF}_2 $ from (\ref{HETSOL}). For a regular solution (${\it e.g.}$, when all four charges are positive), the mass is proportional to the sums of the four absolute values of the charges, ${\it i.e.}$, 
\begin{equation}
M = {1 \over 4} \left( | P^{NK}_1 | + | P^{NF}_1 | + | Q^{NK}_2 | + | Q^{NF}_2 | \right) .
\label{RNSMASS}
\end{equation}
When only three or fewer NS-NS charges are non-zero, the three $\eta$s are chosen in such a way that the mass is proportional to the sum of the absolute values of the charges. The horizon coincides with the singular surface of the black hole, and consequently the black hole has zero thermodynamical entropy. 

I shall study the pattern of supersymmetry breaking in detail 
\footnote{It is interesting to note that while the {\bf {field}} equations depend on the charges continuously, the {\bf spinor} constraints depend on charges discontinuously. The number of supersymmetries preserved depends on the number of non-zero charges, but not on the magnitudes of the charges.}.
When three to four charges are non-zero, one has to fix $\eta_*$ for consistency of the field equations. Half of the spinor degree of freedom thus have to be set to zero. Supersymmetry is then reduced by half. Each of the two spinor constraints in (\ref{NSSC}) reduces the remaining spinor degree of freedom by half. Thus only ${1 \over 8}$ (= ${1 \over 2^3}$) of the original $N=8$ supersymmetry is preserved. 
\footnote{In the heterotic case \cite{CYI}, configurations with three to four non-zero charges also preserve $N=1$ supersymmetry, although the heterotic string only has $N=4$ supersymmetry. The difference between the two cases is made by $\eta_*$. When it is fixed for consistency of the field equations, supersymmetry is reduced from $N=8$ to $N=4$. The patterns of supersymmetry breaking for the two cases are then essentially the same.}.

Consider the case with two charges only. If both charges are of the same type (${\it i.e.}$, both are electric or both are magnetic), $\eta_*$ has to be fixed. But only one of the two constraints in (\ref{NSSC}) remains. Therefore $N=2 (= 8 \times {1 \over 2^2})$ supersymmetry is preserved. When the two charges are of different types (${\it e.g.}$, only $P^{NK}_1$ and $Q^{NK}_2$ are non-zero) both constraints in (\ref{NSSC}) operate, but $\eta_*$ does not need to be fixed. The configuration can be considered as the solution of the KSEs with $\eta_* = 1, {\it i.e.}, (\varepsilon_u, \varepsilon_l) = (\varepsilon^+_l, \varepsilon^-_u)$ as well as the solution of KSEs with $\eta_* = -1, {\it i.e.}, (\varepsilon_u, \varepsilon_l) = (\varepsilon^+_u, \varepsilon^-_l)$. Therefore the supersymmetry transformations of the gravitinos and modulinos in (\ref{STG411}) vanish for the two different choices of sets of Killing spinors: $(\varepsilon_u, \varepsilon_l) = (\varepsilon^+_l, \varepsilon^-_u)$ and $(\varepsilon_u, \varepsilon_l) = (\varepsilon^+_u, \varepsilon^-_l)$ under the bosonic background defined in (\ref{HETSOL}). Therefore, the configuration preserves $N=2 (= 1 + 1)$ supersymmetry.

When only one charge is non-zero, only one of the two constraints in (\ref{NSSC}) remains, and $\eta_*$ need not be fixed. So the solution preserves $N=4$ ( = 8 $ \times {1 \over 2} $ ) supersymmetry.

I can conclude that the specification of each of the $\eta$s breaks ${1 \over 2}$ of the supersymmetry. In the case of only one non-zero charge, ${1 \over 2}$ of the supersymmetry is broken as only $\eta_-$ or $\eta_+$ need to be fixed (to maximize the mass), thus $N=4$ is preserved. In the case of two non-zero charges, only ${1 \over 2^2}$ of supersymmetry is preserved as we need to fix two $\eta$s, ${\it i.e.}$, $\eta_*$ and $\eta_- (\eta_+)$ if both charges are magnetic (electric), $\eta_-$ and $\eta_+$ if the charges are of different types. Therefore $N=2$ supersymmetry is preserved. With three to four non-zero charges, all three $\eta$s need to be fixed, and only ${1 \over 2^3}$ of supersymmetry is preserved, ${\it i.e.}$, $N=1$. 

We would try to find configurations that preserve $3 \over 8$ of the $N=8$ supersymmetry in this paragraph. From supersymmetry algebra, one may conclude that if there exists p different combinations of $\eta$s that give the same physical mass (p central charges coincide), then the corresponding configuration preserves $N=p$ supersymmetry. That is indeed true for the cases of $p=1, 2, 4$. However, it is not true when $p=3$. Consider a configuration with the charges: $ (P^{NK}_1, P^{NF}_1, Q^{NK}_2, Q^{NF}_2) = (P, -P, Q, P)$, where $Q > P > 0$. Such charges satisify the inequalities: $P + Q = Q_R + P_R = Q_R - P_R = Q_L + P_L > ( Q_L - P_L ) > 0$, and leads to singular configurations. The three combinations of $\eta$s that give the same physical mass ($P+Q$) are: $( \eta_*, \eta_q, \eta_p)$ = (1, 1, 1), (1, 1, -1), and (-1, 1, 1). Although these three sets of $\eta$ combinations give the same space-time metric from (\ref{HETSOL}), they have different internal metric fields, $e_1^{\hat 1}$ and $e_2^{\hat 2}$. Therefore they correspond to different configurations. After checking out all cases,  I conclude that there is no solution (static, spherically symmetric, with no axion and off diagonal internal metric elements) that preserves $3 \over 8$ of the $N=8$ supersymmetry
\footnote{A similar situation is found in the case of the heterotic string. When the central charges vanish, one gets massless black hole solutions \cite{CK}. There are two different singular configurations corresponding to the two different choices of $(\eta_+, \eta_-)$, ${\it i.e.}$, $\eta_+ = \eta_-$ or $\eta_+ = - \eta_-$. Each of the two configurations preserves $1 \over 4$ of the original $N=4$ supersymmetry, instead of one configuration preserving $N=2$ supersymmetry.}.

\subsection{Ramond-Ramond sector}

I shall solve the KSEs with only RR charges in this section. In principle, these solutions can be obtained by performing U-duality on the solution with only NS-NS charges. However, by solving the KSEs explicitly, we can study the pattern of supersymmetry breaking in detail. My classical solutions provide a way of verifying the D-brane \cite{POL} \cite{NewCon} \cite{Dnotes} \cite{Pol3} \cite{Leigh} intersection rules \cite{Larsen} \cite{Dnotes} which are defined microscopically.

From the KSEs (\ref{NSRM}) to (\ref{NSRTH}), I find that the KSEs with $\varepsilon^+_u$ on the left hand side have the same form as those with $\varepsilon^-_u$ on the left if we turn off the RR vector charges $Q^{RK}$ and $P^{RK}$. Therefore we shall first consider the case when the charges from the RR vector field vanish (non-zero charges from the RR vector field are considered in the following paragraphs). The KSEs (\ref{NSRM}) - (\ref{NSRTH}) reduces to
\begin{eqnarray}
R \sqrt{\lambda} ( - \ln e_m + {1 \over 2} \sigma )' \varepsilon_{ul} =
{1 \over 2} \left( \pm {\bf{P}}^{RF}_{mi} + i {\bf{Q}}^{mi}_{RF} \right) \gamma^{mi} \varepsilon_{lu}
\nonumber 
\end{eqnarray}
\begin{eqnarray}
R \sqrt{\lambda} \left( \ln {\lambda} + \sigma \right)' \varepsilon_{ul} =\pm {1 \over 2} {\bf P}^{RF}_{ij} \gamma^{ij} \varepsilon_{lu}
\nonumber
\end{eqnarray}
\begin{eqnarray}
R \sqrt{\lambda} \left( \ln {\lambda} - \sigma \right)' \varepsilon_{ul} = - {i \over 2} {\bf Q}^{RF}_{ij} \gamma^{ij} \varepsilon_{lu} ,
\label{RSE}
\end{eqnarray}
where $\varepsilon_{ul} = \varepsilon^{\pm}_{ul}$, and $e_i \equiv e_i^{\hat i}$. In \cite{Sen3}, the $Z_2$ element of the U-duality group mapping all the NS-NS gauge fields to RR gauge fields and vice versa for the IIA superstring compactified on $T^4$ was shown explicitly. The NS-NS charges of the configuration (\ref{HETSOL}) considered in Section IIIA, ${\it i.e.}$, $P^{NF}_1, P^{NK}_1, Q^{NK}_2$ and $Q^{NF}_2$, are mapped to the RR charges : $P_{12}, P_{34}, Q_{23}, Q_{14}$ (up to signs). With this charge assignment, the KSEs are explicitly solved and the solution is:
\begin{eqnarray}
{ \lambda = { r^2 \over { \left[ ( r + \eta_{12} P_{12} ) ( r + \eta_{34} P_{34} ) ( r + \eta_{23} Q_{23} ) ( r + \eta_{14} Q_{14} ) \right]^{1 \over 2} }}}
\nonumber 
\end{eqnarray}
\begin{eqnarray}
{ R = \left[ ( r + \eta_{12} P_{12} ) ( r + \eta_{34} P_{34} ) ( r + \eta_{23} Q_{23} ) ( r + \eta_{14} Q_{14} ) \right]^{1 \over 2} }
\nonumber 
\end{eqnarray}
\begin{eqnarray}
{ e_1 = \left[ {  { (r + \eta_{12} P_{12}) (r + \eta_{23} Q_{23}) } \over { (r + \eta_{34} P_{34}) (r + \eta_{14} Q_{14}) } } \right]^{1 \over 4} }
\nonumber 
\end{eqnarray}
\begin{eqnarray}
{ e_2 = \left[ {  { (r + \eta_{12} P_{12}) (r + \eta_{14} Q_{14}) } \over { (r + \eta_{34} P_{34}) (r + \eta_{23} Q_{23}) } } \right]^{1 \over 4} }
\nonumber 
\end{eqnarray}
\begin{eqnarray}
{ e_3 = \left[ {  { (r + \eta_{34} P_{34}) (r + \eta_{23} Q_{23}) } \over { (r + \eta_{12} P_{12}) (r + \eta_{14} Q_{14}) } } \right]^{1 \over 4} }
\nonumber 
\end{eqnarray}
\begin{eqnarray}
{ e_4 = \left[ {  { (r + \eta_{34} P_{34}) (r + \eta_{14} Q_{14}) } \over { (r + \eta_{12} P_{12}) (r + \eta_{23} Q_{23}) } } \right]^{1 \over 4} }
\nonumber 
\end{eqnarray}
\begin{equation}
{ e_{5,6} = \left[ {  { (r + \eta_{23} Q_{23}) (r + \eta_{14} Q_{14}) } \over { (r + \eta_{12} P_{12}) (r + \eta_{34} P_{34}) } } \right]^{1 \over 4} } ,
\label{RSOL}
\end{equation}
where $\eta_{12} \eta_{34} = \eta_{23} \eta_{14}$, and $\sigma = {\rm ln}\, {\rm det}\, e_m$. The dilaton does not run $( {\it i.e.} \phi = 0)$ as it only couples to the NS-NS charges from (\ref{LIIAE}).

As the RR charge carriers are D-branes in 10 dimensions \cite{POL}, the configuration (\ref{RSOL}) corresponds to the intersection of two D-2-branes and two D-4-branes (${\it i.e.}, 2 \perp 2 \perp 4 \perp 4$)
\footnote{The configuration can also be interpreted as the intersection of two M-2-branes and two M-5-branes (${\it i.e.}, 2 \perp 2 \perp 5 \perp 5$) in 11-d, i.e., as a configuration from intersecting M-branes 
\cite{Tseytlin} \cite{Papa} \cite{Pope}
. Each of the two M-5-branes is parallel to the 11th dimension, while the two M-2-branes are orthogonal to it.}. 
The two electric charges $Q_{23}$ and $Q_{14}$ are carried by two D-2-branes wrapping around the compactified toroidal directions (23) and (14), respectively \cite{Sen3}. The magnetic charges $P_{12}$ and $P_{34}$ are carried by two D-4-branes wrapping around the toroidal directions (3456) and (1256) respectively. With these identification for the directions of the D-branes, one can verify the D-brane intersection rules \cite{Dnotes}. The two D-2-branes intersect each other at a point, while the two D-4-branes intersect each other at a D-2-brane (${\it i.e.}$, (56)). Each of the D-2-branes intersect each of the D-4-branes on a D-1-brane, (${\it i.e.}$, (23) intersects (3456) and (1256) on (3) and (2) respectively, (14) intersects (3456) and (1256) on (4) and (1) respectively). The four D-branes intersect at a point (the origin). Therefore the microscopically-defined D-brane intersection rules are verified with the classical solution (\ref{RSOL}).

Each non-zero charge $P_{12}, P_{34}, Q_{41}, Q_{23}$, creates a spinor constraint,
\begin{equation}
\varepsilon_u = \eta_{12} \gamma^{12} \varepsilon_l, \ \ \ \
\varepsilon_u = \eta_{34} \gamma^{34} \varepsilon_l, \ \ \ \
\varepsilon_u = i \eta_{14} \gamma^{41} \varepsilon_l, \ \ \ \
\varepsilon_u = -i \eta_{23} \gamma^{23} \varepsilon_l ,
\label{RSC}
\end{equation}
respectively. Like the previous case with only NS-NS charges, these spinor constraints are the same as expected from (\ref{GSC}). There are only three independent constraints in (\ref{RSC}), and the four $\eta$s satisfy the relation $\eta_{12} \eta_{34} = \eta_{23} \eta_{14}$. The mass of the black hole is
\begin{equation}
M = {1 \over 4} \left( \eta_{12} P_{12} + \eta_{34} P_{34} + \eta_{23} Q_{23} + \eta_{14} Q_{14} \right) .
\label{RMASS}
\end{equation}
Like the NS-NS black holes, there is no massless solution and it is singular when an odd number of charges are negative. The mass of a regular black hole is equal to the sum of the absolute values of the charges, and the area of the horizon for black holes with four charges are non-zero.

For the pattern of supersymmetry breaking, I note that if three to four charges are non-zero, I have three independent spinor constraints from (\ref{RSC}). Each reduces the spinor degree of freedom by half. Consequently the solution preserves $N=1 (=8 \times {1 \over 2^3})$ supersymmetry. With only two non-zero charges, two of the spinor constraints from (\ref{RSC}) survive, resulting in a configuration with $N=2$ (= $8 \times {1 \over 2^2} $) supersymmetry. With just one non-zero charge, the configuration preserves $N=4$ (= $8 \times {1 \over 2}$) supersymmetry, as only one of the spinor constraints in (\ref{RSC}) survives.  

I shall check whether BPS-saturated states obtained by solving KSEs in RR sector under my working assumptions (i.e., spherical symmetric, static, only have scalars fields from dilaton and the diagonal internal metric, and zero charges from RR vector) can have more than four charges without referring to U-duality. As each magnetic charge leads to a constraint of the form, $\varepsilon_u = \eta_{ij} \gamma^{ij} \varepsilon_l$, while each electric charge creates a constraint of the form, $\varepsilon_u = i \eta_{ij} \gamma^{ij} \varepsilon_l$, I cannot have electric and magnetic charges with the same indices. Suppose I add the fifth charge $P_{13}$. The two spinor constraints from $P_{13}$ and $P_{12}$ imply that ${\varepsilon}_u$ has to be an eigenvector of $\gamma^{23}$. But $\gamma^{23}$ does not commute with $\gamma^{13}$ (from (\ref{RSC}), $\varepsilon_u$ has to be an eigenvector of $\gamma^{13}$), therefore it results in imcompatible spinor constraints. Finally, if I add $P_{56}$, no such imcompatibility occurs. However, the resulting configuration would over-constrain the spinor, ${\it i.e.}$, it only has two spinor degree of freedom. That is because the spinor constraint associated with $P_{56}$ is not derivable from (\ref{RSC}), and so there are four independent spinor constraints, thereby reducing the supersymmetry by a factor of $1 \over 2^4$. Similarly it is also impossible to add a fifth electric charge to the solution (\ref{RSOL}) consistently. Therefore, the BPS-saturated black hole solutions can at most have four non-zero charges from the Ramond-Ramond three-form fields.

\subsection{T-dual configurations}

The configuration (\ref{RSOL}) is T-dual to a more symmetrical configuration corresponding to the intersection of one D-0-brane and three D-4-branes, ${\it i.e.}$, $0 \perp 4 \perp 4 \perp 4$ \cite{Larsen}. By performing T-duality transformations
\footnote{Not the general T-duality transformations. I only consider the particular element of the T-duality group that inverse the radius of the corresponding compactified toroidal dimension.} 
on the 2nd and 3rd toroidal directions, the D-branes : (23),(14),(3456),(1256), are mapped to the D-branes : (),(1234),(2456),(1356). They carry the charges: $Q^{RK}, P_{56}, P_{13}$, and $P_{24}$, respectively. Each of the D-4-brane intersects another D-4-brane on a D-2-brane (${\it i.e.}$, (1234) intersects (2456) at (24), (2456) intersects (1356) at (56), (1234) intersects (1356) at (13)). The three D-2-branes intersect at a point, which couples to the 0-brane. As the above element of the T-duality group maps a configuration with a diagonal internal metric and zero axion to another configuration with diagonal internal metric and zero axion \cite{TDuality}, the configuration corresponding to the D-brane intersection : $0 \perp 4 \perp 4 \perp 4$, should also be a solution of the KSEs.

I have explicitly solved the KSEs with the charge assignment: $( Q^{RK}, P_{56}, P_{13}, P_{24} )$. The space-time metric, $\lambda$, has the expected form:
\begin{equation}
{ \lambda = { r^2 \over { \left[ ( r + \eta_{13} P_{13} ) ( r + \eta_{24} P_{24} ) ( r + \eta_{56} P_{56} ) ( r + \eta_q Q^{RK} ) \right]^{1 \over 2} }}}
\label{MIX1L}
\end{equation}
where $\eta_q$ depends on other $\eta$s as shown below.  The spinor constraints are: 
\begin{equation}
\varepsilon_u = \eta_1 \gamma^{13} \varepsilon_l, \ \ \ \ \
\varepsilon_u = \eta_2 \gamma^{24} \varepsilon_l, \ \ \ \ \
\varepsilon_u = \eta_5 \gamma^{56} \varepsilon_l, \ \ \ \ \
\varepsilon_u = i \eta_e \eta_q \varepsilon_l ,
\label{R2SC}
\end{equation}
where $\eta_e = \pm 1$ for $\varepsilon = \varepsilon^{\pm}$. These constraints agree with (\ref{GSC}) and associate with $P_{13}, P_{24}, P_{56}$ and $Q^{RK}$, respectively. Using the fact that $ \gamma^{13} \gamma^{24} \gamma^{56} = - \gamma^7$, and the relations (\ref{DFESPM}), I find $\eta_q = \eta_1 \eta_2 \eta_5$. There are only three independent spinor constraints in (\ref{R2SC}). Therefore the above solution contains three independent $\eta$s, and preserves $N=1 (=8 \times {1 \over 2^3})$ supersymmetry.

An observation about the electric charge from the RR vector field is made. From the KSEs view point, it is the only charge that can couple to the three RR magnetic charges ($P_{13}, P_{24}, P_{56}$) in a supersymmetric black hole solution. Any additional RR charges from the RR three-form fields would over-constrain the spinor, as has been discussed previously. The magnetic RR vector charge $P^{RK}$ can not replace $Q^{RK}$ consistently because the associated spinor constraint is not consistent with those in (\ref{R2SC}). On the other hand, from the D-brane view point, only the electric D-0-brane can couple consistently with the three intersecting magnetic D-4-branes. The D-0-brane can only carry the electric charge from RR vector, as it has no index. Therefore the KSEs' method provides a consistency check of the intersection rule of D-branes in this case.

There is yet another configuration related to the above two configurations $2 \perp 2 \perp 4 \perp 4$ and $0 \perp 4 \perp 4 \perp 4$ by T-duality. By performing T-duality on the 5th and 6th toroidal directions, the latter configuration: (),(1234),(2456),(1356), is tranformed to: (56),(123456),(24),(13), which carry the charges: $Q_{56}, P^{RK}, Q_{24},$ and $Q_{13}$, respectively. This configuration corresponds to an intersecting D-brane configuration in which three intersecting D-2-branes are all contained in a D-6-brane, i.e., $2 \perp 2 \perp 2 \subset 6$. The spinor constraints of the configuration associated with $Q_{13}, Q_{24}, Q_{56}, P^{RK}$, are the following:
\begin{equation}
\varepsilon_u = i \eta_{13} \gamma^{13} \varepsilon_l, \ \ \ \ \
\varepsilon_u = i \eta_{24} \gamma^{24} \varepsilon_l, \ \ \ \ \
\varepsilon_u = i \eta_{56} \gamma^{56} \varepsilon_l, \ \ \ \ \
\varepsilon_u = - \eta_e \eta_p \varepsilon_l 
\label{R3SC}
\end{equation}
respectively, where $\eta_e = \pm 1$ for $\varepsilon = \varepsilon^{\pm}$. These constranits agree with (\ref{GSC}). Like the previous case, only three independent spinor constraints are in (\ref{R3SC}), and $\eta_p$ = $\eta_1 \eta_2 \eta_5$. The configuration preserves $N=1 (=8 \times {1 \over 2^3})$ supersymmetry.

\subsection{Configurations with both NS-NS and RR charges}

As each NS-NS charge relates the the two-component spinors $\varepsilon^{\pm}_{ul}$ to $\varepsilon^{\mp}_{lu}$ and each RR charge relates $\varepsilon^{\pm}_{ul}$ to $\varepsilon^{\pm}_{lu}$, a configuration with both NS-NS charge(s) and RR charge(s) necessarily involve all the four types of two-component spinors: $\varepsilon^+_u, \varepsilon^+_l, \varepsilon^-_u$, and $\varepsilon^-_l$.

I may consider the relations between the two-component spinors associated with a configuration with both NS-NS charges and RR charges as follows. Starting with $\varepsilon^+_u$, the NS-NS charges relate them with $\varepsilon^-_l$ by the associated spinor constraints from (\ref{GSC}), and the RR charges relate $\varepsilon^+_u$ to $\varepsilon^+_l$. In the same way, I can start with $\varepsilon^-_u$ and put down their relations with $\varepsilon^+_l$ and $\varepsilon^-_l$ from the spinor constraints associated with the NS-NS charges and RR charges. These two sets of relations on the spinors, one starts with $\varepsilon^+_u$, the other starts with $\varepsilon^-_u$, has to be consistent. I thus get a necessary condition of getting consistent KSEs involving both NS-NS charges and RR charges. I shall illustrate such relations among the spinors with a configuration containing the charges: $P^{RF}_{12}, Q^{RF}_{23}, P^{NK}_{1},$ and $Q^{NK}_{3}$.

Starting with $\varepsilon^+_u$, I get the following spinor constraints from (\ref{GSC}) associated with the charges $P^{RF}_{12}, Q^{RF}_{23}, P^{NK}_{1},$ and $Q^{NK}_{3}$:
\begin{equation}
\varepsilon^+_u = \eta_{12} \gamma^{12} \varepsilon^+_l,\ \ \ \ \
\varepsilon^+_u = -i \eta_{23} \gamma^{23} \varepsilon^+_l, \ \ \ \ \
\varepsilon^+_u = i \eta_{1} \gamma^{1} \varepsilon^-_l, \ \ \ \ \
\varepsilon^+_u = - \eta_{3} \gamma^{3} \varepsilon^-_l 
\label{M1SCA}
\end{equation}
respectively. There are only three independent spinor constraints, and $\eta_1 \eta_3 = \eta_{12} \eta_{23}$. I can also start with $\varepsilon^-_u$, and obtain the following spinor constraints from (\ref{GSC}):
\begin{equation}
\varepsilon^-_u = \eta_{12} \gamma^{12} \varepsilon^-_l,\ \ \ \ \
\varepsilon^-_u = -i \eta_{23} \gamma^{23} \varepsilon^-_l, \ \ \ \ \
\varepsilon^-_u = i \eta_{1} \gamma^{1} \varepsilon^+_l, \ \ \ \ \
\varepsilon^-_u = - \eta_{3} \gamma^{3} \varepsilon^+_l .
\label{M1SCB}
\end{equation}
These two sets of spinor constraints, (\ref{M1SCA}) and (\ref{M1SCB}), are consistent with each other. Therefore, I have consistent KSEs with spinor constraints (\ref{M1SCA}) or equivalently (\ref{M1SCB}). The spinor constraints (\ref{M1SCA}) and (\ref{M1SCB}) are the same as what I have found after solving the KSEs explicitly.

With non-zero NK fields and RF fields only, the KSEs (\ref{NSRM}) to (\ref{NSRTH}) reduce to the following form:
\begin{equation}
R \sqrt{\lambda} ( - \ln e_m + \phi + {1 \over 2} \sigma )' \varepsilon^{\pm}_{ul} =
{1 \over 2} \left[ \left( \mp {\bf{Q}}^{NK}_m - i {\bf{P}}^m_{NK} \right) \gamma^m \varepsilon^{\mp}_{lu} + \left( \pm {\bf{P}}^{RF}_{mi} + i {\bf{Q}}^{mi}_{RF} \right) \gamma^{mi} \varepsilon^{\pm}_{lu}  \right] 
\label{MIX1A}
\end{equation}
\begin{equation}
R \sqrt{\lambda} ( \phi + {1 \over 2} \sigma )' \varepsilon^{\pm}_{ul} =
{1 \over 8} \left( \pm {\bf{P}}^{RF}_{ij} + i {\bf{Q}}_{ij}^{RF} \right) \gamma^{ij} \varepsilon^{\pm}_{lu}
\label{MIX1B}
\end{equation}
\begin{equation}
R \sqrt{\lambda} \left( \ln {\lambda} - \sigma \right) ' \varepsilon^{\pm}_{ul} = \mp {\bf Q}^{NK}_i \gamma^i \varepsilon^{\mp}_{lu} - {i \over 2} {\bf Q}_{ij}^{RF} \gamma^{ij} \varepsilon^{\pm}_{lu}  
\label{MIX1C}
\end{equation}
\begin{equation}
R \sqrt{\lambda} \left( \ln {\lambda} + \sigma \right) ' \varepsilon^{\pm}_{ul} = i {\bf P}^{NK}_i \gamma^i \varepsilon^{\mp}_{lu} \pm {1 \over 2} {\bf P}_{ij}^{RF} \gamma^{ij} \varepsilon^{\pm}_{lu} .
\label{MIX1D}
\end{equation}
Solving the KSEs (\ref{MIX1A}) to (\ref{MIX1D}) with the charges: $P^{NK}_1, Q^{NK}_3, P^{RF}_{12}$, and $Q^{RF}_{23}$, I get the following solution:
\begin{eqnarray}
\lambda = { r^2 \over { \left[ ( r + \eta_1 P_{1} ) ( r + \eta_{12} P_{12} ) ( r + \eta_3 Q_{3} ) ( r + \eta_{23} Q_{23} ) \right]^{1 \over 2} }}   
\nonumber 
\end{eqnarray}
\begin{eqnarray}
R = \left[ ( r + \eta_1 P_{1} ) ( r + \eta_{12} P_{12} ) ( r + \eta_3 Q_{3} ) ( r + \eta_{23} Q_{23} ) \right]^{1 \over 2} 
\nonumber 
\end{eqnarray}
\begin{eqnarray}
e^{2 \phi} = \left[ { { r + \eta_1 P_1 } \over {r + \eta_3 Q_3} } \right]^{1 \over 2} 
\nonumber 
\end{eqnarray}
\begin{eqnarray}
e_1 = \left[ {  { (r + \eta_{12} P_{12}) (r + \eta_{23} Q_{23}) } \over { (r + \eta_1 P_{1})^2 } } \right]^{1 \over 4} 
\nonumber 
\end{eqnarray}
\begin{eqnarray}
e_2 = \left[ {  { r + \eta_{12} P_{12} } \over { r + \eta_{23} Q_{23}  } } \right]^{1 \over 4}    
\nonumber 
\end{eqnarray}
\begin{eqnarray}
e_3 = \left[ {  {  (r + \eta_3 Q_{3})^2 } \over { (r + \eta_{12} P_{12}) (r + \eta_{23} Q_{23}) } } \right]^{1 \over 4} 
\nonumber 
\end{eqnarray}
\begin{equation}
e_{4,5,6} = \left[ {  { r + \eta_{23} Q_{23} } \over { r + \eta_{12} P_{12}  } } \right]^{1 \over 4}  ,
\label{MIX1SOL} 
\end{equation}
where $\eta_1 \eta_3 = \eta_{12} \eta_{23}$. 

The 10d interpretation of this configuration is that it is a bound state of a magnetic monople, a D-2-brane which intersects orthogonally to a D-4-brane, and a gravitational wave travelling along the intersection. Upon compactification, the D-2-brane wraps around the toroidal directions (23) and carries the charge $Q_{23}$. The D-4-brane wraps on (3456) and carries the charge $P_{12}$. The wave gives rise to the charge $Q_3^{NK}$, and the monopole carries $P_1^{NK}$. Bound states of fundamental string and D-branes have been studied in \cite{Schwarz} \cite{WitBDb}. 

The mass of the black hole is
\begin{equation}
M = {1 \over 4} \left( \eta_1 P_{1} + \eta_3 Q_{3} + \eta_{12} Q_{12} + \eta_{23} Q_{23} \right) ,
\label{MIX1MA}
\end{equation}
where $\eta_1 \eta_3 = \eta_{12} \eta_{23}$. Like previous cases, there is no massless solution and it is singular when odd number of charges are negative. The mass of a regular black hole is equal to the sum of the absolute values of the charges, and the area of the horizon for black holes with four charges are non-zero.

Now I consider the pattern of supersymmetry breaking. For convenience, I choose to consider the constraints in (\ref{M1SCA}) which relate $\varepsilon^+_u$ with $\varepsilon^-_l$ and $\varepsilon^+_l$, and the first constraint in (\ref{M1SCB}) which relate $\varepsilon^-_u$ with $\varepsilon^-_l$. The two RR constraints determine $\varepsilon^+_l$ from $\varepsilon^+_u$, and force $\varepsilon^+_u$ to be an eigenvector of $\gamma^{13}$. Therefore, there are only four spinor degree of freedom left from the 16 spinor degree of freedom contained in $\varepsilon^+_l$ and $\varepsilon^+_u$. The two NS-NS constraints determine $\varepsilon^-_l$ from $\varepsilon^+_u$, and require $\varepsilon^+_u$ to be an eigenvector of $\gamma^{13}$, again. Therefore, the four spinor constraints in (\ref{M1SCA}) contain one redundant constraints, and imply a relation between the $\eta$s, ${\it i.e.}, \eta_1 \eta_3 = \eta_{12} \eta_{23}$. No spinor degree of freedom are available from $\varepsilon^-_u$, as it is completely fixed by the first constraint in (\ref{M1SCB}) for consistency. Therefore, out of the 32 spinor degree of freedom contained in $\varepsilon^+_u, \varepsilon^+_l, \varepsilon^-_l$ and $\varepsilon^-_u$, only four are unconstrained. Hence the configuration preserves $N=1$ supersymmetry. Of course, I can also find the number of supersymmetries preserved by counting the number of independent $\eta$s, like what I did in the NS-NS black holes.

From the previous two subsections, (3.3.1) and (3.3.2), I conclude that there is no configuration with three or more NS-NS (RR) charges, while still containing at least one RR (NS-NS) charge. That is because three NS-NS (RR) charges already reduce the supersymmetry from $N=8$ to $N=1$. Further constraint from a RR (NS-NS) charge would over-constrain the spinor. The spinor constraint from a NS-NS charge has the form : $\varepsilon_l = C_1 \gamma^i \varepsilon_u$, while that from the RR charge is : $\varepsilon_l = C_2 \gamma^{ij} \varepsilon_u$ for some complex numbers $C_i$. Therefore a NS-NS constraint can never be derivable from pure RR constraints, and vice versa.

\subsection{T-dual configurations}

In this subsection, I study two configurations which are T-dual to the configuration (\ref{MIX1SOL}). The first T-dual configuration is obtained by acting on the first configuration with two T-duality transformations on the 2nd and 3rd toroidal dimensions. The new configuration then corresponds to the bound state of a magnetic monopole with charge $P^{NK}_1$, a D-0-brane with charge $Q^{RK}$, a D-4-brane with charge $P^{RF}_{13}$ and wraps around (2456) upon compactification, and a (winding) fundamental string which wraps on (3) and carries the charge $Q^{NF}_3$ upon compactification. I have solved the KSEs (\ref{NSRM}) to (\ref{NSRTH}) with the above four charges. The spinor constraints associated with these charges respectively are:
\begin{equation}
\varepsilon^+_u = i \eta_{1} \gamma^{1} \varepsilon^-_l,\ \ \ \ \
\varepsilon^+_u = i \eta_q \varepsilon^+_l, \ \ \ \ \
\varepsilon^+_u = \eta_{13} \gamma^{13} \varepsilon^+_l, \ \ \ \ \
\varepsilon^+_u = \eta_{3} \gamma^{3} \varepsilon^-_l ,
\label{M22SC}
\end{equation}
there are only three independent spinor constraints, and the $\eta$s satisfy: $\eta_1 \eta_3 = \eta_q \eta_{13}$. By including the constraint: $\varepsilon^-_u = i \eta_1 \gamma^1 \varepsilon^+_l$, I get a set of consistent spinor constraints relating all the four different types of two-component spinors. As before, the constraints (\ref{M22SC}) which I obtain by explicitly solving the KSEs, is the same as expected from (\ref{GSC}). The configuration preserves $N=1$ supersymmetry.

This configuration with charges $P^{NK}_1, Q^{RK}, P^{RF}_{13}$ and $Q^{NF}_3$ was studied in \cite{Johnson}, where the Bekenstein-Hawking entropy of the corresponding black hole is shown to coincide with the degeneracy of the corresponding stringy states. 

I can get a third configuration T-dual to the above one and also (\ref{MIX1SOL}) by acting on the above configuration with T-duality transformation on the 1st and 3rd toroidal directions. The resulting configuration corresponds to the bound state of a solitonic 5-brane wrapping around (23456) upon compactification, a D-2-brane wrapping around (13) upon compactification, a D-6-brane wrapping on the full six-torus (123456) upon compactification, and a gravitational wave running on the intersection of the solitonic 5-brane and the D-2-brane. These constituents of the bound state carry the charges: $P^{NF}_1, Q^{RF}_{13}, P^{RK}$ and $Q^{NK}_3$ respectively. The spinor constraints with these charges are consistent. The charge constraints associated with the above charges are:
\begin{equation}
\varepsilon^+_u = i \eta_{1} \gamma^{1} \varepsilon^-_l,\ \ \ \ \
\varepsilon^+_u = i \eta_{13} \gamma^{13} \varepsilon^+_l, \ \ \ \ \
\varepsilon^+_u = - \eta_{p} \varepsilon^+_l, \ \ \ \ \
\varepsilon^+_u = \eta_{3} \gamma^{3} \varepsilon^-_l ,
\label{M3SC}
\end{equation}
respectively, where $\eta_p \eta_{13} = \eta_1 \eta_3$. There are only three independent spinor constraints among the four in (\ref{M3SC}). For the consistency of the KSEs, I need the additional constraint on $\varepsilon^-_u$: $\varepsilon^-_u = \eta_p \varepsilon^-_l$. This configuration preserves $N=1$ supersymmetry. It has been studied in \cite{Tseytlin} and \cite{Mald2}.

I conclude this section by recalling from Section IIIC one implication of the spinor constraints stated in (\ref{GSC}). It is that there is no state with both $P^{RK}$ and $P^{NK}$, or both $P^{RK}$ and $Q^{NF}$, under the assumptions of spherical symmetry, time-independence, and only the dilaton and the diagonal internal metric elements are non-zero among the scalar fields. In terms of 10d bound states in string theory, that means  that there is no bound state between a D-6-brane and a monopole, and also no bound state between a D-6-brane and a fundamental string with non-zero winding number, under my assumptions.

\section{Conclusion}

In this chapter, I have found a class of BPS-saturated black hole solutions of the low energy effective supergravity Lagrangian of toroidally compactified IIA superstring in four dimensions. I have solved the Killing spinor equations (KSEs) explicitly under the assumptions of time-independence, spherical symmetry, and have turned off all the scalar fields except the dilaton and the diagonal internal metric elements. I have found a set of spinor constraints associated with different types of charges. In all the configurations considered in this paper, these rules were explicitly verified by solving the corresponding KSEs. They govern the patterns of supersymmetry breaking.

The solutions in general could carry no more than four charges under the assumptions stated above
\footnote{I expect to find solutions with five independent charges when my assumptions are relaxed \cite{CH}. The most general dyonic BPS-saturated black hole solutions are expected to be obtained by performing U-duality on the solutions with five independent charges, ${\it i.e.}$, they are the generating solutions.}.
Configurations with three to four non-zero charges preserved $N=1$ supersymmetry. Those with two charges preserved $N=2$ supersymmetry, and configurations with only one charge preserved $N=4$ supersymmetry. I found no solutions that preserved $3 \over 8$ of the $N=8$ supersymmetry. Configurations with four non-zero charges with an odd number of them negative were singular. The mass of a black hole with no naked singularity was proportional to the sum of the absolute values of the charges, and it had non-zero entropy. Solutions with fewer than four charges had zero entropies, with the horizons coinciding with the singular surfaces. Their masses were also proportional to the sum of the absolute values of their charges. 

There are three different types of solutions differentiated by the origin of their charges. They are the configurations with Neveu-Schwarz-Neveu-Schwarz (NS-NS) charges only, configurations with Ramond-Ramond (RR) charges only, and the configurations with both NS-NS charges and RR charges. The patterns of supersymmetry breaking were carefully studied in each case.

In the first case, the KSEs had the same structure as those of the toroidally compactified heterotic string \cite{CYI}. The KSEs for the IIA superstring related the upper (lower) components of the spinors originating from a ten dimensional spinor with a particular chirality, to the lower (upper) components of other spinors originating from another ten dimensional spinor with opposite chirality. I explicitly gave the map that related the spinors and charges from the compactified IIA superstring to that of the compactified heterotic string. Therefore I saw explicitly the embedding of the $N=4$ supersymmetry of heterotic string to the $N=8$ supersymmetry of IIA superstring.

In the second case, I solved the KSEs with RR charges only. The configuration which was U-dual to the NS-NS configuration found previously, was explicitly obtained. It corresponded to the intersecting D-brane configuration with two D-2-branes and two D-4-branes in ten dimensions. The T-dual configurations, one corresponded to a D-0-brane coupled to the intersection of three orthogonally intersecting D-4-branes, and one corresponded to a D-6-brane containing three intersecting D-2-branes, were also shown to be solutions of the KSEs. I studied the corresponding pattern of supersymmetry breaking. The intersection rules \cite{Dnotes} of the D-branes, which were defined in terms of individual D-brane each of which carried only one unit of charge, were verified with the classical configurations which may contain very large charges.

In the final case, I solved KSEs with both NS-NS charges and RR charges. I found three different solutions. Each contains two NS-NS charges and two RR charges, and is related to the others by T-duality. The first solution corresponded to a bound state of a monopole, a D-2-brane which orthogonally intersects a D-4-brane, and a gravitational wave running along the intersection. The second configuration corresponded to a bound state of a monopole, a D-0-brane, a D-4-brane, and a (winding) fundamental string \cite{Johnson}. The D-4-brane, the fundamental string, and the toroidal direction associated with the gauge field that supported the charge of the magnetic monopole, are all orthogonal to each other. The third configuration corresponded to a bound state of a D-6-brane, a D-2-brane which intersects orthogonally to the solitonic 5-brane, and a gravitational wave running along the intersection \cite{Tseytlin} \cite{Mald2}.

\chapter{Non-orthogonally intersecting BPS states}
\label{chapter:Non-orthogonally intersecting BPS states}

\section{Introduction and Summary}
I have made an extensive study on four dimensional BPS states of type II string theory corresponding to orthogonally intersecting configurations in ten dimensions in the last chapter. In this Chapter, I shall study a class of ten dimensional configurations that does not correspond to orthogonal intersections. 

Starting with an off-diagonal eleven dimensional metric, with no monopole and which can be reduced to a four dimensional black hole solution parametrised by four independent Neveu-Schwarz Neveu-Schwarz charges of toroidally compactified type II superstring theories, I obtained non-threshold static ten dimensional configurations for type II superstring theories preserving $1 \over 8$ of the supersymmetry. One configuration consists of a D-2-brane and a fundamental string both lying {\it partially} on a D-4-brane, and the D-0-brane in type IIA superstring theory. Another configuration consists of a fundamental string and a D-string both lying {\it partially} on a solitonic 5-brane, and the D-instanton in type IIB superstring theory. For the special configurations with one of the four charges removed and preserving $1 \over 4$ of the supersymmetry, the metrics are diagonalizable by a simple rotation. However, the D-2-brane and the fundamental string are still not in general orthogonal in the IIA configuration, and so is the D-string and the fundamental string in the IIB configuration even though they have diagonal static metrics.

The importance of non-perturbative states in string theories is hardly questioned \cite{Cvetictalk}. They are essential in establishing the conjectured duality symmetries of the five superstring theories, and the unifying nature of the underlying eleven dimensional M-theory \cite{MT}. Non-perturbative black hole solutions provide a background for investigating quantum gravity from the string perspective. In particular, one can calculate the Bekenstein-Hawking entropies of certain four and five dimensional black holes in string theory by counting the corresponding string degrees of freedom \cite{Mald1}.

The non-perturbative states are often studied as solutions of ten dimensional supergravity theories, which are the low energy limit of superstring theories, or as solutions of $N=1$ eleven dimensional supergravity \cite{Cremmers}, which is the low energy limit of the eleven dimensional M-theory. In \cite{Gaunt} \cite{HFR}, eleven dimensional metrics of various configurations corresponding to orthogonally intersecting 2-branes and 5-branes were studied. They were interpreted as intersecting ten dimensional string objects as one reduced the eleven dimensional metric on a circle. In \cite{KT}, two of the M-brane configurations were explicitly reduced on a six-torus, and were reduced to four dimensional extremal dyonic black holes with finite area of horizon. In \cite{KLC}, four dimensional black hole solutions of toroidally compactified type IIA superstring corresponding to ten dimensional orthogonally intersecting string configurations were studied extensively. These configurations have a diagonal ten-dimensional metric (after reducing the eleven dimensional theory on a circle)
\footnote{By `diagonal', I mean a diagonal metric describing the space-like isometric dimensions. Metric containing elements of the type $g_{xt}$ but vanishing $g_{ij}$ with $i,j$ being different space-like dimensions are considered as diagonal.},
 and a vanishing four dimensional axion which is dual to Neveu-Schwarz two form fields in four dimensions. The eleven dimensional supergravity also allows solutions corresponding to a 2-brane lying inside a 5-brane \cite{Dymem} \cite{2in5b}. These configurations also have a diagonal metric, but they have non-zero axions when reduced toroidally to four dimensions.

In addition to orthogonally intersecting branes in ten dimensions, string theories also allow configurations corresponding to branes intersecting at an angle. In \cite{Atangles}, configurations with multiple D-branes related by an $SU(N)$ rotations were studied and were shown to preserve supersymmetry. In \cite{Angleb}, these configurations were studied in conjunction with the study of modification of the realization of supersymmetry on D-branes in the presence of the two form fields. These configurations necessarily contain an off-diagonal metric element relating one dimension parallel to a brane and another dimension parallel to another brane, which intersects the previous brane on a third dimension. As the angle between the branes are generated by $SU(N)$ rotation, the off-diagonal metric element can be removed by a symmetry transformation or diffeomorphism. When reduced to lower dimensions, e.g., four dimensions, such an off-diagonal metric will not lead to an independent charge.

It was shown in \cite{Boosted} that by allowing gravitational waves to propagate on a generic cycle of the two torus, various non-threshold new string states could be obtained. In particular, the bound state of a  string and D-string \cite{Schwarz} could be obtained. In \cite{MoreD}, non-threshold D-brane bound states which possessed a difference in dimension of two were constructed by performing T-duality on a rotated axis on the D-brane. In \cite{Costa}, reduction of the eleven dimensional metric along a rotated axis was also considered. All of the configurations considered above were constructed from one to two D-branes. They involved off-diagonal metric elements which, like those in the metric for branes intersecting at angles, could not lead to independent charges when dimensionally reduced on a torus. 

The aim of this Chapter is to investigate the implications of the off-diagonal metric elements, which are {\bf NOT} generated by symmetry transformations or diffeomorphism on type II superstring configurations. On dimensionally reducing the eleven dimensional metric which I start with, I shall obtain a charge associated with the off-diagonal metric element. It is not related to other charges in the configuration by any symmetry
\footnote{One can always diagonalize the `internal' $i$ and $j$ dimension to get rid of the off-diagonal metric element $g_{ij}$. However, such a rotation would affect all the fields of the configuration, and the charge corresponding to $g_{ij}$ before the rotation would appear as modifications on the other charges and a new Neveu-Schwarz Neveu-Schwarz or Ramond-Ramond charge.}.

The organization of this chapter goes as follows. In Section (4.2), I shall present the off-diagonal eleven dimensional metric which I investigate in this paper. The metric describes a M-2-brane intersecting a M-5-brane, with a boost along the intersection. It also contains an off-diagonal metric element relating one direction along the M-2-brane but orthogonal to the M-5-brane, and another isometric direction orthogonal to both M-branes. I assume that the metric elements of the configuration as well as the fields carried by the string objects depend only on three space-like directions. Upon dimensional reduction along the isometric direction orthogonal to the M-branes, I obtained the ten dimensional metric obtained in \cite{CTII} without a monopole
\footnote{Monopole in fact plays a crucial role in determining the singularity structure of the four dimensional black hole after dimensionally reducing the ten dimensional configuration on torus. Comment about the monopole and the work presented here is made in Section (4.4).}.
 Therefore my eleven dimensional off-diagonal metric is guaranteed to generate independent charges, as such a property of the corresponding ten dimensional metric had been proved to possess such property.

In Section (4.3.1), I compactify the eleven dimensional metric along a direction parallel to the M-5-brane but orthogonal to the M-2-brane. After performing two T-duality
\footnote{The $R \rightarrow {1 \over R}$ transformation, not a general one.}
transformations, one on the direction along the M-2-brane but orthogonal to the M-5-brane, and another on the isometric direction orthogonal to both M-branes, I obtained a static ten dimensional configuration of the type IIA superstring. It contains a D-2-brane, a D-4-brane, a fundamental string, and the D-0-brane. The (fields of) the D-2-brane lie in such a way that in addition to a dimension orthogonal to the D-4-brane, the other dimension of the D-2-brane can be resolved into two components. One of the components is parallel to the D-4-brane, the other component is orthogonal to the D-4-brane. The fundamental string lies in a direction (the fields of) which can also be resolved into a component parallel to the D-4-brane, and another component orthogonal to it. I describe that as the string and the D-2-brane lying {\it partially} on the D-4-brane. The fundamental string and the D-2-brane are not perpendicular, though they are in the same hyperplane.

In Section (4.3.2), I compactify the eleven dimensional metric along the isometric direction orthogonal to both M-branes. By T-dualizing the resulting IIA superstring configuration along the intersection of the original M-branes, I obtained a static type IIB configuration. It contains a fundamental string, a D-string, a solitonic 5-brane, and a D-instanton. The strings lie {\it partially} on the solitonic 5-brane in the sense described above. The two strings are not perpendicular. In both Section (4.3.1) and (4.3.2), I shall discuss some special cases in which the metrics can be diagonalized by a simple rotation. However, even in the coordinate systems in which the metrics are diagonal, the D-2-brane and the fundamental string are still not perpendicular in the first configuration. And so are the D-string and the fundamental string in the second configuration. The deviation from orthogonality depends on all the charges of the configurations, as well as the three non-compact space-like dimensions which the metric elements as well as the fields carried by the string objects depend on 
\footnote{It is important to note that I ${\bf define}$ the positions of the string objects as the positions of the sources of the corresponding ${\bf fields}$.}. 
I shall give some concluding comments in Section (4.4).

\section{The Off-diagonal 11 dimensional metric}

In \cite{CTII}, the most general generating solution of the four dimensional black hole solutions in toroidally compactified heterotic string theory were obtained and proved to be exact solutions (to all orders in $\alpha'$) of string theory. Upon dimensional reduction on a six torus of the ten dimensional configuration, it was parametrised by five independent charges. Setting the charge of the monopole to zero
\footnote{I comment on the effect of the monopole in Section (4.4)},
the corresponding ten dimensional metric, Neveu-Schwarz (NS) 2-form field, and the dilaton are:
\begin{eqnarray}
ds^2_{10} &=& f [ f^{-1}F \left( 2 dt dy_2 + K dy_2^2 + 2 A dy_1 dy_2 \right) + dy_1^2
\nonumber \\
&+& f^{-1} \left( dy_3^2 + dy_4^2 + dy_5^2 + dy_6^2 \right) + dx_s dx^s ],
\nonumber 
\end{eqnarray}
\begin{eqnarray}
e^{2 \phi} = Ff,
\nonumber
\end{eqnarray}
\begin{eqnarray}
B_{2t} = -F,\ \ \ \ \ B_{1 \varphi} = -P (1-\cos \theta),\ \ \ \ \ B_{12} = -AF,
\label{hetersol}
\end{eqnarray}
where the metric is in string frame, $P$ is a constant related to $f$ given below, $f, F^{-1}$, and $K$ are harmonic functions depending on $x^s, s=1,2,3$ only, and $A$ satisfies: $\partial_s ( f^{-1} \partial^s A ) = 0$. The Hodge dual is defined only on the three dimensional $x^s$ space. Explicitly, they have the following forms:

\begin{eqnarray}
F^{-1} = 1 + {Q_2 \over r},\ \ \ \ \ K = 1 + {Q_1 \over r},
\nonumber
\end{eqnarray}

\begin{equation}
f = 1 + {P \over r},\ \ \ \ \ A = {q \over r} (1 + { P \over {2r} }),
\label{Functions}
\end{equation}
where $r^2 = x_s x^s$, $Q_1, Q_2, q$ and $P$ are integration constants (the metric is assumed to be asymptotically flat), and become four dimensional charges upon dimensional reduction. In the case of vanishing off-diagonal metric for the $y-$space, {\it i.e.}, $A \rightarrow 0$ or equivalently $q \rightarrow 0$, the configuration (\ref{hetersol}) describes a fundamental string lying on a solitonic 5-brane, with a gravitational wave propagating on it.

Upon dimensional reduction on a six-torus, the above configuration when supplemented with a monopole along $y_1$ gives five independent NS-NS charges. The NS two forms: $B_{2t}, B_{1 \varphi}$ give one electric and one magnetic charge respectively. The metric elements: $g_{2t}, g_{1x_s}$ give one electric and one magnetic Kaluza-Klein charge respectively. These four charges have obvious ten dimensional origin. They are from the fundamental string lying on $y_2$, a solitonic five brane wrapping around (23456)th toroidal directions, a gravitational wave on the intersection of the string and the 5-brane, and the ten dimensional magnetic monopole, respectively. However, the fifth charge does not have a clear ten dimensional interpretation. It comes from $g_{12}$ and $B_{12}$ in ten dimensions, and through the procedures of dimensional reduction, provides two related electric charges. Therefore the fifth charge seems to be an artifact of dimensional reduction only. We shall see that this fifth charge does have a clear ten dimensional origin in Section (4.3).

As the NS sector of the heterotic string is the same as that in the type II superstring, (\ref{hetersol}) is expected to be also a generating solution for the type II superstring. In \cite{CH}, the same configuration (with monopole) was shown to be the generating solution for the general black hole solutions (can have NS charges and Ramond charges) for toroidally compactified type II superstring theories in four dimensions. The corresponding group acting on the generating solution was $SU(8)/[SO(4)_L \times SO(4)_R]$, which has dimension 51 while general four dimensional type II black holes are parametrised by 56 charges. Therefore, the four charges obtained by dimensionally reduced (\ref{hetersol}) are independent. The off-diagonal metric, $G_{12} = AF$
\footnote{I use $G_{ij}$ to denote metric elements of the eleven dimensional metric.}, 
produces an independent charge parameter upon dimensional reduction.

We can interpret the configuration (\ref{hetersol}), which represents a solution for ten dimensional type II supergravities, in 11 dimensions where the conjectured M-theory lives and with the $N=1$ 11 dimensional supergravity as the low energy limit. The 11 dimensional uplift of (\ref{hetersol}) is:
\begin{eqnarray}
ds^2_{11} = F^{-1 \over 3} f^{2 \over 3} [ &F&f^{-1} \left( 2 dt dy_2 + K dy_2^2 + 2 A dy_1 dy_2 \right) + dy_1^2
\nonumber \\
&+& f \left( dy_3^2 + dy_4^2 + dy_5^2 + dy_6^2 \right) + F dy_7^2+ dx_s dx^s ],
\nonumber 
\end{eqnarray}
\begin{equation}
F_4 = 3 \left( dF \wedge dt \wedge dy_2 \wedge dy_7 + *df \wedge dy_1 \wedge dy_7 + dB_{12} dy_1 \wedge dy_2 \wedge dy_7 \right),
\label{Master}
\end{equation}
where $F_4$ is the four form field strength of the three form field which is part of the bosonic fields of the $N=1$ 11 dimensional supergravity. For $q = 0, i.e., A = 0$ from (\ref{Functions}), the configuration describes a M-2-brane orthogonally intersecting a M-5-brane, with a gravitational wave running along the intersection.

Dimensionally reducing (\ref{Master}) on a circle gives ten dimensional type IIA configurations. In particular, I obtain (\ref{hetersol}) by reducing (\ref{Master}) along $y_7$, which is parallel to the M-2-brane and orthogonal to the M-5-brane from 11 dimensional perspective. In the following, I shall reduce (\ref{Master}) along different directions and so obtain different type II superstring configurations. I would then use T-duality transformations to get static configurations
\footnote{This technique of getting static configurations by T-dualization was first used in \cite{CB}.}.
 I shall see that the off-diagonal metric element $G_{12}$ play an important role in these configurations.

\section{Type II configurations}

In this section, I shall obtain static configurations by reducing (\ref{Master}) on different directions and then perform the necessary T-duality transformations. 

\subsection{Reduction along $y_3$}

I obtain a type IIA superstring configuration by compactifying (\ref{Master}) along $y_3$. This direction is parallel to the M-5-brane and orthogonal to the M-2-brane. The ten dimensional IIA configuration is:
\begin{eqnarray}
ds^2_{IIA} &=& f^{1 \over 2} F^{-1 \over 2} [ f^{-1}F \left( 2 dt dy_2 + K dy_2^2 + 2 A dy_1 dy_2 \right) + dy_1^2
\nonumber \\
&+& f^{-1} \left( dy_4^2 + dy_5^2 + dy_6^2 \right) + F dy_7^2 + dx_s dx^s ]
\nonumber
\end{eqnarray}
\begin{eqnarray}
e^{2 \phi_a} = (fF)^{-1 \over 2}
\nonumber
\end{eqnarray}
\begin{eqnarray}
A_{\mu} = 0, \ \ \ \ \ B_{\mu \nu}^a = 0,
\nonumber
\end{eqnarray}
\begin{equation}
A_{27t} = F, \ \ \ \ \ A_{17 \varphi} = P (1 - \cos \theta),\ \ \ \ \ A_{127} = -AF,
\label{y3original}
\end{equation}
where $\phi_a$ is the dilaton, $B_{\mu \nu}^a$ is the NS-NS two form field, $A_{ij...k}$ are Ramond-Ramond one form fields and three form fields, for one subscript and three subscripts, respectively. 

The above configuration is not static, as it contains a non-zero metric element $g_{2t}$. With vanishing $q$, $i.e., A \rightarrow 0$ from (\ref{Functions}), the configuration (\ref{y3original}) describes a D-2-brane orthogonally intersecting a D-4-brane, plus a gravitational wave along the intersection, $i.e., 2_D \perp 4_D + \uparrow$, as expected. The strength of the off-diagonal metric, $i.e., g_{12}$
\footnote{I use $g_{ij}$ to denote the metric elements of ten dimensional IIA configurations.}, 
is always measured by $q$. With a non-zero $q$, the metric for the space ${y_1, y_2, y_4,...,y_7}$ becomes off-diagonal. If compactification is done along these six isometric directions, the compactification space would be a torus with non-trivial complex structure.

To get a static configuration, $i.e.$, to remove $g_{2t}$ in (\ref{y3original}), I perform T-duality along $y_2$ (a technique used in \cite{CB}). The maps for the T-duality transformations relating IIA and IIB configurations were given in \cite{HullT}. The IIB configuration after T-dualizing (\ref{y3original}) along $y_2$ is:
\begin{eqnarray}
ds^2_{IIB} = f^{1 \over 2} F^{-1 \over 2} [&-&f^{-1}F K^{-1} \left( dt^2 + A dt dy_1 \right) + (1-A^2 f^{-1} F K^{-1}) dy_1^2 + K^{-1} dy_2^2
\nonumber \\
&+&f^{-1} \left( dy_4^2 + dy_5^2 + dy_6^2 \right) + F dy_7^2 + dx_s dx^s ],
\nonumber 
\end{eqnarray}
\begin{eqnarray}
e^{2 \phi_b} = F^{-1} K^{-1},
\nonumber
\end{eqnarray}
\begin{eqnarray}
B_{2t}^b = - K^{-1}, \ \ \ \ \ B_{12}^b = A K^{-1},
\nonumber
\end{eqnarray}
\begin{eqnarray}
A_{7t} = F, \ \ \ \ \ A_{17} = A F,
\nonumber
\end{eqnarray}
\begin{equation}
A_{127 \varphi} = P (1 - \cos \theta),\ \ \ \ \ A_{127t} = A F K^{-1}.
\label{y3middle}
\end{equation}
The $B^b$ are NS-NS two form fields, and $A_{ij...k}$ are Ramond-Ramond two forms and four forms for two and four subscripts, respectively.

With $q = 0$, the above configuration describes a non-threshold bound state of a fundamental (winding) string, a D-string, and a D-3-brane. All are orthogonal to each other, ${\it i.e.}, 1_{fs}+1_D+3_D$. When $q$ is turned on, a gravitational wave begins to propagate orthogonally to the strings and the D-3-brane. Another D-3-brane orthogonal to the initial D-3-brane also gets built up. Note that in this configuration, the metric is actually diagonal (except $J_{1t}$
\footnote{I use $J_{ij}$ to denote ten dimensional metric elements of type IIB configurations.}, 
the gravitational wave). We also see that the four dimensional `fifth charge' described in Section (4.2) is now carried by the ten dimensional gravitational wave for non-zero $q$. 

The IIB configuration (\ref{y3middle}) is still non-static, i.e., it contains $J_{1t}$ (because of $g_{12}$ in (\ref{y3original})). I perform a second T-duality transformation along $y_1$ and obtain the following static IIA configuration:
\begin{eqnarray}
ds^2_{IIA} = f^{1 \over 2} F^{-1 \over 2} [ &-& f^{-1} F K^{-1} ( 1 + {\cal W} ) dt^2 + {\cal W} A^{-2} K dy_1^2 - 2 {\cal W} A^{-1} dy_1 dy_2 
\nonumber \\
&+& K^{-1} ( 1 + {\cal W} ) dy_2^2 + f^{-1} ( dy_4^2 + dy_5^2 + dy_6^2 ) + F dy_7^2 + dx_s dx^s ],
\nonumber
\end{eqnarray}
\begin{eqnarray}
e^{2 \phi_a} = {\cal W} A^{-2} f^{1 \over 2} F^{-3 \over 2},
\nonumber
\end{eqnarray}
\begin{eqnarray}
A_{7} = AF, \ \ \ \ \ B_{1t}^a = {\cal W} A^{-1},\ \ \ \ \ B_{2t}^a = - K^{-1} ( 1 + {\cal W} ),
\nonumber
\end{eqnarray}
\begin{equation}
A_{17t} = F (1 - {\cal W}), \ \ \ \ \ A_{27t} = A F K^{-1} (1 + {\cal W}), \ \ \ \ \ A_{27 \varphi} = -P (1 - \cos \theta),
\label{y3IIAfinal}
\end{equation}
where
\begin{equation}
{\cal W} \equiv { 1 \over { f F^{-1} K A^{-2} - 1 } }.
\label{Wdefined}
\end{equation}
When $q=0$, this configuration describes a bound state of a D-2-brane orthogonally intersecting a D-4-brane, with a fundamental string winding orthogonally to them, ${\it i.e.}, 2_D \perp 4_D + 1_{fs}$. When $q$ is turned on, the space of the isometries $y_i$s gets a non-trivial complex structure, and the D-2-brane and the fundamental string are tilted.

I now study the singularity structure of the static configuration (\ref{y3IIAfinal}). In order to have good behavior for $f, F^{-1}, K$ and $A$, I assume $Q_1, Q_2, P$ and $q$ to be positive. Both eigenvalues of the 2$\times$2 symmetric matrix with elements: $(g_{11}, g_{12}, g_{22})$ have to be positive in order to have the right signature for the ten dimensional configuration. That is equivalent to requiring ${\cal W} > 0$. It is true for all values of $P$ only when $Q_1 Q_2 > q^2$. Take $r \ll q \ll Q_1, Q_2, P$, I have ${\cal W} > 0$ only for finite value of $r$, ${\it i.e.}$, $r > { {q^2 P} \over { 4 ( Q_1 Q_2 - q^2 )} }$. Therefore, the configuration (\ref{y3IIAfinal}) is non-singular only when $P = 0$, with $Q_1 Q_2 > q^2$. With a non-zero $P$, the configuration is singular at the origin. Such a singularity structure agrees with the NS-NS configuration considered in \cite{CTII}, which is the same as (\ref{hetersol}) with a monopole supported on $y_1$. 

As the NS-NS two form field associated with the fundamental string has two components $B_{1t}$ and $B_{2t}$, the fundamental string lies at an angle on the $y_1-y_2$ ``plane''. Similarly, one of the two dimensions on which the D-2-brane is situated lies at an angle on $y_1-y_2$ ``plane'', as it has two components $A_{17t}$ and $A_{27t}$. The D-2-brane is also parallel to $y_7$. The D-4-brane is parallel to the (1456)th y-directions, just like the case with $q = 0$. Therefore both the fundamental string and the D-2-brane lie ${\it partially}$ on the D-4-brane, as the corresponding fields have components parallel to the D-4-brane
\footnote{The D-2-brane does not intersect the D-4-brane at an angle in the sense that they do not share a common direction.}
. There is also the field carried by the D-0-brane with the component $A_7$. It carries the charge $q$, as $A_7 \rightarrow {q \over r}$ asymptotically. On top of these configurations, the $y$-space has an off-diagonal metric. 

The two dimensional off-diagonal metric describing the subspace $\lbrace y_1, y_2 \rbrace$ of the space of isometric directions $\lbrace y_i, i=1,2,4,...,7 \rbrace$ can always be diagonalized by an element of $GL(2,R)$. However, only under a few circumstances can it be diagonalized by a simple rotation (an element of $SO(2) \subset GL(2,R)$) on the $y_1-y_2$ plane, implying that the metric describing the isometric directions $y_i$ in (\ref{y3IIAfinal}) is only a metric for a flat six-torus, with a rotation on the orthogonal axis $y_1$ and $y_2$. These circumstances include: (i) $P=0$, (ii) $Q_2 = 0$, (iii) $Q_1=0$, (iv) $r \rightarrow \infty$, and terms of order $({Q \over r})^2$ is neglected where $Q$ is equal to any one of the charges. The cases (i) to (iii) corresponds to configurations preserving ${1 \over 4}$ of supersymmetry. In these four cases, the metric (\ref{y3IIAfinal}) can be written as:
\begin{eqnarray}
ds^2_{IIA} = f^{1 \over 2} F^{-1 \over 2} [ &-& f^{-1} F K^{-1} ( 1 + {\cal W} ) dt^2 + {\cal W} A^{-2} r^{-2} (R_1^2 d {\tilde y}_1^2 + R_2^2 d {\tilde y}_2^2)
\nonumber \\
&+& f^{-1} ( dy_4^2 + dy_5^2 + dy_6^2 ) + F dy_7^2 + dx_s dx^s ],
\label{y3diagonal}
\end{eqnarray}
where
\begin{equation}
R_{1,2}^2 = {1 \over 2} [ r(r+Q_1) + (r+P) (r+Q_2) \mp \sqrt{ \Delta } ],
\label{y3radius}
\end{equation}
where
\begin{equation}
\Delta \equiv r^2 \lbrace [ P-Q_1+ (1+{ P \over r}) Q_2 ]^2 + q^2 (2+ {P \over r})^2 \rbrace .
\label{y3Delta}
\end{equation}
and
\begin{eqnarray}
{\tilde y}_1 = y_1 \cos \alpha  + y_2 \sin \alpha ,
\nonumber
\end{eqnarray}
\begin{equation}
{\tilde y}_2 = - y_1 \sin \alpha  + y_2 \cos \alpha ,
\nonumber
\end{equation}
where
\begin{equation}
\sin 2 \alpha \equiv q { {2 r + P} \over \sqrt{ \Delta } }.
\label{y3alpha}
\end{equation}
Note that in the two cases (i) and (iv), i.e., the cases when $P=0$ and $r \rightarrow \infty$, the angle $\alpha$ is a constant independent of $r$. However, in case (ii) and (iii), i.e., $Q_2 = 0$ and $Q_1=0$, $\alpha$ depends on $r$. Therefore we need to rotate the $y_1, y_2$ axis at different angles with different $r$ in cases (ii) and (iii) in order to bring the off-diagonal metric (\ref{y3IIAfinal}) to the diagonal form (\ref{y3diagonal}). In all four cases (case (i) to case (iv)), the angle between the fundamental string and the ${\tilde y}_1$ axis $\theta_B$ is equal to $\alpha - \theta_B^o$, where 
\begin{equation}
\tan \theta_B^o = - { {2 (r+P) (r+Q_2)} \over {q (2r+P)} }.
\label{y3thetaBo}
\end{equation}
The right hand side is the ratio of $B_{2t}$ and $B_{1t}$ given in (\ref{y3IIAfinal}). Similarly, the angle between the D-2-brane and the ${\tilde y}_1$ axis $\theta_A$ is equal to $\alpha - \theta_A^o$, where
\begin{equation}
\tan \theta_A^o =  { {q (r+P) (r+Q_2) (2r+P)} \over {2r (r+P) (r+Q_2) (r+Q_1) - q^2 (2r+P)^2 } }
\label{y3thetaAo}
\end{equation}
which is the ratio of $A_{27t}$ to $A_{17t}$. Finally, the angle between the solitonic 5-brane and the ${\tilde y}_1$ axis is $\alpha$, given by (\ref{y3alpha}). 

From (\ref{y3thetaBo}) and (\ref{y3thetaAo}), we see that the D-2-brane and the fundamental string in general are not orthogonal to each other, even though the metric of the corresponding configuration has a diagonal form (\ref{y3diagonal}). In particular, $\theta_B - \theta_A = {\pi \over 2} + {\cal O}(q)$ when $q$ is small. The deviation from orthogonality depends on all the charges of the corresponding configuration, as well as $r$. Note that the angles $\theta_B$ and $\theta_A$ were defined as the ratio of the $y_2$ components of the ${\bf fields}$ to that of the $y_1$ components. As $r \rightarrow \infty$, i.e., case (iv), (\ref{y3IIAfinal}) describes a D-2-brane lying orthogonally to a fundamental string. The D-2-brane inclines at angle $\alpha$ to the solitonic 5-brane, and is bounded with a D-0-brane.

As remarked in Section II, the four dimensional `fifth' charge associated with $g_{12}$ in (\ref{hetersol}) has no obvious ten dimensional origin in the NS-NS configuration (\ref{hetersol}). Here we see from the configuration (\ref{y3IIAfinal}) that the ten dimensional Ramond-Ramond vector field, with a component $A_7$ which approaches ${q \over r}$ asymptotically, is the carrier of that charge.

\subsection{Reduction along $y_1$}

I obtain another interesting configuration by starting with the reduction of (\ref{Master}) along $y_1$. From the 11 dimensional perspective, $y_1$ is the isometric direction orthogonal to both the M-2-brane and the M-5-brane
\footnote{In \cite{CTII}, $y_1$ is the direction supporting a magnetic monopole in ten dimensions.}.
 The resulting IIA configuration is:
\begin{eqnarray}
ds^2_{IIA} &=& f F^{-1 \over 2} [ f^{-1}F \left( 2 dt dy_2 + ( K - A^2 f^{-1} F ) dy_2^2 \right)
\nonumber \\
&+& f^{-1} \left( dy_3^2 + dy_4^2 + dy_5^2 + dy_6^2 \right) + F dy_7^2 + dx_s dx^s ],
\nonumber
\end{eqnarray}
\begin{eqnarray}
e^{2 \phi_a} = f F^{-1 \over 2},
\nonumber
\end{eqnarray}
\begin{eqnarray}
B_{27}^a = - A F, \ \ \ \ \ B_{7 \varphi}^a = P(1-\cos \theta),
\nonumber
\end{eqnarray}
\begin{equation}
A_{2} = A f^{-1} F,\ \ \ \ \ A_{27t} = F.
\label{y1original}
\end{equation}
This configuration is not static, as the metric contains a non-zero $g_{2t}$. As $q \rightarrow 0$, the configuration becomes that of describing a D-2-brane intersecting a solitonic 5-brane orthogonally, with a gravitational wave running along the intersection, $i.e., 2_D \perp 5_S + \uparrow$.

I obtain a static type IIB configuration by T-dualizing (\ref{y1original}) along $y_2$. The configuration is:

\begin{eqnarray}
ds^2_{IIB} = f F^{-1 \over 2} [ &-& {\cal W} A^{-2} dt^2 + {\cal W} A^{-2} F^{-1} dy_2^2 + {\cal W} A^{-1} dy_2 dy_7 + F ( 1 + {\cal W} ) dy_7^2 
\nonumber \\
&+& f^{-1} ( dy_3^2 + dy_4^2 + dy_5^2 + dy_6^2 ) + dx_s dx^s ],
\nonumber
\end{eqnarray}
\begin{eqnarray}
e^{2 \phi_b} = {\cal W} f^2 F^{-2} A^{-2},
\nonumber
\end{eqnarray}
\begin{eqnarray}
\chi = -A f^{-1} F, \ \ \ \ \ B_{7 \varphi}^b = P ( 1 - \cos \theta ),
\nonumber
\end{eqnarray}
\begin{eqnarray}
B_{2t}^b = - {\cal W} A^{-2} f F^{-1}, \ \ \ \ \ B_{7 t}^b = - {\cal W} A^{-1} f,
\nonumber
\end{eqnarray}
\begin{equation}
A_{2t} = -A^{-1} {\cal W}, \ \ \ \ \ A_{7t} = F (1 - {\cal W}), \ \ \ \ \ A_{27 t \varphi} = {\cal W} A^{-1} P (1 - \cos \theta),
\label{y1IIBfinal}
\end{equation}
where ${\cal W}$ has been defined in (\ref{Wdefined}). If $q=0$, the configuration describes a fundamental string lying within a solitonic 5-brane, with a D-string lying orthogonally to the 5-brane. 

The metric describing the off-diagonal part of the metric in (\ref{y1IIBfinal}) contains the elements $(J_{22}, J_{27}, J_{77})$. This 2 $\times$ 2 metric is equivalent to that in (\ref{y3IIAfinal}) if we map $(Q_1, Q_2, q)$ in (\ref{y3IIAfinal}) to $(Q_2, Q_1, q/2)$ in (\ref{y1IIBfinal}) (The condition ${\cal W} > 0$ also implies that the other metric elements cause no problem with the singularity structure, e.g., $J_{tt} < 0$ given ${\cal W} > 0$). Consequently, it has a similar singularity structure with that of (\ref{y3IIAfinal}). For non-zero $P$, it is regular only when $4 Q_1 Q_2 > q^2$ and $r > { {q^2 P} \over { 16 ( Q_1 Q_2 - q^2 )} }$ assuming $r \ll q \ll Q_1, Q_2, P$ (regular in the sense that the metric has the right signature). If $P=0$, the only condition required is $Q_1 Q_2 > q^2$. 

The configuration (\ref{y1IIBfinal}) describes a fundamental string and a D-string both lying at an angle to the solitonic 5-brane, and bounded with a D-instanton. As the fields corresponding to the fundamental string and the D-string, ${\it i.e.}$, the NS-NS two form and the Ramond-Ramond two form fields, have components parallel to the 5-brane (the $B_{2t}$ and $A_{2t}$ respectively), I say that the two strings lie {\it partially} in the 5-brane. The configuration also contains a D-instanton carrying the charge from the Ramond-Ramond scalar.

Like the configuration (\ref{y3IIAfinal}), the two dimensional off-diagonal metric with elements $(J_{22}, J_{27}, J_{77})$ in (\ref{y1IIBfinal}) is equivalent to a diagonal metric of a flat torus after a rotation in the $y_2-y_7$ plane in the special cases of (i) $P=0$, (ii) $Q_2 = 0$, (iii) $Q_1=0$, (iv) $r \rightarrow \infty$, and terms of order $({Q \over r})^2$ are neglected where $Q$ is equal to any one of the charges. The angle of the rotation and the two radius of the torus are given by (\ref{y3radius})-(\ref{y3alpha}) with the mapping $(Q_1, Q_2, q)$ in (\ref{y3IIAfinal}) to $(Q_2, Q_1, q/2)$ in (\ref{y1IIBfinal}). The angle between the NS-NS two form field (which is carried by the fundamental string) and the rotated ${\tilde y}_2$ axis is $\theta_B = \alpha - \theta_B^o$ with

\begin{equation}
\tan \theta_B^o = { {q (2r+P)} \over {2r (r+Q_2)} }.
\label{y1thetaBo}
\end{equation}
Similarly, the angle between the Ramond-Ramond two form field (carried by the D-string) and the rotated ${\tilde y}_2$ axis is $\theta_A = \alpha - \theta_A^o$ with

\begin{equation}
\tan \theta_A^o = { {q (2r+P)} \over {r (r+Q_2)} } - { {2 (r+P) (r+Q_1)} \over {q (2r+P)} }.
\label{y1thetaAo}
\end{equation}

Just like the case studied in the previous section IIIA, we see that the D-1-brane and the fundamental string in general are not orthogonal to each other, even though the metric of the corresponding configuration has a diagonal form. In particular, $\theta_B - \theta_A = {\pi \over 2} + {\cal O}(q)$ when $q$ is small. The deviation from orthogonality depends on all the charges of the corresponding configuration, as well as $r$. Again, it is important to note that the angles $\theta_B$ and $\theta_A$ were defined as the ratio of the $y_7$ components of the ${\bf fields}$ to that of the $y_2$ components. As $r \rightarrow \infty$, (\ref{y3IIAfinal}) describes a D-1-brane lying orthogonally to a fundamental string. The D-string inclines at angle $\alpha$ to the solitonic 5-brane, and is bounded with a D-instanton.

The charge of the four dimensional configuration obtained by reducing (\ref{hetersol}) associated with the ten dimensional off-diagonal metric $g_{12}$ has a clear ten dimensional origin in the static configuration (\ref{y1IIBfinal}). It is the charge carried by the D-instanton.

\section{Conclusion and Comments}

I have studied two static ten dimensional configurations for type II superstring theories with an off-diagonal metric. I started with the off-diagonal ten-dimensional metric from \cite{CTII} without the monopole. The charge associated with the off-diagonal element had been shown not to be related to other charges of the configuration by any symmetry after reduction on a six-torus. All charges of that configuration have Neveu-Schwarz Neveu-Schwarz origin. I uplifted that ten dimensional configuration to eleven dimensions, then reduced the eleven dimensional configuration along two different directions. Finally, I performed T-duality transformations to make the two ten-dimensional configurations static. I assumed all fields and metric elements only depend on three non-compact space-like directions.

The first configuration belonged to the type IIA superstring theory. It described a non-threshold bound state of the D-0-brane, the D-2-brane, the D-4-brane, and the fundamental string. I assumed that the extended objects located precisely along the fields they carried. As part of the field carried by the D-2-brane lies along the D-4-brane, I defined that the D-2-brane lies {\it partially} along the D-4-brane. Similarly, the fundamental string also lies {\it partially} along the D-4-brane. The fundamental string in general was not orthogonal to the D-2-brane. In addition, the metric of the isometries has a non-trivial complex structure due to an off-diagonal metric element. The configuration have the right signature $(-1, 1, ...,1)$ only when the radial distance $r$ is larger than certain finite value when the magnetic charge carried by the D-4-brane was non-zero.

The second configuration belonged to the type IIB superstring theory. It described a non-threshold bound state of the D-instanton, the D-string, the fundamental string, and the solitonic 5-brane. The strings were {\it partially} on the 5-brane in the sense defined above. The strings were in general not orthogonal to each other. The metric contained an off-diagonal element, and has the right signature only for finite values of $r$ when the magnetic charge carried by the solitonic 5-brane is non-zero

Removing the magnetic objects, i.e., the D-4-brane in the first configuration and the solitonic 5-brane in the second configuration, made the signature of the two metrics in the configurations right for all values of $r$ (as long as the off-diagonal metric element is not very large, i.e. $Q_1 Q_2 > q^2$). The two dimensional metric containing the off-diagonal metric element could be diagonalized by a simple rotation in the following cases: (i) the magnetic charges are zero, i.e, $P=0$, no M-5-brane to start with, (ii) no gravitational waves at the beginning, i.e., $Q_1=0$, (iii) no M-2-brane at the beginning, i.e., $Q_2=0$. The two dimensional metric in these cases became that of a flat two torus. However, the two electric objects in each case did not intersect orthogonally. The angle between the D-2-brane and the fundamental string in the first configuration and the angle between the D-string and the fundamental string in the second configuration both depended on all the charges in the corresponding configuration, as well as the radial distance. Here I {\it defined} the directions of the string objects to be the directions of the corresponding fields. Only when the off-diagonal metric element vanished, {\it e.g.}, $q=0$, could the two angles be equal to ${\pi \over 2}$.

The two dimensional off-diagonal metric in each case could also be diagonalized by a simple rotation as $r \rightarrow \infty$. In such a limit, the two electric objects in each configuration did intersect orthogonally. These electric objects, however, are still ${\it partially}$ parallel to the magnetic objects in the two configurations.

A few remarks seem appropriate here. In principle, one expects to get a third configuration by reducing (\ref{Master}) along $y_2$, the intersection of the two M-branes in eleven dimensions. Such a reduction does lead to a static configuration with a diagonal metric right away, without any T-duality transformations. However, the resulting configuration has $g_{tt} = 0$, i.e., the time-time component of the metric vanishes. Also, the Neveu-Schwarz Neveu-Schwarz configuration (\ref{hetersol}) cannot be made static by T-duality transformations as both $g_{2t}$ and $B_{2t}^a$ are non-zero, i.e., the gravitational wave propagates along the (winding) fundamental string. Therefore, the two configurations (\ref{y3IIAfinal}) and (\ref{y1IIBfinal}) are the only static configurations one can have starting from (\ref{hetersol}).

I have ignored the ten dimensional monopole in this paper totally, even though it was included in \cite{CTII}. In fact, the monopole can make the configurations with the magnetic objects (the D-4-brane and solitonic 5-brane) regular. I did not include a monopole in this study because it would make the configurations much more complicated. More importantly, it leads to metric elements of the form $g_{xt}$ where $x$ is one of the non-compact dimensions. Therefore I would not have static configurations if I started with an eleven dimensional monopole. I do not need the monopole in showing the importance of the off-diagonal metric elements.

The two configurations (\ref{y3IIAfinal}) and (\ref{y1IIBfinal}) in general preserve ${1 \over 8}$ of the supersymmetry. The configuration (\ref{hetersol}) preserves ${1 \over 4}$ of the ten dimensional $N=1$ supersymmetry of the heterotic string \cite{CTII}. The configuration without the magnetic monopole preserves the same amount of supersymmetry \cite{CYI}. The explicit embedding of the $N=4$ supersymmetry in the toroidally compactified heterotic string into the $N=8$ supersymmetry of the toroidally compactified type IIA superstring was explicitly done in \cite{KLC}. From there, with the fact that T-duality transformations preserve supersymmetry, I conclude that the configurations studied in this chapter preserve ${1 \over 8}$ of the $N=2$ supersymmetry of type II superstrings in ten dimensions.

\chapter{Constant threshold correction to dilatonic electric black holes}
\label{chapter:Constant threshold correction to dilatonic electric black holes}

\section{Introduction and Summary}

In chapters 2 to 4, I have been investigating the non-perturbative aspects of string theories by studying the BPS states. My last two chapters will be on some phenomenological aspects of the non-perturbative behavior of string theories.

In this chapter, I report on my investigation of the effect of a constant threshold correction to a general non-extreme, static, spherically symmetric, electrically charged black hole solution of the dilatonic Einstein-Maxwell Lagrangian, with an arbitrary coupling $\beta$ between the electromagnetic tensor and the dilaton field.

A dilaton, {\it i.e.}, a scalar field without self-interactions, arises naturally in basic theories that unify gravity with other interactions, including certain supergravity theories, and effective theories from superstrings. In general, it couples to the gauge field kinetic energy as well as to the matter potential. It is therefore of interest to address dilatonic topological configurations in such theories. The dilatonic charged black hole solutions without threshold correction are known \cite{Gary} \cite{Dark}. The presence of the coupling between the dilaton and the electromagnetic tensor produces black hole solutions drastically different from the Reissner-Nordstrom solutions.

In addition to the dilaton, there are moduli fields which are generically present in string theories. They are stringy modes in a vacuum associated with compactification of the extra dimensions. In general, they act as threshold corrections to the scalar function that couples to the gauge field kinetic energy \cite{LOUIS} - \cite{Lust}. Such scalar functions determine the strength of the effective gauge coupling constant.

Effects of stringy threshold corrections on charged spherically symmetric dilatonic configurations without gravity have been studied in \cite{CT}. The present work is to generalize the study to charged black hole configurations, i.e., by including gravitational effects.

I found that for a small coupling, i.e., up to first order in $\beta$, an exact analytical solution can be obtained. For an arbitrary $\beta$, a closed form solution, up to first order in the constant threshold correction, of the metric and the dilaton are presented. In the extremal limit, the closed form solution is reduced to an exact analytical form.

\section{The form of the threshold correction}

The most general bosonic Lagrangian for the dilatonic Einstein-Maxwell theory with just one moduli field is of the following form:
\begin{equation}
{\cal L} = \int d^4 x {\sqrt -g} \left[ -R + 2 \partial_{\mu}\phi \partial ^{\mu}\phi + 2 \partial_{\mu}\varphi \partial^{\mu}\varphi + f( \phi , \varphi ) F^2  - V( \phi , \varphi ) \right] ,
\label{GENERAL}
\end{equation}
where $F_{\mu \nu}$ is the electromagentic tensor field strength, and $\phi, \varphi$ are the dilaton field and the moduli field, respectively. The potential $V( \phi, \varphi)$ is expected to be non-perturbative as both the dilaton and the moduli field are flat directions of string theories when studied in terms of perturbation theory.

As a first step, I assume that the threshold correction can be approximated by a small constant, c. Therefore, I neglect the kinetic term of $\varphi$, and take the gauge coupling function as
\begin{equation}
f( \phi , \varphi ) = e^{-2 \beta \phi} + c ,
\label{GCF}
\end{equation}
where $\beta$ is an arbitrary parameter
\footnote{In \cite{Mignemi}, an analytical solution with $\beta = 1$ and a running moduli field, which is subject to a specific relation with the dilaton field, was obtained.}. 
In the compactified supergravity models associated with the low-energy limit of superstring theories, there are several different consistent truncations which give the dilatonic Einstein-Maxwell Lagrangian (of the form (\ref{GENERAL}), but without $V, c$, and $\varphi$) with $\beta = 0, {1 \over \sqrt{3}}, 1$, and $\sqrt{3}$ \cite{DLR} \cite{DR}. Here, I work with an arbitrary $\beta$, and so include more general supergravity theories. Furthermore, I will neglect the potential, $V$, and the terms with higher order derivatives. I assume that the size of the black hole is much bigger than the Planck length
\footnote{Higher order corrections, ${\it i.e.}$, the $\alpha'$ correction, is considered in \cite{Makoto}.}. 

Therefore I work on the Lagrangian 
\begin{equation}
{\cal L} = \int d^4 x {\sqrt -g} \left[ -R + 2 \partial_{\mu}\phi \partial ^{\mu}\phi + f( \phi )  F^2  \right] ,
\label{START}
\end{equation}
with $f$ taken from (\ref{GCF}).

I will study the electrically charged solution only, as I would like to work on situations in which the threshold corrections can always be treated perturbatively. For the electrically charged solution, $\phi (r) \rightarrow - \infty$ as $r \rightarrow \alpha$, where $\alpha$ is the (inner) horizon, when the threshold correction is totally neglected. So I expect that $f$ is dominated by $e^{-2 \beta \phi}$ for all values of r even when the small threshold correction is taken into account, and I can approximate the exact solution with a perturbation series in c 
\footnote{I set the asymptotic value of the dilaton to unity to simplify the formulae.}. 
On the contrary, $\phi (r) \rightarrow \infty$ as $r \rightarrow \alpha$ in a magnetically charged solution when the threshold correction is neglected. Therefore, $f$ is expected to be dominated by the threshold correction as $r$ approaches the horizon when c is non-zero, and the perturbation in c would not be justified. 

It is more convenient to work on a gauge coupling function which is normalized to unity as $r \rightarrow \infty$. Therefore, instead of taking $f$ from (\ref{GCF}), I work with $f = f_N$, with 

\begin{equation}
f_N ( \phi ) \equiv { {e^{-2 \beta \phi} + c} \over {1 + c} } .
\label{NORF}
\end{equation}
I shall first solve the Euler-Lagrange equations from the Lagrangian (\ref{START}), but with the gauge coupling function, $f = f_N$ defined in (\ref{NORF}). Then, I identify the charge of the electromagnetic field, $Q_c$, as:
\begin{equation}
Q_c = Q_o \sqrt{1+c} ,
\label{QC}
\end{equation}
where $Q_o$ is the charge when $c=0$, to get the solution of the Lagrangian (\ref{START}) with the gauge coupling function, $f$ given by (\ref{GCF}).

\section{Perturbed black holes}

I take a static, spherically symmetric ansatz for the metric:
\begin{equation}
ds^2 = - {\lambda}^2(r) dt^2 + {\lambda}^{-2}(r) dr^2 + R^2(r) d{\Omega} .
\label{METRIC}
\end{equation}
The dilaton depends on $r$ only from spherical symmetry. The electromagnetic tensor for an electric solution is:
\begin{equation}
F = {Q_c \over {R^2 f_N ( \phi ) } }  dt \ \ \bigwedge \ \ dr .
\label{TENSORN}
\end{equation}
As $f_N$ is normalized, and I expect $R^2 \rightarrow r^2$ asymptotically, so $Q_c$ is the physical electric charge of the solution. That is the advantage of taking $f = f_N$, instead of taking $f$ from (\ref{GCF}) directly. The equations of motion are
\begin{equation}
{\lambda}^2 R^2 = r^2 ( 1 - a x ) ( 1 - x ) ,
\label{LAMXR}
\end{equation}
\begin{equation}
( 1 - x ) ( 1 - a x ) \left[ ( 1 - x ) ( 1 - a x ) \varphi ' \right] ' = 2 ( 1 + c ) { Q_c^2 \over \alpha^2 }{ {Z  e^{ (1 + \beta^2) \varphi }} \over ( 1 + c e^{\beta^2 \varphi} )^2 } ,
\label{MASONE}
\end{equation}
\begin{equation}
\left[ ( 1 - x ) ( 1 - a x ) { Z' \over Z } \right]' = c e^{\beta^2 \varphi} \left[ ( 1 - x ) ( 1 - a x ) \varphi ' \right]' ,
\label{MASTWO}
\end{equation}
where
\begin{equation}
\varphi \equiv { {2 \phi} \over \beta } ,
\label{VARPHI}
\end{equation}
\begin{equation}
{Z} = {\lambda}^2 {\rm e}^{-{ \varphi }} ,
\label{DEFZ}
\end{equation}
and $ x \equiv {\alpha \over r} $, $ a \equiv {r_+ \over r_-} $, $ \alpha \equiv r_- $, where $r_-, r_+$ are respectively the inner and outer horizon. Note that $\varphi$ in the above equations are not the moduli fields mentioned at the beginning of this paper. Here $\varphi$ is defined in (\ref{VARPHI}).

For small $\beta$, {\it i.e.}, when $\beta^2$ is ignored, the equations (\ref{MASONE}) and (\ref{MASTWO}) are exactly solvable.
\begin{equation}
\lambda^2 = ( 1 - x ) ( 1 - a x ) ,
\label{RNMETRIC}
\end{equation}
\begin{equation}
\phi = { \beta \over { 1 + c } } \ln ( 1 - x ) ,
\label{RNPHI}
\end{equation}
\begin{equation}
\alpha^2 = { Q_c^2 \over a } ,
\label{RNALPHA}
\end{equation}
With $Q_c$ given by (\ref{QC}), the above is the solution of the Lagrangian (\ref{START}) with the gauge coupling function $f$ given by (\ref{GCF}). The mass of the black hole is: $M = { {1 + a} \over {2 \sqrt{a}} } Q_c$. The inner horizon $r_-$ is given by: $r_- = \alpha = { Q_c \over \sqrt{a} }$, and the outer horizon ,$r_+$, is given by: $r_+ = a r_-$. 

Therefore the mass of the black hole increases and the horizons are pushed outward when the threshold correction is taken into account. In the limit as $\beta \rightarrow 0$ with arbitrary c, I recover the expected Reissner-Nordstrom solution with a vanishing dilaton, as required by the no-hair theorem. It should be noted that as long as $\beta$ is non-zero, the singularity is still at the inner horizon (${\it i.e.}, r = r_-$). The dilaton diverges as $x \rightarrow 1$ no matter how big $c$ is, in spite of the fact that the metric $\lambda$ has the same form as the Reissner-Nordstrom metric. That is because I have neglected the terms with the order of $\beta^2$. 

I now consider the case when $\beta^2$ can be arbitrary. I use first order perturbation theory in $c$ to study the change of the metric and the dilaton by the constant threshold correction. 

I expand the metric $Z$ and the scaled dilaton $ \varphi $ around $ c = 0 $ up to 1st order
\begin{equation}
\varphi = \varphi_o + c \varphi_1 ,
\label{PERVP}
\end{equation}
\begin{equation}
Z = Z_o + c Z_1 ,
\label{PERZ}
\end{equation}
\begin{equation}
\lambda^2 = \lambda^2_o + c \lambda^2_1 ,
\label{PERLAM}
\end{equation}

From the zeroth order equations of (\ref{MASONE}) and (\ref{MASTWO}), and use (\ref{DEFZ}) to relate $Z_o$ to $\lambda^2_o$, I get the expected GHS solution
\begin{equation}
{\lambda}^{2}_0 = ( 1 - a x ) ( 1 - x )^{{1 - \beta ^2}\over {1 + \beta ^2}} ,
\label{GHSLAM}
\end{equation}
\begin{equation}
{\rm e}^{2 \phi_0} = ( 1 - x )^{{2 \beta}\over {1 + \beta ^2}} ,
\label{GHSPHI}
\end{equation}
\begin{equation}
\alpha ^2 = {1 \over a} ( 1 + \beta ^2) Q^2 ,
\label{GHSAL}
\end{equation}
\begin{equation}
Q = Q_o .
\label{GHSQ}
\end{equation}

Note that as I work on the first order corrections for the metric and the dilaton in the following, I do not consider the dependence on $c$ of the electric charge, $Q_c$, as explained in previous paragraphs. After I have gotten the solutions, I `reinsert' the c-dependence of $Q_c$ from (\ref{QC}). From (\ref{MASTWO}), the first order correction of $Z$ is:
\begin{equation}
{Z}_1(x) = {{K_1(1-ax)}\over {(1-a)(1-x)}} \ln {{1-ax}\over{1-x}} - {{2a(1-ax)}\over{(1+3 \beta^2)(1-x)}} \int^x_0 dy \left[{ {(1-y)^{{2\beta^2}\over{1+\beta^2}}} \over {1-ay}} \right] ,
\label{ZONE}
\end{equation}
One of the two integration constants expected from the 2nd order linear differential equation (\ref{MASTWO}) has been fixed by requiring that $Z_1 \rightarrow 0$ as $x \rightarrow 0$. Such a boundary condition is equivalent to requiring both $\lambda^2_1 \rightarrow 0$ and $\varphi \rightarrow 0$ asymtotically. In fact, from (\ref{DEFZ}), and (\ref{PERZ}), (\ref{PERLAM}), I find
\begin{equation}
{Z}_1 = {\rm e}^{- \varphi_o} \left( \lambda^2_1 - \varphi_1 \lambda^2_0 \right) .
\label{Z1LAM1}
\end{equation}

To fix $K_1$, I require that as $x \rightarrow {1 \over a}$, $\lambda^2_1 \over \lambda^2_o$ converges (though both $\lambda_o$ and $\lambda_1$ vanish at the horizons), which should be reasonable for perturbation theory to be applicable. So I get
\begin{equation}
K_1 = {{2(a-1)}\over {1+3\beta^2}} \left({ {a-1}\over a} \right)^{{2\beta^2}\over{1+\beta^2}} .
\label{K1FIXED}
\end{equation}
The first order equation from (\ref{MASONE}) is
\begin{equation}
{\varphi}''_1 + {{2ax-1-a}\over {(1-x) (1-ax)}} \varphi'_1 - {{2a}\over {(1-x) (1-ax)}} \varphi_1 = R(x) ,
\label{VAR2ND}
\end{equation}
where
\begin{equation}
R(x) = {2 a \over {(1+\beta^2) (1-ax)^2}} \left[ {\cal Z}_1 - 2 c (1-ax) (1-x)^{{\beta^2-1}\over {\beta^2+1}}  + { {1-a x} \over {1-x}} \right] .
\label{SOURCE}
\end{equation}
This equation has the closed form solution 
\begin{equation}
\varphi_1 (x) = K_3 \varphi^h_1 (x) + K_4 \varphi^h_2 (x) + \varphi_p (x) ,
\label{VARMAS}
\end{equation}
where $\varphi^h$ are solutions of the homogeneous equation, given by
\begin{equation}
\varphi^h_1(x) = {1 \over 2} ( 1 + a - 2 a x ) ,
\end{equation}
\begin{equation}
\varphi^h_2(x) = {6 \over(1-a)^3}\left[2(a-1)+(1+a-2ax)\ln{{1-ax}\over{a(1-x)}} \right] .
\end{equation}
They have been normalized so that as $a \rightarrow 1$, they reduce to the forms: $ 1-x $ and $ 1/(1-x)^2 $. The particular solution is
\begin{equation}
\varphi_p = \varphi^o_p - {1 \over {1 + \beta^2} } ,
\label{VARPMAS}
\end{equation}
where
\begin{equation}
\varphi_p^o (x) = {{2a} \over {3(1+\beta^2)}} \left[ - \varphi^h_1(x) \int ^x_0 dy \left( h(y) \varphi^h_2 (y) \right) + \varphi^h_2 (x) \int^x_0 dy \left( h(y) \varphi^h_1(y) \right) \right] ,
\label{VARO}
\end{equation}
\begin{equation}
h(x) = {{1-x}\over {1-ax}} \left[ {\cal Z}_1 (x) - 2c(1-ax)(1-x)^{{\beta^2-1}\over {\beta^2+1}} \right] .
\end{equation}
Requiring $\varphi( x \rightarrow 0 ) \rightarrow 0$ relate $K_3$ and $K_4$ as
\begin{equation}
K_3 = K_4 {{12}\over{(1-a)^3 (1+a)}} \left[ (1+a)\ln a - 2(a-1) \right] + { 2 \over { ( 1 + a ) ( 1 + \beta^2 )} } .
\label{K3K4}
\end{equation}
As $x = { 1 \over a }$ is equivalent to $ r = a \alpha $, which is a horizon, but not a singularity, I expect $\varphi_1$, like $\varphi_o$, does not diverge there. This fixes $K_4$ as
\begin{equation}
K_4 = {-a\over {3(1+\beta^2)}} \int ^{1\over a}_0 dx ( 1 + a - 2 a x ) \left[ { {1 - y} \over {1 - a y} } Z_1(y) - 2 ( 1 - y )^{ {2 \beta^2} \over {1 + \beta^2}} \right] .
\label{K4CLOSE}
\end{equation}
So I have obtained the first order correction in c of the dilaton from (\ref{VARMAS}) to (\ref{VARO}), while the first order correction to the metric is from (\ref{ZONE}) to (\ref{K1FIXED}). The inner horizon, $\alpha$, is still given by (\ref{GHSAL}), but with $Q = Q_c$, instead of $Q = Q_o$ in (\ref{GHSQ}). The outer horizon is given by: $r_+ = a r_-$.

In the extremal limit, {\it i.e.}, $a \rightarrow 1$, the above equations for $Z_1, \varphi_1$ have exact analytical forms as follows:
\begin{equation}
\varphi_1 = { {( 1 + \beta^2) (1 - 3 \beta^2)} \over { \beta^2 (1 - \beta^2) (1 + 3 \beta^2)} } \left[ 1 - ( 1 - x )^{ {2 \beta^2} \over {1 + \beta^2}} \right] + \left( { {2 \beta^2} \over { (1+\beta^2) (1-\beta^2)} } \right) x ,
\label{VAR1A1}
\end{equation}
\begin{equation}
Z_1 = { {1+\beta^2} \over { \beta^2 (1+3 \beta^2)} } \left[ (1-x)^{ {2 \beta^2} \over {1+\beta^2} } - 1 \right] .
\label{Z1A1}
\end{equation}
They imply
\begin{equation}
\lambda^2_1 = (1-x)^{2 \over {1+\beta^2}} \left[ { {2 \beta^2} \over { (1+\beta^2)(1-\beta^2) } } x - { {2(1+\beta^2)} \over {(1+ 3 \beta^2)(1-\beta^2)}} \left( 1-(1-x)^{ {2 \beta^2} \over {1+\beta^2}} \right) \right] .
\label{LAM1A1}
\end{equation}
In the limit of $\beta^2 \rightarrow 0$, the first order dilaton and metric from (\ref{VAR1A1}), (\ref{Z1A1}), and (\ref{LAM1A1}) agree with that obtained by expanding the metric and dilaton from (\ref{RNMETRIC}) and (\ref{RNPHI}) with $a \rightarrow 1$.

\section{Comments}

It is well known that in the extremal (BPS) limit, the non-perturbative states satisfy a ``no-force'' condition and solutions with multiple BPS states are always possible. I shall show that the extremal perturbed black holes also satisfy this condition. 

With the analytical forms (\ref{VAR1A1}) and (\ref{LAM1A1}), I find that the mass, $M$, of the black hole and the charge of the dilaton, $D$, are:
\begin{equation}
M = \alpha \left[ {1 \over { 1+\beta^2 }} + c { {\beta^2} \over { (1+3 \beta^2)(1+\beta^2) }} \right] ,
\label{MA1}
\end{equation}
\begin{equation}
D = \alpha \left[ {\beta \over { (1+\beta^2) }} - c { { \beta} \over { (1+3 \beta^2)(1+\beta^2) }} \right] ,
\label{DA1}
\end{equation}
respectively. Recall that the physical electric charge is given by $Q_c$ from (\ref{QC}), and $\alpha$ is given by (\ref{GHSAL}) with $Q \equiv Q_c$. Up to first order in $c$, I find: $ Q_1 Q_2 - M_1 M_2 - D_1 D_2 = 0 $, where $Q_i$, $M_i$, and $D_i$ are the physical electric charge, mass, and dilatonic charge of an extremal electrically charged black hole labelled by $i$, respectively. Therefore the repulsive force between any two black holes exactly balances the attractive forces from gravity and that produced by the dilaton fields. Multi-black hole solutions are thus possible in the extremal limit, just like the case with no threshold correction \cite{Dark}. 

Therefore, the dilatonic extremal black hole perturbed by a small constant threshold correction has its mass increased and its horizon pushed outward. Like the case without the threshold correction, the singularity surface coincides with the horizon, and so the black hole has zero entropy.

A direction for further investigation is to include moduli-dependence on the threshold correction, i.e., replace $f(\phi,\varphi)$ as assumed in \ref{GCF} by a more general function with explicit $\varphi$ dependence. 

I would also like to work on the magnetically charged black holes. The dilaton blows up to positive infinity at the origin in this case, making perturbation theory impossible for the small $r$ region. However, it also suggests that the space-time properties of a magnetic black hole near the origin is that of Reissner-Nordstrom black holes, as $f(\phi,\varphi)$ is dominated by the threshold correction $c$.
\chapter{Phenomenology of a superpotential with non-trivial moduli dependence}
\label{chapter: Phenomenology of a superpotential with non-trivial moduli dependence}

\section{Introduction and Summary}

In this chapter, I shall report on a study of the phenomenological and cosmological implications of a specific superpotential in a four dimensional $N=1$ supergravity theory. Such a supergravity theory can be considered as a consistent truncation and compactification of a string theory, ${\it e.g.}$, the $E_8 \times E_8$ heterotic string compactified on a $Z_3$ orbifold.

Despite being the most promising candidate for the theory of unification, string theory(ies) suffers notable phenomenological and cosmological difficulties. On the phenomenological side \cite{Quevedo}, we need mechanism(s) to break supersymmetry and lift the vacuum degeneracy (${\it i.e.}$, to fix the flat directions as well as choosing the right compactification scheme). On the cosmological side, the most popular scenario for supersymmetry breaking, gaugino condensation, has at least two serious problems. Firstly, the dilaton potentials in general cannot give inflation. Secondly, there is the cosmological moduli problem. It means that if the moduli ($S$ and $T$) are fixed by the supersymmetry breaking mechanism, then the moduli acquire masses at the electroweak scale. Then they may either overclose the universe by being stable, or destroy nucleosynthesis with their late decay (they only have gravitational interactions with the observable sector) by being unstable \cite{PJS}. I shall study the implications of a particular superpotential on these problems. 

The superpotential studied here depends on the dilaton $S$ and also a moduli $T$ which parametrizes the size of (a subspace of) the internal space of compactification. As both $S$ and $T$ are flat directions in perturbative string theory, the superpotential is expected to arise from non-perturbative effects. This study is therefore an effort to further understand the non-perturbative behavior of string theories, in phenomenological terms. 

I constrained the dependence of the superpotential on $T$ by requiring the $N=1$ theory to respect $T$-duality \cite{TDuality} \cite{CF}. The $T$-duality originates from the isometry of the compactification space. There is no fine tunning of any parameters in my superpotential. It is completely fixed by the symmetry expected from string theory. 

I made a detailed analysis of the scalar potential which depends on four variables: $Re S$, $Re T$, $Im S$ and $Im T$. With the constraints imposed from symmetry, the analysis of the extremum can be confined to a function depending on two variables only. This investigation thus generalizes previous works \cite{Font} \cite{BdeCar}-\cite{LalakNill}.

The result of the study shows that with the chosen superpotential, a supersymmetric ground state with zero cosmological constant can be obtained without any fine tunning of parameters. It raises the possibility of giving masses to the moduli fields from string theory in a supersymmetric vacuum, at a scale unrelated to the supersymmetry breaking scale. Also, the scalar potential vanishes for all values of $S$ when $T$ settles at the self-dual points, ${\it e.g.}$, the vacuum degeneracy is only partially lifted. Any further non-perturbative effects that lift the flat $S$ direction may lead to weak scale inflation suggested in \cite{Lisa}. Both findings suggest possible solution to the cosmological moduli problem \cite{PJS}. The mass of the moduli can be much larger than the weak scale, or the density of the moduli is diluted by a weak scale inflation arising from non-perturbative effect on the dilaton.

In the case when $T$ did not settle at the self-dual points, I numerically analysed the scalar potential and found that the potential had the runaway behavior common to many string theory inspired models.

\section{Motivation for the superpotential}

The superpotential I studied has the following form,
\begin{equation}
{W(S,T) = {\Omega (S) {H(T) \over{{\eta (T)}^6}}}} ,
\label{HG}
\end{equation}
where $\eta (T)$ is the the Dedekind function and $H(T)$ is an arbitrary modular invariant function which in general is a rational function of the absolute modular invariant $j(T)$ \cite{CF}. The function ${\Omega (S)}$ depends on the dilaton $S$ only, I shall discuss its functional form latter. 

I took the Kahler potential to be the model independent tree-level form \cite{WitDiRe},
\begin{equation}
{K(S,S^*,T,T^*) = - \log (S + S^*) - 3 \log (T + T^*)}
\label{KF}
\end{equation}
and neglect non-perturbative effects that might cause additional contributions to the Kahler potential. 

I shall discuss the motivation for choosing the superpotential $W(S,T)$ as in (\ref{HG}) in this section. The dependence on $S$ is not surprising. It represents the non-perturbative effect which stablilizes the dilaton and hopefully it gives the correct value for the unified gauge coupling constant, as the inverse square of the coupling is equal to the real part of $S$ at tree level. A popular scenario for the non-perturbative effect is gaugino condensation. The fact that $W$ depends on $T$ as well as $S$ deserves special attention. 

In \cite{CF}, the $T$-dependence of the superpotential was constructed so as to reproduce the expected singularities of the threshold correction, ${\it i.e.}$, the one-loop contribution to the gauge coupling function. One of the reasons for the singularities is that as $T \rightarrow \infty$, which implies the radius of compactification becomes large, the Kaluza-Klein states (produced in compactification) become light. Therefore, the effective theory which was constructed by integrating out the `massive' modes, gets contributions from these Kaluza-Klein states. These `unexpected' light states produce a linear divergence. Additional charged states which become massless at finite value of $T$ causes logarithmic divergence. It was shown in \cite{CF} that superpotential of the form (\ref{HG}) reproduces these divergences. 

The dependence of the superpotential $W$ on $T$ is also supported by the M-theory interpretation of the heterotic string theory. As described in Chapter 1, the strongly coupled $E_8 \times E_8$ heterotic theory is equivalent to the eleven dimensional M-theory compactified on an interval $S^1 / Z_2$
\footnote{The M-theory interpretation of the heterotic theory gives more support to the claim that gaugino condensation is the mechanism for supersymmetry breaking. The resulting gaugino masses are comparable with the gravitino mass, thus avoiding the problem of small gaugino mass in perturbation theory \cite{NillesYam}.}
. In \cite{LalakTom}, the effective scalar potential for gaugino condensation constructed from compactifying M-theory on $M_4 \times X \times S^1/Z_2$ was compared with that expected from the conventional approach which starts from the $N=1$ four dimensional theory. The comparison indicated that the superpotential should depend on the moduli $T$ as well as $S$. 

The superpotential $W$ with the well-motivated form (\ref{HG}) has been studied by various authors with different choices for $\Omega (S)$ and $H(T)$. In \cite{CF}, a large class of $H$ was studied, though $\Omega (S)$ is left arbitrary. With a fine tuning in a constant parameter, a supersymmetry-breaking minimum (assuming $S$ field break supersymmetry) with zero cosmological constant can be obtained. In \cite{BdeCar}, $H(T)$ is set to be a constant. Resonable vacuum expectation values for $S$ and $T$ can be obtained with a negative cosmological constant. Specific gauge groups participating in the condensation mechanism were assumed. In \cite{Moore}, $H$ was again set to be a constant. S-duality was assumed and by requiring that the potential received the same asymptotic form as that from the conventional gaugino condensation mechanism, a selection rule about the possible gauge group was found. Stabilization of the dilaton was achieved even with just one gauge group. The vacuum also had a negative cosmological constant. In \cite{LalakNill}, both S- and T-duality were assumed. By fine tuning a constant, these dualities were broken. A vacuum with zero cosmological constant and reasonable values of the vev's of $S$ and $T$ were obtained.

My study of the superpotential $W(S,T)$ is a generalization of some of the above works, in that I study the variation of the potential depending on both $S$ and $T$. Without any fine tuning, I found a supersymmetric vacuum with zero cosmological constant.

\section{The Superpotential}

I shall study the superpotential $W(S,T)$ from (\ref{HG}), with 
\begin{equation}
{H(T) = j(T) ( j(T)-1728 )} .
\label{HM}
\end{equation}
This form of T-dependence is motivated by \cite{LustMoh}. I have assumed a symmetrical form of the superpotential with respect to the zero's of the $j$ function ( $H(T)$ with $j^n$ and $(j-1728)^n$, share the same features for $n \ge 1$). 

In addition to the motivation from \cite{CF} and \cite{LustMoh}, there are other reasons to study (\ref{HM}). Firstly, the superpotential is entirely generic with no fine tunning. Secondly, as H(T) from (\ref{HM}) has multiple zeros at self-dual points ( $\it i.e.,$ $T = 1, {\rm e}^{i {\pi \over 6}}$ ), one might have a SUSY preserving vacuum with zero cosmological constant. Thirdly, dynamics of the 4 fields including Re$S$, Im$S$, Re$T$, Im$T$ points to the possibility of having one field carries out inflation, while another carries out reheating. It may be possible for Im$S$ or Im$T$ to be the source of natural inflation as they are always arguments of trigonometric functions \cite{NatInf}, while Re$S$ or Re$T$ are the source of reheating. One can stabilize S-field with the same gaugino condensation mechanism proposed in \cite{BdeCar}. With $H(T)$ from (\ref{HM}), the contribution from the $T$-sector in the scalar potential may help make a non-negative cosmological constant while stabilizing the dilaton. 

From (\ref{HG}) the scalar potential is:
\begin{equation}
{V(S,S^*,T,T^*) = {{ {{\mid \Omega (S) \mid}^2} \over {S_r {T_r}^3}} {{\mid H(T) \mid}^2 \over {\mid \eta (T) \mid}^{12}} \{ {\mid S_r {{\Omega}_s \over \Omega} - 1\mid}^2 + { {{T_r}^2 \over 3} {\mid {H' \over H } + { 3 \over { 2 \pi}} {\widehat{G}}_2 } }\mid}^2 -  3  \}   }  ,
\label{MA}
\end{equation}
with H(T) taken from (\ref{HM}). I defined $S_r \equiv S + S^*$, $T_r \equiv  T + T^*$.

\subsection{With arbitrarily fixed $T$ and $T^*$}

With a fixed $T$ and $T^*$, the extremizing condition for the $S$ field \cite{CF} is ${{\partial V} \over {\partial S}} = 0$. This gives either
\begin{equation}
{S_r {\Omega}_s - \Omega = 0   },
\label{SC0}
\end{equation}
or
\begin{equation}
{ {S_r}^2 {\Omega}_{ss} = \widetilde{E} e^{2 i \gamma} {\Omega}^*  },
\label{SCE}
\end{equation}
where
\begin{equation}
{ \widetilde{E} \equiv 2 - { {{T_r}^2 \over 3} {\mid {H' \over H} + {3 \over {2 \pi}} {\widehat{G}}_2  \mid }^2 }   },
\label{ET}
\end{equation}
and
\begin{equation}
{\gamma \equiv \arg (S_r {\Omega}_s - \Omega) },
\label{GA}
\end{equation}
Condition (\ref{SC0}) is equivalent to $D_s W = 0$, $\it i.e.$, SUSY is not broken by $S$ under this condition. As V is real, and $S$ is complex, each of the 2  conditions gives 2 equations in x ( $\equiv$ Re$S$ ) and y ( $\equiv$ Im$S$ ). Because of the choice for H from (\ref{HM}), the $S$ value that satisfies (\ref{SC0}) is ${\it not}$ always the minimum of $V$. For an explicit illustration, I assume a double gaugino condensation mechanism as described in \cite{BdeCar}, and take
\begin{equation}
{\Omega (S) = A e^{- \alpha S} + B e^{- \beta S}}
\label{SF}
\end{equation}
With this form for $\Omega (S)$, I found that both (\ref{SC0}) and (\ref{SCE}) give the same y,
\begin{equation}
{y = N {\pi \over {\beta - \alpha}}}
\label{YM}
\end{equation}
where N is an odd integer. This y in fact always minimizes V along y direction. That can be seen by substituting (\ref{SF}), (\ref{HM}) into (\ref{MA}). 
\begin{equation}
{V(x,y,T,T^*) = F(x,T,T^*) \left[ Q(x,T,T^*) + R(x,T,T^*) \cos \left( ( \beta - \alpha ) y  \right)   \right]}
\label{YV}
\end{equation}
where 
\begin{equation}
{F(T,T^*) = { {A^2 \over {2 x} } e^{-2 \alpha x}}{{\mid H(T) \mid}^2 \over {{T_r}^3 {\mid \eta (T) \mid}^6}}} 
\label{YF}
\end{equation}
\begin{equation}
{Q(x,T,T^*) = { \left( {B \over A} e^{-( \beta - \alpha ) x} \right) }^2 \left( (1 + 2 \beta x )^2 - \widetilde{E} - 1 \right) + (1 + 2 \alpha x)^2 - \widetilde{E} - 1 }
\label{YQ}
\end{equation}
\begin{equation}
{R(x,T,T^*) = 2 {B \over A} e^{- ( \beta - \alpha ) x} \left[ ( 1 + 2 \beta x ) ( 1 + 2 \alpha x ) - \widetilde{E} - 1  \right]}
\label{YR}
\end{equation}

Therefore as long as
\begin{equation}
{R(x,T,T^*) > 0}
\label{YC}
\end{equation}
(\ref{YM}) would give a y that minimize V. (\ref{YC}) is easily satisfied if we have $x,\alpha , \beta \approx O(1)$, as is the case required to stabilize x reasonably. So I fix y from (\ref{YM}) hereafter.

A few words about minimization in general seems appropriate here. In order that a specific point be a minimum of a certain function of two variables, the Hessian has to be negative, $\it i.e.$,
\begin{equation}
{{f_{xy}}^2 - f_{xx} f_{yy} < 0}
\label{HES}
\end{equation}
for a 2-variable function f(x,y), and $f_{ij} \equiv {{\partial}^2 f \over {\partial x^i \partial x^j}}$. In my case, with y fixed from (\ref{YM}), we have $V_{xy} = 0$ at any particular $T$ and $T^*$. We also know from the previous argument that y from (\ref{YM}) minimizes V along the y direction. Therefore, whenever x satisfies the x equation from either (\ref{SC0}) or (\ref{SCE}), and that x makes ${{\partial}^2 V \over {\partial x^2}} > 0$, then that x, with y from (\ref{YM}) would minimize V at that particular fixed $T$ and $T^*$. Therefore, I only need to check the sign of ${\partial}^2 V \over {\partial x}^2$.

The second derivatives at $S$ ,which satifies (\ref{SC0}), is, 
\begin{equation}
{ {1 \over 2} {{{\partial}^2 V} \over {\partial x}^2} = P(x,T,T^*) \left[ \widetilde{E} + {{{\left( \alpha + \beta + 2 x \alpha \beta  \right)}^2 - \left( 1 + \alpha + \beta + 2 x \alpha \beta \right)} \over {1 + \alpha + \beta + 2 x \alpha \beta - {1 \over {4 x^2}}}} \right] }
\label{XSD}
\end{equation}
where P is a positive definite function,
\begin{equation}
{P(x,T,T^*) = { {{\mid H(T) \mid}^2} \over {{T_r}^3 {\mid \eta (T) \mid}^6 }} 2 x A^2 e^{-2 \alpha x} {\left( {{\alpha - \beta} \over {1 + 2 x \beta}} \right)}^2 \left( 1 + \alpha + \beta + 2 x \alpha \beta - {1/{4 x^2}} \right)}
\label{PRED2X}
\end{equation}
RHS in (\ref{XSD}) is obviously positive when $\widetilde{E}$ is small. However, as $\widetilde{E}$ goes to negative infinity when $T$ approaches the self-dual points from (\ref{ET}), ${{\partial}^2 V} \over {\partial x}^2$ becomes negative. So the $S$ that satisfies (\ref{SC0}) ${\it cannot}$ always corresponds to a minimum of V. On the contrary, if I have assumed H to be a constant, (\ref{SC0}) would always be the minimization condition \cite{BdeCar}.

The x-equations from (\ref{SC0}) and (\ref{SCE}) are surprisingly simple. Let $x_0$ and $x_E$ be the x's that satisfy the 2 conditions, respectively. It is shown that
\begin{equation}
{ {B \over A} e^{-(\beta - \alpha) x_0} = { {\alpha + {1 \over {2 x_0}}} \over {\beta + {1 \over {2 x_0}}} }}
\label{X0E}
\end{equation}
and
\begin{equation}
{ {B \over A} e^{-(\beta - \alpha) x_E} = { {{\alpha}^2 - {\widetilde{E} \over {4 {x_E}^2} }} \over { {\beta}^2 - {\widetilde{E} \over {1 \over {4 {x_E}^2}}} }}}
\label{XEE}
\end{equation}
As $\mid \widetilde{E} \mid$ approaches to infinity ( when $T$ $\mapsto$ self-dual points ), $x_E \mapsto x_{BE}$
\begin{equation}
{x_{BE} = {-1 \over {\beta - \alpha}} {\log {A \over B}}}
\label{XBEE}
\end{equation}
while for small $\mid \widetilde{E} \mid$ ( actually  $2 \geq \widetilde{E} > - \infty $ from (\ref{ET}) ), $x_E \mapsto x_{SE}$
\begin{equation}
{x_{SE} = {-1 \over {\beta - \alpha}} \log { {A {\alpha}^2} \over {B {\beta}^2}}}
\label{XSEE}
\end{equation}
It is interesting to note from (\ref{XEE}) that if x is sufficiently small, $x_E$ could also be given by (\ref{XBEE}) even if $\mid \widetilde{E} \mid$ is small. In fact, when $\mid \widetilde{E} \mid$ is small, V has two maxima close to $x_{BE}$ and $x_{SE}$ and a minimum at $x_0$. With $\mid B \mid > \mid A \mid$ and $\beta > \alpha$, one always has $x_{BE} < x_0 < x_{SE}$.

As a numerical example, I chose the group (non-simple) for gaugino condensation \begin{equation}
{G = SU(N_1)_{k_1=1} \bigotimes SU(N_2)_{k_2=1}}
\label{GR}
\end{equation}
with $\left( N_1, M_1 \right)$ = (7,1), $\left( N_2, M_2 \right)$ = (8,7), where $M_i$ is the number of flavors associated with the i th group. Following \cite{BdeCar}, I found ($\alpha$, $\beta$, A, B) = (11.8435, 13.9336, -0.0144438, -2.43852). I then have ( $x_{BE}, x_0, x_{SE}$ ) = ( 2.45397, 2.53047, 2.60949 ). It is important to note that all 3 values are reasonably close to the desirable value, $\it i.e.$, 2, which would give an acceptable grand unified coupling constant. I found numerically that as $T$ approaches to the self-dual point, the minimum of V shifts from $x_0$ to $x_{BE}$, while the maximum shifts from $x_{SE}$ to $x_0$. In other words, the $S$ from condition (\ref{SC0}) changes from being the minimum of V to a saddle point as $T \mapsto e^{i {\pi \over 6}}$. 

\subsection{With arbitrarily fixed $S$ and $S^*$}

I have been considering the scalar potential from (\ref{MA}) with a fixed $T$. Now I am going to consider $V$ with a fixed $S$ and $S^*$. Without doing any derivatives, one learned that the {\it extremum of $V$ lies on the boundary of the fundamental domain on the $T$-plane.}

As shown in \cite{CF}, the fact that $V$ respects $SL(2,Z)$ which acts on $T$ implies that the extrema of V with respect to $T$ must lie on the boundary of the fundamental domain. Furthermore, since V $\mapsto \infty$ as $T$ $\mapsto$ 0 or $\infty$ ( these two points are considered the same under $SL(2,Z)$ ), I would only consider those $T$ on the curve $T = e^{i \varphi}$, with $0 \leq \varphi \leq {\pi \over 6}$. In fact, superpotentials similar to ours have been  proven to have this curve as the geodesics of the scalar potentials \cite{MCRey} (the lines with Im$T$ fixed on the boundary of the fundamental domain are not considered, as V is expected to vary insignificantly along them).

\subsection{Reducing the number of variables}

Before looking into the physical content of the scalar potential, it is interesting to see how the 4-field dependence of $V$ reduced. First I assumed $y$ ( $\equiv$ Im$S$ ) to be stabilized according to (\ref{YM}). Then with $SL(2,Z)$, I reduced the dependence on $T$ and $T^*$ to $\varphi$. Thus I have in effect reduced the independent variables from 4 to 2. A second look on conditions (\ref{SC0}) and (\ref{SCE}) tells us that these extremal conditions of $S$ are also invariant under transformation of $T$ by $SL(2,Z)$. If I write $S$ as function of $T$ and $T^*$ according to (\ref{SC0}) or (\ref{SCE}), whichever gives a smaller value of $V$, I can have $V$ as a (sectionally) $SL(2,Z)$ invariant function of only one variable, $\varphi$. However I would prefer to take $V$ as function of $x$ ( $\equiv$ Re$S$ ) and $\varphi$ ( or $N_t$ ). It is simpler and more illustrative.

\section{Cosmological implications of the scalar potential}

The most direct physical implication of the scalar potential $V$ from (\ref{MA}) is that it possesses a SUSY-preserved ground state $\it with$ zero cosmological constant. Without specifying $\Omega$, and with the fact that $j(T)$ has a triple zero at $T$ = $e^{i {\pi \over 6}}$, while $(j(T) - 1728)$ has a double zero at $T$ = 1, I have from (\ref{HM}) that, 
\begin{equation}
{H(T=1,e^{i {\pi \over 6}}) = 0} ,
\label{CR1}
\end{equation}
\begin{equation}
{H'(T=1,e^{i {\pi \over 6}}) = 0} .
\label{CR2}
\end{equation}
Consequently, the scalar potential vanishes if $T$ settles at the self-dual points. At these points, the potential lies at a minimum as well. Consider a small neighbourhood of any one of the self-dual points in T-space. The second term inside the curly bracket in (\ref{MA}) becomes very large, as H' approaches zero slower than H near the self-dual points. Thus as long as the neighbourhood is not very big, $\it e.g.$, small enough so that the second term is bigger than 3, then the curly bracket gives a positive number. 

Supersymmetry is broken if one of the auxilliary fields:
\begin{equation}
{h^i = { {\mid W \mid}  e^{{1 \over 2} K}  { \left( K_i + {W_i \over W} \right)} }  }
\label{SU}
\end{equation}
has a non-zero vev, where $i=S,T$ and the subscript means that a partial derivative is taken. Therefore (\ref{CR1}) and (\ref{CR2}) show that when $T$ = 1, $e^{i {\pi \over 6}}$, supersymmetry is preserved ( by $\it Both$ $S$ and $T$ ). As emphasized in \cite{PJS}, in order to have a supersymmetric ground state with vanishing cosmological constant, one has to solve n+1 equations for n unknowns, where n is the number of (complex scalar) fields involved, and that is a nongeneric condition. It is rather interesting that there $\it is$ such a superpotential with a clear stringy origin. 

Like most scalar potentials from supergravity, the globle minimum of $V$ $\it could$ be negative. That does not keep us from further investigating the local minimum associated with self-dual points in $T$ space. As shown in \cite{PJS}, a zero energy ground state has a lifetime much longer that the age of the Universe for decaying into hypothetical negative energy states. 

Implications of such a SUSY preserving vacuum with zero cosmological constant were discussed extensively in \cite{PJS}. With such a SUSY vacuum, moduli can get masses which do not have to be related to the SUSY breaking scale, and so the moduli problem can possibly be avoided. My scalar potential from (\ref{MA}) implies that $S$ becomes a flat direction when $T$ settles at the self-dual points. It implies that SUSY $must$ be broken in the present Universe if all flat directions are to be lifted up. The reason is the following. After $T$ and $T^*$ have settled, the scalar potential which is generated by another non-perturbative mechanism for stabilizing $S$ would get no help from the $T$ fields. Therefore $S$ cannot settle at a value that makes $h^s$ from (\ref{SU}) vanishes, otherwise that new scalar potential would settle at a vacuum with negative cosmological constant, which the equations from cosmology dynamically reject. In short, stabilizing the flat $S$ direction when $T$ settles at the self-dual points implies that SUSY must be broken. Appearence of the flat $S$ direction suggests that this unconstrained $S$ may serve as an inflaton for weak-scale inflation \cite{Lisa}
\footnote{The scalar potential is too steep to be the inflation potential.}
. 

Away from the self-dual points, the scalar potential has a global minimum at $T$ = $e^{0.215 i}$ and $S$ satisfies (\ref{SC0}), $\it i.e.$, Re$S$ = 2.53, Im$S$ = 1.50 x (odd integer) from (\ref{YM}). Though Re$S$ was stabilized at a desirable value, $h^T$ at that global minimum vanishes. Therefore both $S$ and $T$ do not break supersymmetry. The corresponding vacuum also has a negative cosmological constant equal to -4 x $10^{-20}$ in Planckian units. 

Away from the global minimum and the self-dual points, I found numerically that the minimum of the potential at fixed $T$ always occurs at a desirable value of $S \approx 2$, and decreased as $T$ approached the self-dual points. Once $T$ reaches the self-dual points, $S$ becomes a flat direction.  

\chapter{Conclusion and Outlook}
\label{chapter:Conclusion and Outlook}

The non-perturbative behavior of string theories has always been important, both theoretically and phenomenologically. Recent progress has made us realize that the five superstring theories are actually non-perturbatively connected. Therefore we see a unique candidate for the theory of unification of particle physics and gravitation, the M-theory.

This thesis reports on my efforts to further understand the non-perturbative aspects of string theories. It contains two sections. The first section (Chapter 1 to 4) is a study on the non-perturbative Bogomol'nyi-Prasad-Sommerfield (BPS)-saturated states. It occupies the major part of the thesis. The second section (Chapter 5 to 6) is a study on some phenomenological aspects of the non-perturbative behavior of string theories. In this last chapter of the thesis, I shall summarize the main results of my investigations and give an outlook to future directions.

\section{BPS-saturated States}

I presented a review on the status and the theory behind the non-perturbative Bogomol'nyi-Prasad-Sommerfield (BPS)-saturated states in Chapter 1. My investigations on BPS-saturated states went from Chapter 2 up to Chapter 4.

A special class of four dimensional BPS-saturated states in the toroidally compactified heterotic string was studied in Chapter 2 \cite{CK}. These BPS states are labelled by the charge vectors associated with the gauge fields of the compactified heterotic string. I identified the electrically charged BPS states which preserve $1\over 2$ of  $N=4$ supersymmetries. They become massless along the hyper-surfaces of enhanced gauge symmetry of the two-torus moduli sub-space. As the mass spectrum of the BPS states is S-invariant, {\it i.e.}, the same spectrum can be found in the strongly-coupled theory, the gauge symmetry enhancement at special points (or hyper-surfaces) of moduli space happens perturbatively as well as non-perturbatively when the theory is strongly-coupled. I also identified the dyonic BPS states which preserve ${1\over 4}$ of $N=4$ supersymmetries. They become massless at two points with the maximal gauge symmetry enhancement. It suggested the possibility of supersymmetry enhancement.

The possibility of supersymmetry enhancement has to be studied in more detail. For the particular two-torus moduli subspace of the toroidally compactified heterotic string, there are four dyonic massless BPS states at the point $T=U=1$, suggesting a supersymmetry enhancement from $N=1$ to $N=5$. At the point $T=U=e^{i {\pi \over 6}}$, there are 12 massless dyonic states, suggesting a supersymmetry enhancement from $N=1$ to $N=13$. A four dimensional supergravity theory with $N > 8$ supersymmetries contains particles with spin greater than the spin of graviton (which is equal to two).

To understand such a large increase of supersymmetry, I have studied the stability of the dyonic BPS states. From consideration of the mass formula alone, I found that the dyonic states are stable against decaying into pure electric and pure magnetic states when they are not on a specific hyper-surface in the moduli space. They are metastable along that specific hypersurface. Further study of stability would be required to address the question of supersymmetry enhancement.

In Chapter 3, I presented a systematic study of four dimensional BPS-saturated states in the toroidally compactified type II string \cite{KLC}. The Killing spinor equations were explicitly solved. These BPS-saturated states are four dimensional black holes. They correspond to orthogonally intersecting ten dimensional BPS-saturated states in the uncompactified theory, often with Kaluza-Klein monopoles. They are parametrized by four charges and preserve ${1 \over 8}$ of the $N=8$ supersymmetries. 

I presented a set of simple relations between each charge of the black holes and the corresponding spinor constraint. This set of relations allows us to associate directly the pattern of supersymmetry breaking with the types of charges the black hole contains. 

The embedding of the $N=4$ supersymmetries of the toroidally compactified heterotic string into the $N=8$ supersymmetries of the type II superstring was explicitly shown. I showed that the $N=4$ supersymmetries are embedded in such a way that both types of Killing spinors with different chiralities from the ten dimensional point of view are involved, in contrast to the obvious guess that only spinors of certain ten dimensional chirality are involved.

The fact that there is no massless black hole in the type II theory was explained in terms of the amount of supersymmetries in the theory. Comparing with the heterotic $N=4$ theory, which has massless black holes at points of symmetry enhancement in the corresponding moduli space, the $N=8$ supersymmetries of the type II theory bring a new degree of freedom ($\eta_* = \pm 1$) in the black hole configurations. That keeps the theory from having massless black holes.

I presented the configurations with only Ramond-Ramond (R-R) charges, configurations with both Neveu-Schwarz-Neveu-Schwarz (NS-NS) and R-R charges, and also configurations with only NS-NS charges. I gave in detail the ten dimensional interpretations of these states. The R-R configurations are intersections of D-branes, while the configurations with both NS-NS and R-R charges are intersections of different BPS-saturated states in the string theory.

While the calculation of five dimensional black hole entropy from stringy degrees of freedom is well-studied, it remains a challenge to calculate in detail the entropy of four dimensional black holes from string theory. The black holes with four R-R charges have a non-zero area at the horizon in four dimensions. They correspond to orthogonal intersections of D-branes, which have simple and exact descriptions in conformal field theory. Therefore, they are candidates for detailed black hole entropy calculations. Three such configurations were studied in Chapter 3. The first one is the intersection of two D-2-brane and two D-4-branes, {\it i.e.}, $2 \perp 2 \perp 4 \perp 4$. The other two are $0 \perp 4 \perp 4 \perp 4$ and $2 \perp 2 \perp 2 \subset 6$. 

Another direction for the investigation is to relax my working assumptions and find the most general generating solution for the four dimensional black hole solutions of the $N=8$ theory. The most general generating solution is parametrized by five independent charges, as expected from the symmetry argument. It is quite plausible that the only assumption I have to relax is the assumption of zero axion.

The most general generating solution of the four dimensional BPS-saturated configurations for type II strings, which is parametrized by five independent NS-NS charges, has been found in \cite{CTII}. In Chapter 4, I removed the Kaluza-Klein monopole of the five-charge configuration and obtained static four-charge BPS-saturated configurations in ten dimensions after performing duality transformations. The configurations correspond to non-orthogonal intersections of BPS-saturated states. One of the four charges parametrized the deviation from non-orthogonality. 

My first configuration describes the bound state of a D-4-brane, a D-2-brane, a fundamental string, and a D-0-brane of the type IIA string. The D-2-brane intersects the fundamental string non-orthogonally on a plane with non-trivial complex structure. They both have a component along the D-4-brane. The charge which signified a tilted torus in the NS-NS configuration studied in \cite{CTII}, is the charge carried by the D-0-brane in my configuration.

My second configuration describes the bound state of a fundamental string, a D-1-brane, a D-instanton ({\it i.e.,}D-(-1)-brane), and a solitonic five-brane of type IIB string. The fundamental string and the D-string intersect non-orthogonally on a plane with non-trivial complex structure. They both have a component along the solitonic five-brane. The D-instanton carries the charge which signifies the tilted torus in \cite{CTII}.

Upon removal of one further charge, the metrics of the three-parameter configurations could be diagonalized by a $SO(2)$ rotation on a plane in ten dimensions. However, by identifying the states as the sources of the gauge fields, the BPS-saturated configurations still composed of various BPS-saturated states intersecting non-orthogonally, even though the metrics are diagonal. 

The ten dimensional configurations I obtained in Chapter 4 are static, yet they are not the most general generating solutions. That is because I removed the magnetic monopole at the beginning of the construction. I found that with the magnetic monopole, no static configurations can be found. The configurations upon toroidal compactification become rotating black holes. My configurations with only four charges are at best the most general generating solution in five dimensions upon compactifying the ten dimensional theory on a five torus. 

However, given that the five parameter solution studied in \cite{CTII} is the most general generating solution in four dimensions, it is still worthwhile to study the corresponding rotating black holes with the magnetic monopole in place. An even more challenging question is to find or disprove the existence of a static five parameter solution. Also, my configurations obtained in Chapter 4 always contain fundamental strings. It remains to check whether there exists configurations with only five independent R-R charges.

\section{A few aspects in the phenomenology of non-perturbative behavior}

My report on the study of BPS-saturated states ends at Chapter 4. Chapter 5 to 6 describe my study on the phenomenological consequences of the non-perturbative effects of the moduli fields, which specify the compactification scale of the corresponding string theories. 

The moduli fields in string theories lie on flat directions in perturbation theory, {\it i.e.}, the ground state energy in perturbative string theories do not depend on the vacuum expectation values of these moduli fields. Non-perturbative effects arise as threshold correction and are expected to lift the degeneracy. 

The gauge coupling function, which is the coefficient of the kinetic terms of the gauge fields, depends on both the dilaton and moduli. Therefore it is affected by the non-perturbative effects significantly. In Chapter 5, I took a step towards understanding the effect of threshold corrections to the electric dilatonic-Einstein-Maxwell black holes \cite{KLCthres}, which could be considered as a solution to a consistent truncation of the compactified string theory. As a first step, I approximated the threshold correction by a constant. 

The coupling $\beta$ between the electromagnetic tensor and the dilaton field could be arbitrary. For a small $\beta$, an exact analytical solution was obtained. For an arbitrary $\beta$, a closed form solution, up to first order in the constant threshold correction, of the metric and the dilaton were presented. In the extremal limit, the closed form solution is reduced to an exact analytical form. I showed that the mass of an extremal dilatonic black hole was increased, and its horizon was pushed outward by the threshold correction. Like the case with no threshold correction, the perturbed black holes exerted no force on each other. Therefore multi-black hole solution could be constructed.

An immediate generalization of the work presented in Chapter 5 is to give the threshold correction a functional form depending on the moduli. The functional form can be constrained by $T$-duality.

Another direction to be explored, perhaps even more interesting than the generalization mentioned in the previous paragraph, is to investigate the effect of threshold corrections on magnetically charged dilatonic black holes. As the dilaton blows up to positive infinity at the horizon in the theory without threshold corrections, a non-zero threshold correction may dominate the gauge coupling function near the horizon. Assuming higher order corrections (in $\alpha '$) are not significant in the region we are interested in, the space-time structure near the horizon would become Reissner-Nordstrom type, which is very different from that of the dilaton black holes with no threshold correction. However, we also have to study the quantum corrections which increase with the blowing up dilaton near the horizon.

I investigated the cosmological and phenomenological consequences of a non-perturbatively induced superpotential of a $N=1$ supergravity theory in Chapter 6. Such a supergravity theory could be a consistent truncation of a compactified string theory. The superpotential depends on both the dilaton and the moduli. I used symmetry expected from their string theory origin to constrain the functional form. In particular, I assumed that the superpotential respects $T$-duality (isometry of the compactification space).

I carried out a detailed analysis, both analytically and numerically, on the scalar potential obtained from the superpotential which depends on two complex fields, the dilaton and the moduli. At the self-dual points of the moduli fields, the scalar potential vanished, leading to a supersymmetric vacuum with zero cosmological constant. The dilaton field was a flat direction with the moduli field at the self-dual points. These suggest possible solutions to the cosmological moduli problem. The mass of the moduli can be obtained at a scale much larger than the weak scale, and the density of the moduli can be diluted by a weak scale inflation arising from non-perturbative effect on the dilaton.

When the moduli field does not settle at the self-dual points, I have checked numerically that the minimum of the scalar potential at a fixed vev of $T$ always occurs at a desirable value of the dilaton $S \approx 2$ (which is expected as the unification coupling constant), and decreased as $T$ approached the self-dual points. Once $T$ reaches the self-dual points, $S$ becomes a flat direction.

In order to study in detail the possibility of solving the cosmological moduli problem with the scalar potential, we have to find the correct scale at which the scalar potential is induced non-perturbatively. Once the scale is obtained from cosmological consideration, {\it i.e.}, it is tailored to avoid the cosmological moduli problem, one can work backward to study the possible non-perturbative mechanism responsible for the scalar potential.

\printindex

\end{document}